\documentclass[submission,copyright,creativecommons]{eptcs}
\providecommand{\event}{QPL 2026}

\usepackage{amsmath}
\usepackage{amsfonts}
\usepackage{amssymb}
\usepackage{amsthm}
\usepackage{framed} 
\usepackage{thmtools} 
\usepackage{subcaption}

\usepackage{tikz}
\usetikzlibrary{quantikz2}

\usepackage{algorithm} 
\usepackage[noend]{algpseudocode} 

\usepackage{xfrac} 
\usepackage{lmodern} 
\usepackage{anyfontsize} 
\usepackage{braket} 
\usepackage{xspace} 

\usepackage{xcolor} 

\usepackage{hyperref}
\usepackage[capitalise]{cleveref}

\DeclareMathOperator{\spn}{span}

\DeclareFontFamily{U}{mathx}{\hyphenchar\font45}
\DeclareFontShape{U}{mathx}{m}{n}{
      <5> <6> <7> <8> <9> <10>
      <10.95> <12> <14.4> <17.28> <20.74> <24.88>
      mathx10
      }{}
\DeclareSymbolFont{mathx}{U}{mathx}{m}{n}
\DeclareFontSubstitution{U}{mathx}{m}{n}

\DeclareMathSymbol{\bigboxtimes}{1}{mathx}{"D2}

\newcommand{\HqcHDef}{\text{E(1)}\xspace}
\newcommand{\HqcTargDef}{\text{E(2)}\xspace}
\newcommand{\HqcOPhaseDef}{\text{E(3)}\xspace}
\newcommand{\HqcTargMDef}{\text{E(4)}\xspace}
\newcommand{\HqcSwapDef}{\text{E(5)}\xspace}
\newcommand{\HqcEuler}{\text{E(6)}\xspace}
\newcommand{\HqcPOrd}{\text{E(7)}\xspace}
\newcommand{\HqcCtrlId}{\text{E(8)}\xspace}
\newcommand{\HqcCtrlComp}{\text{E(9)}\xspace}
\newcommand{\HqcCtrlSwap}{\text{E(10)}\xspace}
\newcommand{\HqcFreePhase}{\text{E(11)}\xspace}
\newcommand{\HqcCtrlDag}{\text{E(12)}\xspace}
\newcommand{\HqcCtrlPair}{\text{E(13)}\xspace}
\newcommand{\HqcCtrlExpand}{\text{E(14)}\xspace}
\newcommand{\HqcCtrlMerge}{\text{E(15)}\xspace}
\newcommand{\HqcPowZero}{\text{E(16)}\xspace}
\newcommand{\HqcPowId}{\text{E(17)}\xspace}
\newcommand{\HqcPowAdd}{\text{E(18)}\xspace}
\newcommand{\HqcPowIdem}{\text{E(19)}\xspace}
\newcommand{\HqcPowDag}{\text{E(20)}\xspace}
\newcommand{\HqcPowExpand}{\text{E(21)}\xspace}
\newcommand{\HqcPowCong}{\text{E(22)}\xspace}
\newcommand{\HQCCtrlCong}{\text{E(23)}\xspace}

\theoremstyle{plain}
\newtheorem{theorem}{Theorem}[section]
\newtheorem{lemma}[theorem]{Lemma}
\newtheorem{proposition}[theorem]{Proposition}
\newtheorem{corollary}[theorem]{Corollary}

\theoremstyle{definition}

\newtheorem{example}[theorem]{Example}
\newtheorem{remark}[theorem]{Remark}

\usetikzlibrary{external}
\usepackage{stmaryrd} 
\SetSymbolFont{stmry}{bold}{U}{stmry}{m}{n}
\DeclarePairedDelimiter\sem{\llbracket}{\rrbracket}

\usepackage{iftex}
\ifpdf
  \usepackage{underscore}
  \usepackage[T1]{fontenc}
\else
  \usepackage{breakurl}
\fi

\def\titlerunning{A Complete Equational Theory for Quantum Circuits with Generalized Control}
\title{\titlerunning}

\author{
    William Schober
    \institute{Università della Svizzera italiana\\Lugano, Switzerland}
    \email{william.schober@usi.ch}
    \and
    Scott Wesley
    \institute{Dalhousie University\\Halifax, Canada}
    \email{scott.wesley@dal.ca}
}

\def\authorrunning{W. Schober \& S. Wesley}

\begin{document}

\newcommand{\Ob}{\operatorname{Ob}}
\newcommand{\Mor}{\operatorname{Mor}}
\newcommand{\dom}{\operatorname{dom}}
\newcommand{\cod}{\operatorname{cod}}
\newcommand{\Log}{\operatorname{Log}}
\newcommand{\CatProp}{\mathbf{Prop}}
\newcommand{\CatSig}{\mathbf{Sig}}
\newcommand{\CatUnitary}{\mathbf{Unitary}}
\newcommand{\CatCQC}{\mathbf{CQC}}
\newcommand{\CatHQC}{\mathbf{HQC}}
\newcommand{\exparr}{\!\uparrow\!}
\newcommand{\Core}{\mathbf{Core}}
\newcommand{\CoreExt}{\mathbf{CExt}}

\newcommand{\encoder}{\mathsf{Enc}}
\newcommand{\decoder}{\mathsf{Dec}}

\newcommand{\Reduce}{\mathsf{Reduce}}
\newcommand{\Drop}{\mathsf{Drop}}
\newcommand{\DiagDrop}{\mathsf{DiagDrop}}
\newcommand{\Expand}{\mathsf{Expand}}
\newcommand{\DiagExpand}{\mathsf{DiagExpand}}
\newcommand{\LambdaExpand}{\mathsf{LambdaExpand}}
\newcommand{\Diag}{\mathsf{Diag}}
\DeclarePairedDelimiter{\ceil}{\lceil}{\rceil}

\newcommand{\Power}{\mathsf{Power}}
\newcommand{\Tower}{\mathsf{Tower}}

\newcommand{\circimg}[1]{\raisebox{-0.5\height}{\includegraphics{#1}}}

\maketitle

\begin{abstract}
    Controlled subcircuits are commonplace in quantum algorithms as they represent the quantum analogue of branching statements (\texttt{if then}).
    We introduce a generalization of controlled subcircuits in the form of a binary operation which sends pairs of unitary quantum circuits to a generalized controlled operation, whose semantics are based on matrix exponentials and logarithms.
    The family of quantum circuits with generalized controlled operations and circuit exponentiation is denoted $\CatHQC$ for Hierarchical Quantum Circuits.
    We formalize $\CatHQC$ as a modified prop category using a novel category-theoretic construction, and provide a sound and complete equational theory for diagrammatic reasoning in $\CatHQC$.
\end{abstract}

\section{Introduction}

As quantum computing hardware continues to advance, so too does its software.
While quantum circuits remain the de facto standard language for low-level quantum programming, they are cumbersome when designing large-scale programs with tens of thousands of gates.
Higher-level abstractions of quantum circuits are on the horizon. 
Diagrammatic reasoning~\cite{Coecke2011} is a natural fit for reasoning about such abstractions since quantum circuits are diagrams. 
Diagrammatic reasoning has already been applied to a variety of circuit-related tasks, including optimization~\cite{Kissinger2020}, simulation~\cite{Sutcliffe2024,Sutcliffe2025}, equivalence checking~\cite{Peham2022}, and error correction~\cite{Khesin2025,Kissinger2024,Rodatz2024}.
Equational theories are collections of equations that serve as toolboxes for performing diagrammatic reasoning.

Control is one such ubiquitous abstraction in quantum algorithm design. Controlled subcircuits appear in essentially every powerful quantum algorithm~(e.g.,~\cite{Kitaev1995,Shor1997,Grover1996,Harrow2009}). 
We generalize the existing notion of control by allowing a subcircuit to be controlled not just by one qubit in a particular basis, but by an arbitrary basis on any number of qubits.
This notion of control is captured by a new binary operation $( \odot )$ called the \textit{control product} which joins two subcircuits together to create a generalized controlled operation.
The simplest example is the $\textsf{CNOT}$, which is created by joining the Pauli $Z$ and $X$ gates together as $Z \odot X = \textsf{CNOT}$.
However unlike the normal notion of control, the control product allows any two subcircuits $U$ and $V$ of arbitrary sizes to control each other as $U \odot V$.
Examples of how such a product can aid in circuit design and verification are found in~\cite{Gidney2017} and~\cite{Schober2024}, respectively.

We present the control product formally in the language of symmetric monoidal categories, as part of a universal algebraic (see~\cite{BaaderNipkow1998}) extension of a prop category~\cite{MacLane1965,Bergholm2011,Clement2023}. We then provide a sound and complete equational theory for performing diagrammatic reasoning on the category $\CatHQC$ of \emph{hierarchical quantum circuits}, that is, quantum circuits with generalized control and circuit exponentiation.
We note that the definition of circuit exponentiation used in this paper is consistent with the power modifier in OpenQASM~3~\cite{Cross2022}.

\vspace{1em}
\noindent
\textbf{Related work}.
The first complete equational theory for uncontrolled quantum circuits was presented in~\cite{Clement2023} and improved in~\cite{Clement2024Extensions,Clement2024Minimal}. 
Two recent works have provided equational theories for standard controlled quantum circuits using controlled props~\cite{Delorme2025} and rig categories~\cite{Heunen2025}. 
Another recent work introduced a related methodology for constructing controlled unitaries via subspace selection and phasing~\cite{Heunen2026}.
The control product we introduce here can be seen as a way to efficiently package exponentially many such selection-and-phasing statements together into one circuit object.
Controlled operations have also recently been studied in the context of quantum control flow~\cite{Yuan2024,Valiron2022} for its applications to quantum computer architectures with a quantum program counter. 

\vspace{1em}
\noindent
\textbf{Structure and contents}.
All proofs are found in the appendices.
\cref{Sect:Background} provides the necessary mathematical background.
\cref{Sect:Semantics} introduces the semantics of control and exponentiation on unitary matrices and provides a characterization in terms of control functors.
\cref{Sect:Hierarchical} constructs the hierarchical quantum circuit language $\CatHQC$.
We describe the inductive construction that is used to generate $\CatHQC$ via closures over an increasing family of nested sub-languages, provide its semantics, and present a sound equational theory.
In \cref{Sect:Completeness} we prove completeness of the equational theory for $\CatHQC$ via a reduction to a known complete equational theory for controlled quantum circuits found in~\cite{Delorme2025}. 
This reduction includes a transpilation procedure which transforms hierarchical circuits into quantum circuits with standard controls.

\section{Background}
\label{Sect:Background}

We assume familiarity with symmetric monoidal categories (see~\cite{MacLane2010}) and introductory functional analysis (see~\cite{Hall2015}).
Given a functor $F: \mathcal{C} \to \mathcal{D}$, we write $\Ob( F )$ for underlying mapping on objects and $\Mor( F )$ for the underlying mapping on morphisms.
We write $\{ X, Y, Z \}$ for the standard Pauli matrices, $\mathbb{I}_n$ for the $n \times n$ identity matrix, and $\mathbb{O}_n$ for the $n \times n$ zero matrix.

\vspace{1em}
\noindent
\textbf{Matrix Exponentials}.
If $M$ is a matrix, then $\exp( M ) = \sum_{n=0}^{\infty} \frac{M^n}{n!}$.
Clearly $\exp( \mathbb{O}_n ) = \mathbb{I}_n$ and $\exp( M \otimes \mathbb{I}_n ) = \exp( M ) \otimes \mathbb{I}_n$.
If $H$ is a skew Hermitian matrix, then $U = \exp( H )$ is a unitary matrix and $H$ is a \emph{logarithm} of $U$.
For each unitary matrix $U$, there exists a unique logarithm $H = \Log( U )$ of $U$ such that all eigenvalues of $H$ fall within $i ( -\pi, \pi ]$.
If $H$ is an $n$-dimensional Hermitian matrix with eigenvalues $\{ \alpha_j \}_{j=1}^n$ and an associated orthonormal eigenbasis $\{ b_j \}_{j=1}^n$, then $\exp( iH ) = \sum_{j=1}^n e^{i\alpha_j} \ket{b_j} \bra{b_j}$.
The matrix exponential enjoys several useful properties~\cite{Hall2015}.
\begin{align}
    &\text{If } MN = NM \text{, then } \exp( N + M ) = \exp( N ) \exp( M ).\label{Eqn:SimulDiag}
    \\
    &\text{If } U \text{ is unitary, then } \exp( i{U^\dagger}HU) = U^\dagger \exp( iH ) U \text{ and } \Log( i{U^\dagger}HU) = U^\dagger \Log( iH ) U.\label{Eqn:UnitaryConj}
\end{align}

\begin{figure}[t]
    \begin{subfigure}[b]{0.13\textwidth}
        \centering
        \caption{$0: 0 \to 0$.}
    \end{subfigure}
    \begin{subfigure}[b]{0.13\textwidth}
        \centering
        \scalebox{0.65}{\circimg{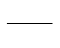}}
        \vspace{-0.5em}
        \caption{$1: 1 \to 1$.}
    \end{subfigure}
    \begin{subfigure}[b]{0.16\textwidth}
        \centering
        \scalebox{0.65}{\circimg{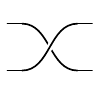}}
        \vspace{-0.5em}
        \caption{$\sigma: 2 \to 2$.}
    \end{subfigure}
    \begin{subfigure}[b]{0.16\textwidth}
        \centering
        \scalebox{0.65}{\circimg{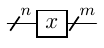}}
        \vspace{-0.5em}
        \caption{$x: n \to m$.}
    \end{subfigure}
    \begin{subfigure}[b]{0.17\textwidth}
        \centering
        \scalebox{0.65}{\circimg{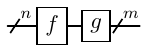}}
        \vspace{-0.5em}
        \caption{$g \circ f: n \to m$.}
    \end{subfigure}
    \begin{subfigure}[b]{0.22\textwidth}
        \centering
        \scalebox{0.65}{\circimg{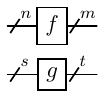}}
        \vspace{-0.5em}
        \caption{$f \boxtimes g: n + s \to m + t$.}
    \end{subfigure}
    \caption{A graphical language for the morphisms in $P( \Sigma )$. Two morphisms in $P( \Sigma )$ are equal if and only if their diagrams are isomorphic~\cite{Selinger2010}. A complete set of graphical rewrite rules to determine if two morphisms are isomorphic can be found in~\cite{Clement2023}.}
    \label{Fig:Circuits}
\end{figure}

\noindent
\textbf{Prop Categories}.
A \emph{prop category} is a strict symmetric monoidal category $( \mathcal{C}, \boxtimes, \mathbb{I} )$ in which $( \Ob( \mathcal{C} ), \boxtimes )$ is the monoid of natural numbers~\cite{MacLane1965}.
A functor between two prop categories is a symmetric monoidal functor $F: \mathcal{C} \to \mathcal{D}$ such that $\Ob( F )( 1 ) = 1$.
A \emph{prop signature} is a collection of symbols $\Sigma$ equipped with a domain operator $\dom: \Sigma \to \mathbb{N}$ and a codomain operator $\cod: \Sigma \to \mathbb{N}$.
Prop signatures form a category in which the morphisms from $\Sigma$ to $\Gamma$ are the functions $\tau: \Sigma \to \Gamma$ which satisfy the equations $\dom( \tau( x ) ) = \dom( x )$ and $\cod( \tau( x ) ) = \cod( x )$ for each $x \in \Sigma$.
The underlying signature of a prop category $\mathcal{C}$ is the prop signature $( \Mor( \mathcal{C} ), \dom, \cod )$ such that $f \in \Mor( \dom( f ), \cod( f ) )$ for each $f \in \Mor( \mathcal{C} )$.
There exists an underlying signature functor $U: \CatProp \to \CatSig$ which sends each prop category to its underlying signature and each prop functor $F: \mathcal{C} \to \mathcal{D}$ to $\Mor( F )$.
For each prop signature $\Sigma$, there exists a prop category $P( \Sigma )$ such that the set of morphisms in $P( \Sigma )$ is the minimal solution to the following set of equations where $( \circ )$, $( \boxtimes )$, $( 0 )$, $( 1 )$, and $( \sigma )$ are formal symbols~\cite{Clement2023}.
\begin{itemize}
\item \textbf{Structure}.
      $0 \in P( \Sigma )( 0, 0 )$, $1 \in P( \Sigma )( 1, 1 )$ and $\sigma \in P( \Sigma )( 2, 2 )$.
\item \textbf{Generators}.
      If $x \in \Sigma$, then $g \in P( \Sigma )( \dom( x ), \cod( x ) )$.
\item \textbf{Sequential Comp}.
      If $f \in P( \Sigma )( n, x )$ and $g \in P( \Sigma )( x, m )$, then $( g \circ f ) \in P( \Sigma )( n, m )$.
\item \textbf{Parallel Comp}.
      If $f \in P( \Sigma )( n, m )$ and $g \in P( \Sigma )( s, t )$, then $( f \boxtimes g ) \in P( \Sigma )( n + s, m + t )$.
\end{itemize}
The identity morphisms in $P( \Sigma )$ are $1_0 = 0$ and $1_{n+1} = 1 \boxtimes 1_{n}$, and $\sigma$ is the generating permutation.
In particular, $\sigma_{1,1} = \sigma$ and $\sigma_{1,k+1} = ( \sigma \boxtimes 1_k ) \circ ( 1 \boxtimes \sigma_{1,k} )$ defines the family of permutations which swap $1$ with $k$.
The morphisms in $P( \Sigma )$ are subject to the following relations~\cite{Clement2023}.
\begin{itemize}
\item \textbf{Sequential Unitality}.
      $1_m \circ f = f = f \circ 1_n$ for each $f: n \to m$.
\item \textbf{Parallel Unitality}.
      $1_0 \boxtimes f = f = f \boxtimes 1_0$ for each $f: n \to m$.
\item \textbf{Sequential Associativity}.
      $( h \circ g ) \circ f = h \circ ( g \circ f )$ for each $f: n \to x$ and $g: x \to m$.
\item \textbf{Parallel Associativity}.
      $( f \boxtimes g ) \boxtimes h = f \boxtimes ( g \boxtimes h )$ for each $f, g, h \in \Mor( P( \Sigma ) )$.
\item \textbf{Bifunctoriality}.
      $( k \circ h ) \boxtimes ( g \circ f ) = ( k \boxtimes g ) \circ ( h \boxtimes f )$ for each $n \xrightarrow{f} x \xrightarrow{g} m$ and $s \xrightarrow{h} y \xrightarrow{k} t$. 
\item \textbf{Symmetry}.
      $\sigma \circ \sigma = 1_2$ and $( f \boxtimes 1 ) \circ \sigma_{1,k} = \sigma_{1,k} \circ ( 1 \boxtimes f )$ for each $f: n \to m$.
\end{itemize}
Note that from the \textbf{Symmetry} relations, it is possible to derive the standard permutation relations as defined~\cite{Lafont2003}.
Obviously, there exists an inclusion $i: \Sigma \to U( P( \Sigma ) )$ which maps each generator in $\Sigma$ to its corresponding element in the free prop.
Moreover, $P( \Sigma )$ is free in the sense that for each $\mathcal{C} \in \Ob( \CatProp )$ and each $\tau \in \CatSig( \Sigma, U( \mathcal{C} ) )$, there exists a unique functor $F: P( \Sigma ) \to \mathcal{C}$ such that $U( F ) \circ i = \tau$~\cite{Clerc2015}.
Concretely, $F$ is defined by the following equations where $x \in \Sigma$.
\begin{align*}
    F( 1 ) &= 1
    &
    F( \sigma ) &= \sigma
    &
    F( x ) &= f( x )
    &
    F( g \circ f ) &= F( g ) \circ F( f )
    &
    F( f \boxtimes g ) &= F( f ) \boxtimes F( g )
\end{align*}
The morphisms in $P( \Sigma )$ can be understood as circuit diagrams~\cite{Hotz1965}, as depicted in~\cref{Fig:Circuits}.

\vspace{1em}
\noindent
\textbf{Dagger Props}.
A functor $F: \mathcal{C} \to \mathcal{C}^\textbf{op}$ is said to be \emph{involutive} if $F \circ F = 1_{\mathcal{C}}$.
A \emph{dagger prop} is a prop category $\mathcal{C}$ with an involutive monoidal functor $\dagger: \mathcal{C} \to \mathcal{C}^{\text{op}}$.
The application of $\dagger$ to $f \in \mathcal{C}( n, m )$ is denoted $f^\dagger \in \mathcal{C}( m, n )$.
Since $( \dagger )$ is contravariant, then $( f \circ g )^\dagger = g^\dagger \circ f^\dagger$.
If $f^\dagger$ is a two-sided inverse to $f$, then $f$ is \emph{unitary}.

\begin{figure}[t]
    \begin{subfigure}[b]{0.14\textwidth}
        \centering
        \scalebox{0.6}{\circimg{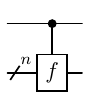}}
        \caption{$C( f )$.}
        \label{Fig:ControlDefn:Notation}
    \end{subfigure}
    \begin{subfigure}[b]{0.20\textwidth}
        \centering
        \scalebox{0.6}{\circimg{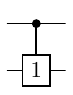}}
        =
        \scalebox{0.6}{\circimg{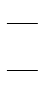}}
        \caption{$C( 1 ) = 1_2$.}
        \label{Fig:ControlDefn:Units}
    \end{subfigure}
    \begin{subfigure}[b]{0.37\textwidth}
        \centering
        \scalebox{0.6}{\circimg{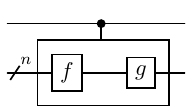}}
        =
        \scalebox{0.6}{\circimg{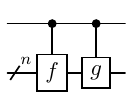}}
        \vspace{-0.5em}
        \caption{$C( f \circ g ) = C( f ) \circ C( g )$.}
        \label{Fig:ControlDefn:Composition}
    \end{subfigure}
    \begin{subfigure}[b]{0.26\textwidth}
        \centering
        \scalebox{0.6}{\circimg{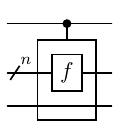}}
        =
        \scalebox{0.6}{\circimg{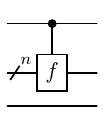}}
        \vspace{-0.5em}
        \caption{$C( f \boxtimes 1 ) = C( f ) \boxtimes 1$.}
        \label{Fig:ControlDefn:IdTensor}
    \end{subfigure}
    \\
    \begin{subfigure}[b]{0.43\textwidth}
        \centering
        \scalebox{0.6}{\circimg{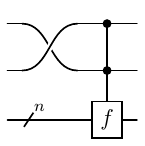}}
        =
        \scalebox{0.6}{\circimg{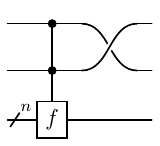}}
        \vspace{0.3em}
        \caption{$C( f ) \circ ( \sigma \boxtimes 1_n ) = ( \sigma \boxtimes 1_n ) \circ C( f )$.}
        \label{Fig:ControlDefn:ContrlSym}
    \end{subfigure}
    \begin{subfigure}[b]{0.55\textwidth}
        \centering
        \scalebox{0.6}{\circimg{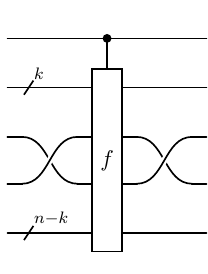}}
        =
        \scalebox{0.6}{\circimg{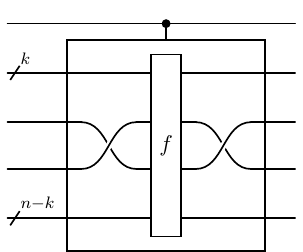}}
        \vspace{-0.5em}
        \caption{$( 1 \boxtimes \gamma_{n,k} ) \circ C( f ) \circ ( 1 \boxtimes \gamma_{n,k} ) = C( \gamma_{n,k} \circ f \circ \gamma_{n,k} )$.}
        \label{Fig:ControlDefn:DataSym}
    \end{subfigure}
    \caption{The definition of a control functor stated in terms of graphical equations.}
    \label{Fig:ControlDefn}
\end{figure}

\vspace{1em}
\noindent
\textbf{Controlled Props}.
If $\mathcal{C}$ is a category, then $\mathcal{C}_{\mathbf{endo}}$ denotes the subcategory of endomorphisms in $\mathcal{C}$.
A \emph{controlled prop}~\cite{Delorme2025} is a prop category $\mathcal{C}$ with an ordinary functor $C: \mathcal{C}_{\mathbf{endo}} \to \mathcal{C}_{\mathbf{endo}}$ such that $\Ob( C )( n ) = n + 1$ and the following equations hold where $\gamma_{n,k} = 1_k \boxtimes \sigma \boxtimes 1_{n-k}$.
\begin{enumerate}
\item $C( f \boxtimes 1 ) = C( f ) \boxtimes 1$ for each $f \in \mathcal{C}( m, m )$ and $n \in \mathbb{N}$.
\item $( \sigma \boxtimes 1_n ) \circ C( C( f ) ) = C( C( f ) ) \circ ( \sigma \boxtimes 1_n )$ for each $f \in \mathcal{C}( n, n )$.
\item $( 1 \boxtimes \gamma_{n,k} ) \circ C( f ) \circ ( 1 \boxtimes \gamma_{n,k} ) = C( \rho_{n,k} \circ f \circ \gamma_{n,k} )$ for each $f \in \mathcal{C}( n + 2, n + 2 )$ and $0 \le k \le n$.
\end{enumerate}
A controlled prop is said to be \emph{$(u,v)$-pointed}~\cite{Delorme2025} when there exists points $u, v \in \mathcal{C}( 0, 1 )$ such that $C( f ) \circ ( u \boxtimes 1_n ) = u \boxtimes f$ and $C( f ) \circ ( v \boxtimes 1_n ) = v \boxtimes 1_n$ for each $f \in \mathcal{C}( n, n )$.
A \emph{dagger controlled prop}~\cite{Delorme2025} is a dagger prop $( \mathcal{C}, \dagger )$ with a choice of control functor $C: \mathcal{C}_{\mathbf{endo}} \to \mathcal{C}_{\mathbf{endo}}$ satisfying the equation $C \circ \dagger = \dagger \circ C$.
A \emph{conjugated controlled prop}~\cite{Delorme2025} is a dagger controlled prop $( \mathcal{C}, \dagger, C )$ which satisfies the equation $C( g^\dagger \circ f \circ g ) = ( 1 \boxtimes g^\dagger ) \circ C( f ) \circ ( 1 \boxtimes g )$ for each $f \in \mathcal{C}( n, n )$ and $g \in \mathcal{C}( m, n )$.
The control functor $C( - )$ is said to be \emph{conjugated}.
Given a prop signature $\Sigma$, it is possible to extend the construction of $P( \Sigma )$ to obtain a controlled prop $P_C( \Sigma )$.
The morphisms in $P_C( \Sigma )$ must also satisfy the equation that $C( f ) \in P_C( \Sigma )( n + 1, n + 1 )$ for each $f \in P_C( \Sigma )( n, n )$.
The morphisms in $P_C( \Sigma )$ are also subject to the following equations.
\begin{itemize}
\item \textbf{C-Functoriality}.
      $C( 1 ) = 1_{2}$ and $C( g \circ f ) = C( g ) \circ C( f )$ for each $f: n \to n$ and $g: n \to n$.
\item \textbf{Controlled Identity}.
      $C( f \boxtimes 1 ) = C( f ) \boxtimes 1$ for each $f: m \to m$.
\item \textbf{Control Symmetry}.
      $( \sigma \boxtimes 1_n ) \circ C( C( f ) ) = C( C( f ) ) \circ ( \sigma \boxtimes 1_n )$ for each $f: n \to n$.
\item \textbf{Data Symmetry}.
      $( 1 \boxtimes \gamma_{n,k} ) \circ C( f ) \circ ( 1 \boxtimes \gamma_{n,k} ) = C( \gamma_{n,k} \circ f \circ \gamma_{n,k} )$ for each $f: n + 2 \to n + 2$.
\end{itemize}
Just as $P( \Sigma )$ is free with respect to strict monoidal functors, it is straight-forward to show that the category $P_C( \Sigma )$ is free with respect to strict monoidal functors that preserve controls.
A graphical language for $P_C( \Sigma )$ can be found in~\cref{Fig:ControlDefn}.

\vspace{1em}
\noindent
\textbf{Rewriting in Props}.
Let $\Sigma$ be a prop signature.
A \emph{$\Sigma$-equation} is a pair $( f, g )$ of morphisms from $\Mor( P( \Sigma ) )$ such that $\dom( f ) = \dom( g )$ and $\cod( f ) = \cod( g )$.
A set of $\Sigma$-equations $E$ is called a \emph{prop congruence relation} if it is transitive, reflexive, symmetric, and closed under both sequential and parallel composition.
In particular, the following congruence conditions hold for each $f, g \in P( \Sigma )( n, m )$.
\begin{enumerate}
\item If $( f, g ) \in E$ with $h: s \to n$ and $k: t \to m$, then $( k \circ f \circ h, k \circ g \circ h ) \in E$.
\item If $( f, g ) \in E$ with $h: s \to t$ and $k: x \to y$, then $( h \boxtimes f \boxtimes k, h \boxtimes g \boxtimes k ) \in E$.
\end{enumerate}
A set of $\Sigma$-equations $E$ from $\Mor( P_C( \Sigma ) )$ is called a \emph{controlled prop congruence relation} if it also satisfies $( C( f ), C( g ) ) \in E$ for each $( f, g ) \in E$.
Given a set of $\Sigma$-equations $E$, there exists a minimal congruence relation $R$ such that $E \subseteq R$.
There then exists a unique prop $P( \Sigma ) / E$ with a unique projection $\pi_E: P( \Sigma ) \to P( \Sigma ) / E$ such that the following equations hold (see~\cite{Clerc2015}).
\begin{enumerate}
\item $\pi_E( f ) = \pi_E( g )$ if and only if $( f, g ) \in R$.
\item If $F: P( \Sigma ) \to \mathcal{C}$ is a prop functor and $F( f ) = F( g )$ for each $( f, g ) \in E$, then there exists a unique prop functor $F_E: P( \Sigma ) / E \to \mathcal{C}$ such that $F_E \circ \pi_E = P( F )$.
\end{enumerate}
This extends readily to controlled prop categories and their functors.
The pair $( \Sigma, E )$ is said to be a \emph{prop presentation} for the class of props isomorphic to $P( \Sigma ) / E$.
In particular, if $F: P( \Sigma ) \to \mathcal{C}$ is surjective, then a presentation for $\mathcal{C}$ is a pair $( \Sigma, E )$ such that $( f, g ) \in E$ if and only if $F( f ) = F( g )$.
Often $E$ is called a \emph{complete equational theory} for $\mathcal{C}$ and $F$ is called a \emph{semantic interpretation} and is denoted $\sem{-}_\mathcal{C}$.
We write $f \approx_E g$ when $( f, g )$ is in the congruence relation induced by $E$.

\vspace{1em}
\noindent
\textbf{Euler Relations}.
The Euler relation is a complex generalization of the Euler angle decomposition for 3D-rotations, which describes how rotations on a sphere can be decomposed into rotations around any two chosen axes.
There are many equivalent variations of this relation in the literature, varying in the choice of axes, the exact number of parameters involved, and their corresponding algebraic relations to one another.
Use of an Euler angle relation was shown to be necessary for completeness when working with quantum circuits~\cite{Clement2024Minimal,Vilmart2021}.
In~\cite{Delorme2025}, a certain two parameter version of the Euler relation was introduced.
Define a function $( \beta_0, \beta_1, \beta_2, \beta_3 ) = \mathsf{Euler}( \alpha_1, \alpha_2 )$ as follows, for $\alpha_1, \alpha_2 \in \mathbb{R}$.
\begin{equation*}
    \begin{array}{l}
      u := -\sin\left(\frac{\alpha_1+\alpha_2}{2}\right)+i\cos\left(\frac{\alpha_1-\alpha_2}{2}\right)\\
      v := +\cos\left(\frac{\alpha_1+\alpha_2}{2}\right)-i\sin\left(\frac{\alpha_1-\alpha_2}{2}\right)
    \end{array}
\end{equation*}
\begin{center}
  \begin{tabular}{c|c|c|c}
    & $\beta_1$ & $\beta_2$ & $\beta_3$ \\
    \hline
    if $v=0$ & $2\arg(u)$ & $0$ & $0$\\
    \hline
    if $u=0$ & $2\arg(v)$ & $\pi$ & $0$\\
    \hline
    otherwise \hspace{0.2em}&\hspace{0.2em} $\arg(u)+\arg(v)$ \hspace{0.2em}&\hspace{0.2em} $2\arg\left(i+\left\lvert\tfrac{u}{v}\right\rvert\right)$ \hspace{0.2em}&\hspace{0.2em} $\arg(u)-\arg(v)$
  \end{tabular}
\end{center}
\begin{equation*}
    \beta_0 = \frac{(\pi+\alpha_1+\alpha_2-\beta_1-\beta_2-\beta_3)}{2}
\end{equation*}
Then $H Z( \alpha_2 ) H Z( \alpha_1 ) H = e^{i\beta_0} Z( \beta_3 ) H Z( \beta_2 ) H Z( \beta_1 )$ where $H$ is the Hadamard gate and $Z( - )$ is the $Z$-rotation $Z( \theta ) = \exp( i \ket{0} \bra{0} \theta )$.
It should be noted that $H Z( \theta ) H$ is an $X$-rotation, giving rise to the second axis of rotation.
For more details about Euler relations beyond what is listed here, we refer the readers to~\cite{Vilmart2021,Clement2024Minimal,Clement2024Extensions,Delorme2025}.

\begin{figure}[t]
    \begin{framed}
        \vspace{-1.5em}
        \begin{align*}
            \scalebox{0.58}{\circimg{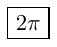}}
            &\overset{Q(1)}{=}
            \scalebox{0.58}{\circimg{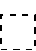}}
            & 
            \scalebox{0.58}{\circimg{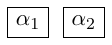}}
            &\overset{Q(2)}{=}
            \scalebox{0.58}{\circimg{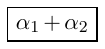}}
            & 
            \scalebox{0.58}{\circimg{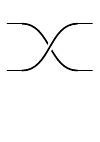}}
            &\overset{Q(3)}{=}
            \scalebox{0.58}{\circimg{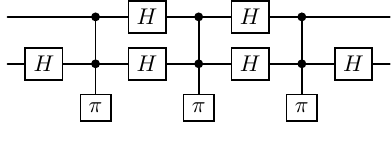}}
            & 
            \scalebox{0.58}{\circimg{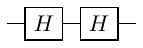}}
            &\overset{Q(4)}{=}
            \scalebox{0.58}{\circimg{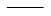}}
        \end{align*}
        \vspace{-2em}
        \begin{align*}
            \scalebox{0.58}{\circimg{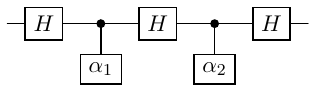}}
            &\overset{Q(5)}{=}
            \scalebox{0.58}{\circimg{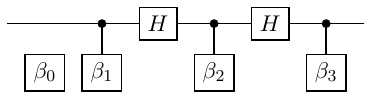}}
            & 
            \scalebox{0.58}{\circimg{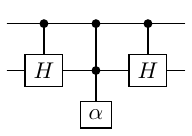}}
            &\overset{Q(6)}{=}
            \scalebox{0.58}{\circimg{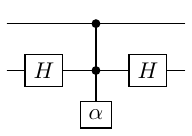}}
        \end{align*}
        \vspace{-2em}
        \begin{align*}
            \scalebox{0.58}{\circimg{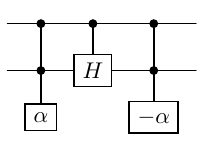}}
            &\overset{Q(7)}{=}
            \scalebox{0.58}{\circimg{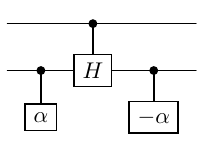}}
            & 
            \scalebox{0.58}{\circimg{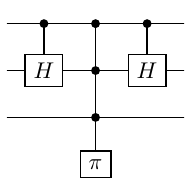}}
            &\overset{Q(8)}{=}
            \scalebox{0.58}{\circimg{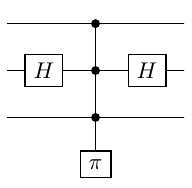}}
            & 
            \scalebox{0.58}{\circimg{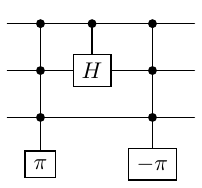}}
            &\overset{Q(9)}{=}
            \scalebox{0.58}{\circimg{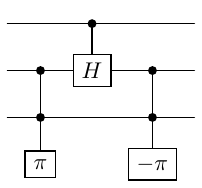}}
        \end{align*}
        \vspace{-1.5em}
    \end{framed}
    \caption{A known equational theory~\cite{Delorme2025} for the category of controlled quantum circuits, where $\alpha, \alpha_1, \alpha_2 \in \mathbb{R}$ and $( \beta_0, \beta_1, \beta_2, \beta_3 ) = \mathsf{Euler}( \alpha_1, \alpha_2 )$.}
    \label{Fig:CQCEqnTheory}
\end{figure}

\vspace{1em}
\noindent
\textbf{Quantum Control}.
Let $\CatUnitary$ denote the prop category of unitary matrices and $\{ \ket{0}, \ket{1} \}$ denote the computational basis for $\mathbb{C}^2$.
It is well known that $\CatUnitary$ is a dagger prop with $\dagger: \CatUnitary \to \CatUnitary^\text{op}$ sending each matrix $U$ to its conjugate transpose $U^\dagger$.
For each $d \in \mathbb{N}$, there is a prop subcategory $\CatUnitary_d$ of $\CatUnitary$ such that $\CatUnitary_d( n, n ) = \CatUnitary( d^n, d^n )$.
It was shown in \cite{Delorme2025} that $C( f: n \to n ) = \ket{0} \bra{0} \otimes \mathbb{I}_n + \ket{1} \bra{1} \otimes f$ is a conjugated control functor for $\CatUnitary_2$ with points $( \ket{0}, \ket{1} )$.
It was further shown that $\CatUnitary_2$ admits a controlled prop presentation $\CatCQC$ with generators
$   \Sigma_{\CatCQC}
    =
    \{
        \scalebox{0.65}{\circimg{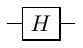}}
    \} \cup \{
        \scalebox{0.65}{\circimg{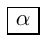}}
        :
        \alpha \in \mathbb{R}
    \}
$
and relations $Q$ as depicted in~\cref{Fig:CQCEqnTheory}.
The semantic interpretation for this theory is generated by 
$\sem{
\scalebox{0.65}{\circimg{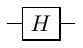}}
}_C = H$
and 
$\sem{
\scalebox{0.65}{\circimg{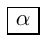}}
}_C = e^{i\alpha}$ where $H$ is the Hadamard matrix. 

\section{The Semantics of Generalized Controls}
\label{Sect:Semantics}

This section introduces an operation $( \odot )$ called the \emph{control product} which provides a symmetric generalization of controlled operations in unitary quantum circuits.
We begin by defining an assignment $\odot : \CatUnitary \times \CatUnitary \to \CatUnitary$ on unitary matrices.
If $U$ and $V$ are two unitary matrices, then we define $U \odot V = \exp( \Log(U) \otimes \Log(V) / ( i\pi ) )$.
The normalization factor $1 / ( i\pi )$ restricts the eigenvalues of the Hermitian matrix to $( -\pi, \pi ]$, and consequently $( \odot )$ is associative.

\begin{theorem}
    \label{Thm:ControlAssoc}
    The control product is unital and associative with unit $( -1 )$.
    Moreover, given a sequence $U_1, U_2, \ldots, U_n \in \Mor( \CatUnitary )$, $\bigodot_{j=1}^n U_j = \exp\left( i\pi \cdot \bigotimes_{j=1}^{n} \frac{\Log( U_j )}{i\pi} \right)$.
\end{theorem}

Unfortunately, the control product is not a bifunctor since $U \odot ( V \circ W ) \ne ( U \odot V ) \circ ( U \odot W )$ in general.
However, the control product does enjoy many desirable properties.
For example, if the first coordinate is fixed to $Z$, then for each $U \in \CatUnitary( n, n )$, $Z \odot U = \ket{0} \bra{0} \otimes \mathbb{I}_n + \ket{1} \bra{1} \otimes U$.
That is to say, $Z \odot( - )$ is the standard canonical control functor on $\CatUnitary_2$.
It follows that $Z \odot X$ is a \textsf{CNOT} gate and $Z \odot Z$ is a \textsf{CZ} gate.
Since $X \odot ( - )$ is also a control functor, then this formalization suggests that the \textsf{CNOT} gate is also a $Z$-gate controlled in the $X$-basis.
This insight can be used to explain phase kickback~\cite{OssorioCastillo2023}.
Of course, it is natural to ask for which $U \in \Mor( \CatUnitary )$ will $U \odot ( - )$ be a control functor.
This is answered by the following theorem.

\begin{theorem}
    \label{Thm:GenCtrlNF}
    Let $H$ be an $n$-dimensional Hermitian matrix with eigenvalues $\{ \alpha_j \}_{j=1}^n$ in $( -1, 1 ]$ and orthonormal eigenbasis $\{ b_j \}_{j=1}^n$.
    If $U = \exp( iH\pi )$, then $U \odot V = \sum_{j=1}^n \ket{b_j} \bra{b_j} \otimes V^{\alpha_j}$ for each $V \in \Mor( \CatUnitary )$.
    In particular, $Z \odot V = \ket{0} \bra{0} \otimes \mathbb{I}_m + \ket{1} \bra{1} \otimes V$ for $V \in \CatUnitary( m, m )$.
\end{theorem}

\begin{corollary}
    \label{Cor:GenCtrlFunc}
    If $U \in \CatUnitary( d, d )$ and $F( - ) = U \odot ( - )$, then the following conditions hold.
    \begin{itemize}
    \item \textbf{C1}.
          $F( \mathbb{I}_n ) = \mathbb{I}_{dn}$ for each $n \in \mathbb{N}$.
    \item \textbf{C2}.
          $F( V \otimes \mathbb{I}_n ) = F( V ) \otimes \mathbb{I}_n$ for each $n \in \mathbb{N}$ and $V \in \Mor( \CatUnitary )$.
    \item \textbf{C3}.
          $F( F( V ) ) \circ ( \sigma_{d,d} \otimes \mathbb{I}_n ) = ( \sigma_{d,d} \otimes \mathbb{I}_n ) \circ F( F( V ) )$ for each $V \in \CatUnitary( n, n )$ where $\sigma_{d,d}$ is the monoidal symmetry $\mathbb{C}^d \otimes \mathbb{C}^d \cong \mathbb{C}^d \otimes \mathbb{C}^d$.
    \item \textbf{C4}.
          $F( P^\dagger \circ V \circ P ) = ( \mathbb{I}_d \otimes P^\dagger ) \circ F( V ) \circ ( \mathbb{I}_d \otimes P )$ for each $V, P \in \CatUnitary( n, n )$.
    \end{itemize}
    The assignment $F( - )$ is a functor if and only if $U$ is Hermitian.
    Moreover, if $F( - )$ is a functor, then $F( - )$ restricts to a conjugated control functor on $\CatUnitary_d$.
    This control functor is $( u, v )$-pointed if and only if $U\ket{u} = -\ket{u}$ and $U\ket{v} = \ket{v}$.
\end{corollary}

\cref{Thm:GenCtrlNF} establishes a relation between control products and matrix powers.
Moreover, if $\alpha \in (-1, 1]$, then $U^\alpha = U \odot e^{i\alpha\pi}$.
Unfortunately, matrix powers are only well-behaved for small choices of $\alpha$, as illustrated by the following theorem.
In general, the properties of exponentiation will depend on the eigenvalues of $U$.
To this end, we will say that $\lambda_j$ is \emph{$( \alpha, \beta )$-admissible} if either $\alpha \lambda_j \in ( -1, 1  ]$ or $\beta \in \mathbb{Q}$ with reduced denominator dividing $\ceil*{( \alpha \lambda_j - 1) / 2}$.

\begin{theorem}
    \label{Thm:PowAsCtrl}
    For $\alpha \in \mathbb{R}$, $U \odot e^{i\alpha\pi} = U^\alpha$ for all $U \in \Mor( \CatUnitary )$ if and only if $\alpha \in ( -1, 1 ]$.
    If $U \in \Mor( \CatUnitary )$ is Hermitian, then $U \odot e^{i\alpha\pi} = U^\alpha$ for all $\alpha \in \mathbb{R}$.
\end{theorem}

\begin{theorem}
    \label{Thm:AddPow}
    Let $U \in \CatUnitary( n, n )$ with $\{ \lambda_j \}_{j=1}^{n}$ the eigenvalues of $\Log( U ) / (i\pi)$.
    For each $\alpha \in \mathbb{R}$ and $\beta \in \mathbb{R}$, $( U^\alpha )^\beta = U^{\alpha \beta}$ if and only if for each $j \in \{ 1, 2, \ldots, n \}$, $\lambda_j$ is $( \alpha, \beta )$-admissible.
    This means that $\mathbb{Z}$ and $( -1, 1 ]$ are maximal submonoids of $( \mathbb{R}, \cdot )$ for which $( U, \alpha ) \mapsto U^\alpha$ defines a monoid action on $\Mor( \CatUnitary )$.
\end{theorem}

It was shown in~\cite{Schober2024} that generalized controls emerge naturally when studying quantum algorithms using the techniques of~\cite{Gidney2017}.
Moreover, \cref{Thm:GenCtrlNF} and \cref{Thm:PowAsCtrl} show that generalized controls are inherently connected with the matrix powers already used in quantum programming languages such as OpenQASM 3~\cite{Cross2022}.
Ideally, one would hope that the intuition for standard controls and scalar powers would generalize to this setting.
It is clear from \cref{Cor:GenCtrlFunc} and \cref{Thm:AddPow} that neither of these hopes hold true.
For example, integer exponentiation (i.e., repeating and inverting circuits) does not interact well with unit fraction exponentiation (i.e., computing roots of circuits), despite these two operations being fundamentally related.
Given these limitations, one is left to seek out techniques to manipulate control products and matrix powers, even in the cases where the standard intuition fails (e.g., when $\alpha > 1$).
This motivates the study of complete equational theories, which could be used to work soundly with these constructs in the context of circuit optimization, simulation, and equivalence checking.

\begin{example}[Circuit Exponentiation in IBM Heron]
    \label{Ex:RZZ}
    Matrix powers are present in many quantum programming languages, whether explicitly or implicitly, in the form of parametrized rotation gates.
    With the advent of new quantum hardware, multi-qubit rotation gates have been gaining attention.
    For example, the IBM Heron r3 qubit processor released in 2025~\cite{IBMProcessors} offers a multi-qubit $ZZ$-rotation~\cite{IBMRZZ} in addition to the more standard one-qubit $X$- and $Z$-rotation gates, which is intended to reduce execution times and error rates when running quantum simulations and variational algorithms~\cite{IBMFractional}.
    The semantics of IBM Heron r3's parametrized gates are as follows. On the right we include their equivalent semantics in terms of matrix powers.
    \begin{align*}
        R_X(\theta) &:= \exp( -i \theta X / 2 ) = X^{\theta / ( 2\pi )} \cdot (-X)^{-\theta / ( 2\pi )} \\
        R_Z(\theta) &:= \exp( -i \theta Z / 2 ) = Z^{\theta / ( 2\pi )} \cdot (-Z)^{-\theta / ( 2\pi )} \\
        R_{ZZ}(\theta) &:= \exp( -i \theta Z \otimes Z / 2 ) = e^{-i\theta/2} ( Z \otimes Z )^{ \theta / \pi}
    \end{align*}
    A sound and complete equational theory for circuit exponentiation and generalized control would be able to prove the following decomposition of $R_{ZZ}( \theta )$, without the need for new axioms specific to the $ZZ$-rotation.
    \begin{equation*}
        R_{ZZ}( \theta )
        =
        \sem*{\scalebox{0.65}{\circimg{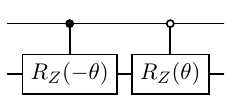}}}
        =
        \sem*{\scalebox{0.65}{\circimg{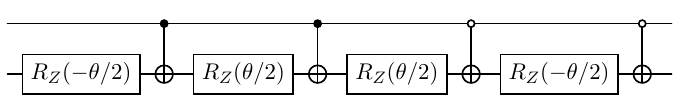}}}
    \end{equation*}
\end{example}

\section{Props with Exponentiation and Generalized Control}
\label{Sect:Hierarchical}

Hierarchical quantum circuits build upon the prop formalism by introducing two new connectives, namely $( \odot )$ and $( \exparr )$.
The $( \odot )$ connective can be thought of as an abstraction of generalized control, and the $( \exparr )$ connective can be thought of as an abstraction of circuit exponentiation.
As outlined in the previous section, the $( \odot )$ connective should be associative, though few assumptions can be made about the $( \exparr )$ connective.
In particular, neither $( \odot )$ nor $( \exparr )$ are functorial in general.
This means that unlike a controlled prop, it does not suffice to simply complete a standard prop under the application of two functors.
Instead, the hierarchical language will be constructed in levels, with the lowest level corresponding to a standard prop.

The basic idea of this construction is that given a free prop $P( \Sigma )$, the powers and generalized controls can be freely constructed as purely syntactic objects.
Since circuits in $P( \Sigma )$ are equal up to diagram isomorphism, then their exponentiated and controlled forms will also be equal up to isomorphism of their subcircuit constituents.
These syntactic objects are then combined with the gates in $\Sigma$ to generate a strictly larger prop containing $P( \Sigma )$, and the process repeats.
More formally, the construction proceeds as follows.

\begin{figure}[t]
    \begin{subfigure}[b]{0.17\textwidth}
        \centering
        \scalebox{0.65}{\circimg{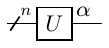}}
        \caption{$U \exparr \alpha$.}
        \label{Fig:Extended:Power}
    \end{subfigure}
    \begin{subfigure}[b]{0.17\textwidth}
        \centering
        \scalebox{0.65}{\circimg{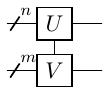}}
        \caption{$U \odot V$.}
        \label{Fig:Extended:BiCtrl}
    \end{subfigure}
    \begin{subfigure}[b]{0.22\textwidth}
        \centering
        \scalebox{0.65}{\circimg{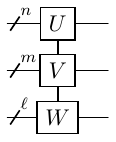}}
        \caption{$U \odot V \odot W$.}
        \label{Fig:Extended:TriCtrl}
    \end{subfigure}
    \begin{subfigure}[b]{0.39\textwidth}
        \centering
        \scalebox{0.65}{\circimg{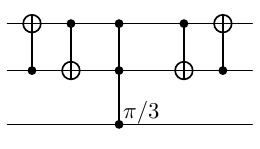}}
        \caption{$U \in \CatHQC[2]( 3, 3 ) \setminus \CatHQC[1]( 3, 3 )$.}
    \end{subfigure}
    \caption{A graphical language for the morphisms in $\CatHQC$. In these diagrams we assume that $U: n \to n$, $V: m \to m$, and $W: \ell \to \ell$.}
    \label{Fig:Extended}
\end{figure}

\vspace{1em}
\noindent
\textbf{Exponentiation}.
Let $\mathcal{C}$ be a prop category.
If $U \in \Mor( \mathcal{C} )$ and $\alpha \in \mathbb{R}$, then there should be a unique term $U \exparr \alpha$ which is thought of as exponentiating $U$ by $\alpha$.
Since we do not require exponentiation to satisfy any further properties, then $\Power( \mathcal{C} ) := \Mor( \mathcal{C} ) \times \mathbb{R}$.
We write $U \exparr \alpha$ for an element $( U, \alpha ) \in \Power( \mathcal{C} )$.
Of course, $\dom( U \exparr \alpha ) = \dom( U )$ and $\cod( U \exparr \alpha ) = \cod( U )$.
A term of the form $U \exparr \alpha$ is denoted by raising $U$ to the power of $\alpha$, as depicted in \cref{Fig:Extended:Power}.

\vspace{1em}
\noindent
\textbf{Generalized Controls}.
Let $\mathcal{C}$ be a prop category.
Since $( \odot )$ is known to be associative, then it suffices to generate the free monoid $\Mor( \mathcal{C} )^*$ on $\Mor( \mathcal{C} )$.
We write $U_1 \odot U_2 \odot \cdots \odot U_n$ for an element $( U_1, U_2, \ldots, U_n ) \in \Mor( \mathcal{C} )^*$.
Of course, only terms of length two or more correspond to valid generalized controls in the circuit language.
We write $\Tower( \mathcal{C} ) := \{ T \in \Mor( \mathcal{C} )^* : |T| \ge 2 \}$.
If $T \in \Tower( \mathcal{C} )$, then $\dom( T ) = \sum_{j=1}^n \dom( T_j )$ and $\cod( T ) = \sum_{j=1}^n \cod( T_j )$ where $n = |T|$.
An $n$-ary $(\odot)$-product is illustrated by a vertical wire joining the $n$ factors, as in \cref{Fig:Extended:BiCtrl} and \cref{Fig:Extended:TriCtrl}.

\vspace{1em}
\noindent
\textbf{Constructing the Hierarchy}.
The construction begins with a primitive gate set $\Sigma_{\mathbf{Prim}}$.
In the case of this paper,
$\Sigma_{\mathbf{Prim}} = \{
        \scalebox{0.65}{\circimg{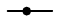}},
        \scalebox{0.65}{\circimg{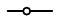}},
        \scalebox{0.65}{\circimg{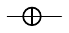}},
        \scalebox{0.65}{\circimg{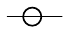}},
        \scalebox{0.65}{\circimg{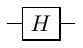}},
        \scalebox{0.65}{\circimg{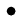}}
\}$.
This gives rise to a free prop $\CatHQC[0] = P( \Sigma_{\CatHQC}^0 )$ where $\Sigma_{\CatHQC}^0 = \Sigma_{\mathbf{Prim}}$.
The construction then proceeds inductively as follows.
For each $n \in \mathbb{N}$, define a signature $\Sigma_{\CatHQC}^{n+1} = \Sigma_{\CatHQC}^n \cup \Tower( \CatHQC[n] ) \cup \Power( \CatHQC[n] )$ and a prop category $\CatHQC[n+1] = P( \Sigma_{\CatHQC}^{n+1} )$.
Since $\Sigma_{\CatHQC}^n \subseteq \Sigma_{\CatHQC}^{n+1}$, then this construction can be summarized by the following diagram.
\begin{center}
    \includegraphics{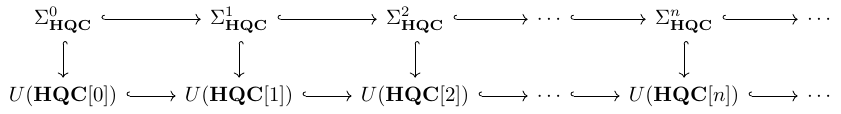}
\end{center}

Since $\Mor( \CatHQC[n] ) \subseteq \Mor( \CatHQC[n+1] )$, then there is an identification of monoid generators such that $\Tower( \CatHQC[n] ) \le \Tower( \CatHQC[n+1] )$ in the sense of submonoids.
We will assume this identification, such that if $T \in \Tower( \CatHQC[n] )$ with $k = |T|$ and $S \in \Tower( \CatHQC[n+1] )$, then $T \odot S = T_1 \odot ( T_2 \odot ( \cdots \odot ( T_k \odot S ) ) )$.
The categorical colimit of this construction is denoted $\CatHQC$, so that $\CatHQC = P( \Sigma_{\CatHQC} )$ where $\Sigma_{\CatHQC} = \bigcup_{n=0}^\infty \Sigma_{\CatHQC}^n$.
The following lemmas characterize how to map freely out of $\CatHQC$ where $\sqcup$ denotes the disjoint union of sets.

\begin{lemma}
    \label{Lem:FuncDefFam}
    Let $\Sigma_0^* = \Sigma_{\CatHQC}^0$ and $\Sigma_{n+1}^* = \Sigma_{\CatHQC}^{n+1} \setminus \Sigma_{\CatHQC}^n$ for each $n \in \mathbb{N}$.
    If $\left\{ \tau_n: \Sigma^*_{n} \to U( \mathcal{C} ) \right\}_{n=0}^{\infty}$ is a family of prop signature morphisms, then there exists a unique prop functor $F: \CatHQC \to \mathcal{C}$ such that $F( g ) = \tau_n( g )$ for each $n \in \mathbb{N}$ and $g \in \Sigma_n^*$.
\end{lemma}

\begin{lemma}
    \label{Lem:FuncDefPart}
    $
        \Sigma_{\CatHQC}
        =
        \Sigma_{\mathbf{Prim}}
        \sqcup \Sigma_{\mathbf{Ctrl}}
        \sqcup
        \Sigma_{\mathbf{Pow}}
    $
    where
    $
        \Sigma_{\mathbf{Ctrl}}
        =
        \{ U \odot V : U, V \in \Mor( \CatHQC ) \}
    $
    and
    $
        \Sigma_{\mathbf{Pow}}
        =
        \{
            U \exparr \alpha
            :
            U \in \Mor( \CatHQC )
            \text{ and }
            \alpha \in \mathbb{R}
        \}
    $.
\end{lemma}

It should come as no surprise that the standard semantic interpretation for $\CatHQC$ is defined in terms of the generalized controls and exponentiation in $\CatUnitary$.
It follows from \cref{Lem:FuncDefFam} that the following set of equations defines a unique prop functor $\sem{-}_H: \CatHQC \to \CatUnitary$.
Note that this is well-defined since $( \odot )$ is also associative for $\CatUnitary$ by \cref{Thm:ControlAssoc}.
\begin{align*}
    s_0\left( \scalebox{0.65}{\circimg{tikz-cache/dagger__term1.pdf}} \right)
    &=
    Z
    &
    s_0\left( \scalebox{0.65}{\circimg{tikz-cache/dagger__term2.pdf}} \right)
    &=
    -Z
    &
    s_0\left( \scalebox{0.65}{\circimg{tikz-cache/dagger__term3.pdf}} \right)
    &=
    X
    &
    s_0\left( \scalebox{0.65}{\circimg{tikz-cache/dagger__term4.pdf}} \right)
    &=
    -X
    &
    s_0\left( \scalebox{0.65}{\circimg{tikz-cache/dagger__term8.pdf}} \right)
    &= -1
\end{align*}
\vspace{-2em}
\begin{align*}
    s_0\left( \scalebox{0.65}{\circimg{tikz-cache/dagger__term5.pdf}} \right)
    &=
    H
    &
    s_{n+1}( U_1 \odot \cdots \odot U_k )
    &=
    s_n( U_1 ) \odot \cdots \odot s_n( U_k )
    &
    s_{n+1}\left( U \exparr \alpha \right)
    &=
    s_n( U )^\alpha
\end{align*}
The following set of equations are also well-defined by \cref{Lem:FuncDefPart}, and define a unique prop functor $\dagger: \CatHQC \to \CatHQC^\text{op}$ which sends each circuit in $\CatHQC$ to its syntactic adjoint.
\begin{align*}
    d \left( \scalebox{0.65}{\circimg{tikz-cache/dagger__term1.pdf}} \right)
        &= \scalebox{0.65}{\circimg{tikz-cache/dagger__term1.pdf}}
    &
    d \left( \scalebox{0.65}{\circimg{tikz-cache/dagger__term2.pdf}} \right)
        &= \scalebox{0.65}{\circimg{tikz-cache/dagger__term2.pdf}}
    &
    d \left( \scalebox{0.65}{\circimg{tikz-cache/dagger__term3.pdf}} \right)
        &= \scalebox{0.65}{\circimg{tikz-cache/dagger__term3.pdf}}
    &
    d \left( \scalebox{0.65}{\circimg{tikz-cache/dagger__term4.pdf}} \right)
        &= \scalebox{0.65}{\circimg{tikz-cache/dagger__term4.pdf}}
    &
    d \left( \scalebox{0.65}{\circimg{tikz-cache/dagger__term5.pdf}} \right)
        &= \scalebox{0.65}{\circimg{tikz-cache/dagger__term5.pdf}}
\end{align*}
\vspace{-2em}
\begin{align*}
    d \left( \scalebox{0.65}{\circimg{tikz-cache/dagger__term8.pdf}} \right)
        &= \scalebox{0.65}{\circimg{tikz-cache/dagger__term8.pdf}}
    &
    d \left( \scalebox{0.65}{\circimg{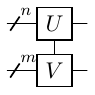}} \right)
        &= \scalebox{0.65}{\circimg{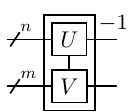}}
    &
    d \left( \scalebox{0.65}{\circimg{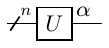}} \right)
        &= \scalebox{0.65}{\circimg{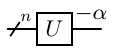}}
\end{align*}

\begin{theorem}
    \label{Thm:DagSem}
    If $U \in \Mor( \CatHQC )$, then $\sem{U^\dagger}_H = \sem{U}_H^\dagger$.
\end{theorem}

Given the syntactic adjoint map $( \dagger )$, it is then possible to define syntactic diagonalization.
Fix some $n \in \mathbb{N}$ which will denote the number of wires in these circuits.
Then for each $x \in \{ 0, 1 \}^n$ and $\alpha \in \mathbb{R}$, there exists a circuit $\lambda( x, \alpha ) \in \CatHQC( n, n )$ such that $\sem{ \lambda( x, \alpha ) }_H = \exp( i H )$ where $H = \alpha \bigotimes_{j=0}^{n-1} \ket{x_j} \bra{x_j}$.
Moreover, $\lambda( x, \alpha )$ can be constructed entirely from NOT-gates and global phase gates with zero or more $Z$-controls (see~\cref{App:DiagGate}).
A circuit $\Lambda$ is in \emph{diagonal form} if there exists a bijection $f: \{ 1, 2, \ldots, 2^n \} \to \{ 0, 1 \}^n$ and function $\alpha: \{ 1, 2, \ldots, 2^n \} \to ( -\pi, \pi ]$ such that $\Lambda = \lambda( f( 1 ), \alpha( 1 ) ) \circ \cdots \circ \lambda( f( 2^n ), \alpha( 2^n ) )$.
Then a syntactic diagonalization of $U \in \CatHQC( n, n )$ is a choice of diagonal form $\Lambda \in \CatHQC( n, n )$ and $P \in \CatHQC( n, n )$ such that $\sem{U}_H = \sem{P^\dagger \circ \Lambda \circ P}_H$.

\subsection{A Sound Equational Theory with Unitary Controls}
\label{Sect:Background:Theory}

\begin{figure}[t]
    \begin{framed}
        \textbf{Collection 1: Ordinary Circuit Relations}
        \vspace{-1em}
        \begin{align*}
            \scalebox{0.65}{\circimg{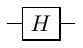}}
            &\overset{\HqcHDef}{=}
            \scalebox{0.65}{\circimg{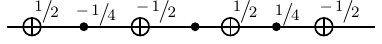}}
            & 
            \scalebox{0.65}{\circimg{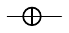}}
            &\overset{\HqcTargDef}{=}
            \scalebox{0.65}{\circimg{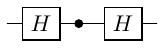}}
            & 
            \scalebox{0.65}{\circimg{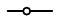}}
            &\overset{\HqcOPhaseDef}{=}
            \scalebox{0.65}{\circimg{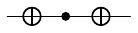}}
        \end{align*}
        \vspace{-2.5em}
        \begin{align*}
            \scalebox{0.65}{\circimg{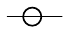}}
            &\overset{\HqcTargMDef}{=}
            \scalebox{0.65}{\circimg{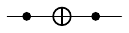}}
            & 
            \scalebox{0.65}{\circimg{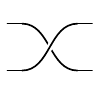}}
            &\overset{\HqcSwapDef}{=}
            \scalebox{0.65}{\circimg{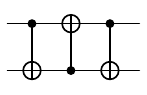}}
            & 
            \scalebox{0.65}{\circimg{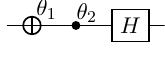}}
            &\overset{\HqcEuler}{=}
            \scalebox{0.65}{\circimg{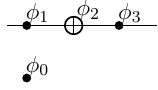}}
            & 
            \scalebox{0.65}{\circimg{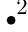}}
            &\overset{\HqcPOrd}{=}
            \scalebox{0.65}{\circimg{tikz-cache/cqc__eqn1_rhs.pdf}}
        \end{align*}
        
        \textbf{Collection 2: Generalized Control Relations}
        \vspace{-0.5em}
        \begin{align*}
            \scalebox{0.65}{\circimg{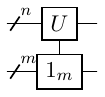}}
            &\overset{\HqcCtrlId}{=}
            \scalebox{0.65}{\circimg{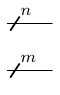}}
            & 
            \scalebox{0.65}{\circimg{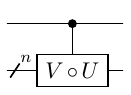}}
            &\overset{\HqcCtrlComp}{=}
            \scalebox{0.65}{\circimg{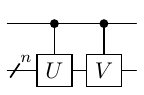}}
            & 
            \scalebox{0.65}{\circimg{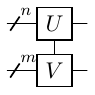}}
            &\overset{\HqcCtrlSwap}{=}
            \scalebox{0.65}{\circimg{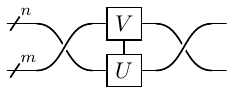}}
            & 
            \scalebox{0.65}{\circimg{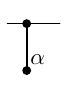}}
            &\overset{\HqcFreePhase}{=}
            \scalebox{0.65}{\circimg{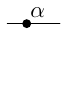}}
        \end{align*}
        \vspace{-1.5em}
        \begin{align*}
            \scalebox{0.65}{\circimg{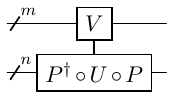}}
            &\overset{\HqcCtrlDag}{=}
            \scalebox{0.65}{\circimg{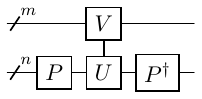}}
            & 
            \scalebox{0.65}{\circimg{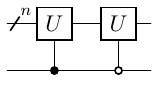}}
            &\overset{\HqcCtrlPair}{=}
            \scalebox{0.65}{\circimg{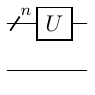}}
            & 
            \scalebox{0.65}{\circimg{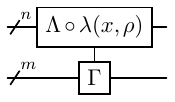}}
            &\overset{\HqcCtrlExpand}{=}
            \scalebox{0.65}{\circimg{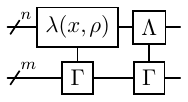}}
        \end{align*}
        \vspace{-1.5em}
        \begin{align*}
            \scalebox{0.65}{\circimg{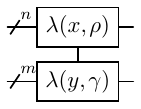}}
            &\overset{\HqcCtrlMerge}{=}
            \scalebox{0.65}{\circimg{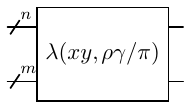}}
        \end{align*}
        
        \textbf{Collection 3: Generalized Exponential Relations}
        \vspace{-1em}
        \begin{align*}
            \scalebox{0.65}{\circimg{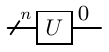}}
            &\overset{\HqcPowZero}{=}
            \scalebox{0.65}{\circimg{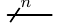}}
            & 
            \scalebox{0.65}{\circimg{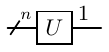}}
            &\overset{\HqcPowId}{=}
            \scalebox{0.65}{\circimg{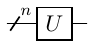}}
            & 
            \scalebox{0.65}{\circimg{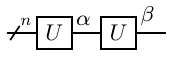}}
            &\overset{\HqcPowAdd}{=}
            \scalebox{0.65}{\circimg{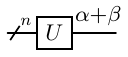}}
            & 
            \scalebox{0.65}{\circimg{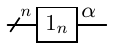}}
            &\overset{\HqcPowIdem}{=}
            \scalebox{0.65}{\circimg{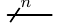}}
        \end{align*}
        \vspace{-2.5em}
        \begin{align*}
            \scalebox{0.65}{\circimg{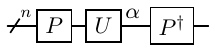}}
            &\overset{\HqcPowDag}{=}
            \scalebox{0.65}{\circimg{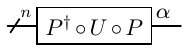}}
            & 
            \scalebox{0.65}{\circimg{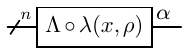}}
            &\overset{\HqcPowExpand}{=}
            \scalebox{0.65}{\circimg{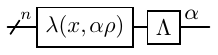}}
        \end{align*}
        
        \textbf{Collection 4: Congruence Relations (Inductively Defined)}
        \vspace{-1em}
        \begin{equation*}
            \scalebox{0.65}{\circimg{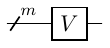}}
            \approx_E
            \scalebox{0.65}{\circimg{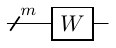}}
            \qquad\Rightarrow\qquad
            \scalebox{0.65}{\circimg{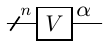}}
            \overset{\HqcPowCong}{=}
            \scalebox{0.65}{\circimg{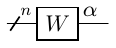}}
            \qquad\text{and}\qquad
            \scalebox{0.65}{\circimg{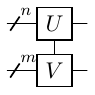}}
            \overset{\HQCCtrlCong}{=}
            \scalebox{0.65}{\circimg{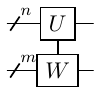}}
        \end{equation*}
        
        \vspace{-1.5em}
    \end{framed}

    \caption{An equational theory for the category of hierarchical quantum circuits in which $\alpha, \beta, \theta_1, \theta_2 \in \mathbb{R}$, $( \phi_0\pi, \phi_1\pi, \phi_2\pi, \phi_3\pi ) = \mathsf{Euler}( \theta_1\pi, \theta_2\pi )$, $x \in \{ 0, 1 \}^n$, $y \in \{ 0, 1 \}^m$, and $\rho, \gamma \in (-\pi, \pi]$.
    The circuits $\Lambda$ and $\Gamma$ are prefixes of diagonal forms such that a lambda generator with binary string $x$ does not appear in $\Lambda$.}
    \label{Fig:HQCEqnTheory}
\end{figure}

The equational theory $E$ for hierarchical quantum circuits can be found in~\cref{Fig:HQCEqnTheory}.
Some care must be taken when defining this equational theory, since \cref{Fig:HQCEqnTheory} also includes congruence relations, such as $( V \approx_E W ) \Rightarrow ( U \odot V \approx_E U \odot W )$.
To avoid cyclic reasoning, $E$ must be defined inductively.
First, let $E_0$ denote the set of relations described by $\HqcHDef$ through $\HqcPowExpand$.
Then for each $n \in \mathbb{N}$, define $E_{n+1}$ as follows.
\begin{align*}
    E_{n+1}
    =
    E_n
    &\cup
    \{ ( V \exparr \alpha, W \exparr \alpha ) : \alpha \in \mathbb{R} \text{ and } V \approx_{E_n} W \}
    &
    (\text{see \HqcPowCong})
    \\
    &\cup
    \{ ( U \odot V, U \odot W ) : U \in \Mor( \CatHQC ) \text{ and } V \approx_{E_n} W \}
    &
    (\text{see \HQCCtrlCong})
\end{align*}
Then $E = \bigcup_{n=0}^\infty E_n$, so that $U \approx_E V$ if and only if there exists some $n \in \mathbb{N}$ such that $U \approx_{E_n} V$.
It follows from $E(1-4)$ that all of $\CatHQC / E$ can be generated using only two of the three gates in
$\{
        \scalebox{0.65}{\circimg{tikz-cache/dagger__term1.pdf}},
        \scalebox{0.65}{\circimg{tikz-cache/dagger__term3.pdf}},
        \scalebox{0.65}{\circimg{tikz-cache/dagger__term5.pdf}}
\}$ together with
$\{
    \;\scalebox{0.65}{\circimg{tikz-cache/dagger__term8.pdf}}
\}$.
Moreover, the relations $\HqcHDef$ and $\HqcTargDef$ are redundant and are only included to make these observations more evident (see~\cref{App:Redundancies}).

\begin{theorem}[Soundness]
    \label{Thm:Soundness}
    If $U \approx_E V$, then $\sem{U}_H = \sem{V}_H$.
\end{theorem}

\begin{example}[Conversion Between Adder Circuits]
    The equational theory in~\cref{Fig:HQCEqnTheory} allows one to perform rewrites to parametrized and controlled subcircuits, even when the controls are non-standard (e.g., controlled on the state $\ket{-}$) or involve nested controlled subcircuits.
    Exponentiation subsumes the notion of parametrized circuits. 
    We argue that all of these notions occur naturally in existing circuits.
    ~\cref{App:Adders} shows an explicit conversion between two implementations of a quantum-quantum adder circuit on 4 qubits. 
    The QFT-based adder proposed by Draper \cite{Draper2000} involves controlled $Z$-rotation gates and quantum Fourier transforms, while the ripple-carry adder proposed by Takahashi \cite{Takahashi2009} uses only CNOT and Toffoli gates. 
    While it is clear from their output that both implement the same unitary, it is not easy to convert one to the other using prior existing equational theory techniques.
    \begin{equation*}
        \begin{array}{c}
             \text{Draper} \\
             \text{QFT-based} \\
             \text{adder } \cite{Draper2000}
        \end{array}
        :=
        \scalebox{0.65}{\circimg{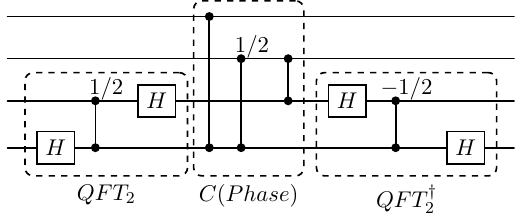}}
        \overunderset{\text{rewrites to}}{\text{using Fig. }\ref{Fig:HQCEqnTheory}}{\longleftrightarrow}
        \scalebox{0.65}{\circimg{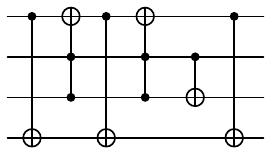}}
        =: 
        \begin{array}{c}
             \text{Takahashi} \\
             \text{ripple-carry} \\
             \text{adder } \cite{Takahashi2009}
        \end{array}
    \end{equation*}
\end{example}

Recall from the previous section that even though $( \dagger )$ mapped each circuit to its intended adjoint, it was not a dagger functor on $\CatHQC$.
This is because $\CatHQC$ was freely generated, and therefore unaware of the intended groupoid structure on $\CatHQC$.
However, it can be shown that $U \circ U^\dagger \approx_E 1_n \approx_E U^\dagger \circ_E U$ for each $U \in \CatHQC( n, n )$.
In categorical terms, this means that $( \dagger )$ lifts to a dagger functor on the quotient category $\CatHQC / E$.
Unfortunately, defining a dagger functor on a quotient of a free prop category can be tricky.
To this end, \cref{Thm:PropInv} first introduces a general procedure which works in the case where all morphisms are intended to be unitary.
The relations ensure that $P( \Sigma ) / E$ is a groupoid and that $H$ maps each morphism in $P( \Sigma ) / E$ to its inverse.
This theorem is then used in \cref{Thm:HQCInv} to show that $( \dagger )$ lifts to a dagger functor.

\begin{theorem}
    \label{Thm:PropInv}
    Let $\Sigma$ be a prop signature, $\tau: \Sigma \to U( P( \Sigma )^\text{op} )$ a prop signature morphism, $E$ a collection of $\Sigma$-equations, and $F: P( \Sigma ) \to P( \Sigma )^\text{op}$ the unique prop functor satisfying the equation $\tau = U( F ) \circ i$.
    If $F( g ) \circ g \approx_E 1_{\dom(g)}$ and $g \circ F( g ) \approx_E 1_{\cod(g)}$ for each $g \in \Sigma$, then there exists a unique prop functor $H: P( \Sigma ) / E \to ( P( \Sigma ) / E )^\text{op}$ such that $H \circ \pi_E = \pi_E^\text{op} \circ F$.
    Moreover, $H$ is a dagger functor and every morphism in $P( \Sigma ) / E$ is unitary.
\end{theorem}

\begin{lemma}
    \label{Lem:InvPow}
    If $U \in \CatHQC( n, n )$ and $\alpha \in \mathbb{R}$, then
    $
        \scalebox{0.65}{\circimg{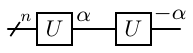}}
        \overset{\HqcPowAdd}{\approx}
        \scalebox{0.65}{\circimg{tikz-cache/hqc__eqn9_lhs.pdf}}
        \overset{\HqcPowZero}{\approx}
        \scalebox{0.65}{\circimg{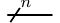}}
    $.
\end{lemma}

\begin{theorem}
    \label{Thm:HQCInv}
    $\pi_E^\text{op} \circ \dagger$ lifts to a unique dagger functor $\overline{\dagger}: \CatHQC / E \to ( \CatHQC / E )^\text{op}$ and every morphism in $\CatHQC / E$ is unitary with respect to $( \overline{\dagger} )$.
\end{theorem}

It can then be shown that mapping each circuit $U$ to a circuit $\scalebox{0.65}{\circimg{tikz-cache/dagger__term1.pdf}} \odot U$ defines a dagger control functor on $\CatHQC / E$.
This will be necessary to compare the prop presentation $( \Sigma_{\CatHQC}, E )$ with the controlled prop presentation $( \Sigma_{\CatCQC}, Q )$.
To simplify this construction, it is first shown that every control functor on a category such as $\CatHQC / E$ must be unitary (\cref{Lem:UnitaryCtrl}).
Using this lemma, a general procedure is then introduced for defining control functors on prop presentations (\cref{Thm:DefCtrl}), which is then applied to $\CatHQC$ with $\tau_n( U: n \to n ) = \scalebox{0.65}{\circimg{tikz-cache/dagger__term1.pdf}} \odot U$ in \cref{Thm:HQCCtrl}.
Note that all conditions except for (4) correspond directly to relations in $E$.

\begin{lemma}
    \label{Lem:UnitaryCtrl}
    If $( \mathcal{C}, \dagger )$ is a dagger prop such that every morphism in $\mathcal{C}_{\mathbf{endo}}$ is unitary, then every control functor $C: \mathcal{C}_{\mathbf{endo}} \to \mathcal{C}_{\mathbf{endo}}$ is a dagger control functor.
\end{lemma}

\begin{theorem}
    \label{Thm:DefCtrl}
    Let $\Sigma$ be a prop signature consisting of endomorphic generators, $E$ a collection of $\Sigma$-equations, and $\{ \tau_n: P( \Sigma )( n, n ) \to P( \Sigma )( n + 1, n + 1 ) \}_{n \in \mathbb{N}}$ be a family of ordinary functions.
    Consider the following list of properties which $\{ \tau_n \}_{n \in \mathbb{N}}$ could satisfy.
    \begin{enumerate}
    \item If $n \in \mathbb{N}$, then $\tau_n( 1_n ) \approx_E 1_{n+1}$.
    \item If $f: n \to n$ and $g: n \to n$, then $\tau_n( g \circ f ) \approx_E \tau_n( g ) \circ \tau_n( f )$.
    \item If $f: n \to n$ and $( f, g ) \in E$, then $\tau_n( f ) \approx_E \tau_n( g )$.
    \item If $f: n \to n$, then $\tau_{n+1}( f \boxtimes 1 ) \approx_E \tau( n ) \boxtimes 1$.
    \item If $f: n \to n$, then $\tau_{n+1}( \tau_n( f ) ) \approx_E ( \sigma \boxtimes 1_n ) \circ \tau_{n+1}( \tau_n( f ) ) \circ ( \sigma \boxtimes 1_n )$.
    \item If $f: n + 2 \to n + 2$ and $0 \le k \le n$, then $\tau_{n+2}( \gamma_{n,k} \circ f \circ \gamma_{n,k} ) \approx_E (1 \boxtimes \gamma_{n,k} ) \circ \tau_{n+2}( f ) \circ ( 1 \boxtimes \gamma_{n,k} )$.
    \end{enumerate}
    If conditions (1) and (2) hold, then there exists a unique ordinary functor $F: P( \Sigma ) \to P( \Sigma ) / E$ such that $\Ob( F )( n ) = n + 1$ and $F( f: n \to n ) = \pi_E( \tau_n( f ) )$.
    If condition (3) also holds, then there exists a unique ordinary functor $H: P( \Sigma ) / E \to P( \Sigma ) / E$ such that $H \circ \pi_E = F$.
    If conditions (4) through to (6) also hold, then $H$ is a control functor.
\end{theorem}

\begin{theorem}
    \label{Thm:HQCCtrl}
    There exists a unique conjugated control functor $C: \CatHQC / E \to \CatHQC / E$ satisfying the equation $C( \pi_E( U ) ) = \pi_E( \scalebox{0.65}{\circimg{tikz-cache/dagger__term1.pdf}} \odot U )$ for each $U \in \Mor( \CatHQC )$.
\end{theorem}

\section{Equational Completeness}
\label{Sect:Completeness}

This section proves that $E$ is a complete equational theory for $\CatHQC$ with respect to the semantic interpretation $\sem{-}_H$.
The proof begins by showing that the equational theory is complete for the sub-language $\Core$ of $\CatHQC$ consisting of global phase gates, single qubit rotations, and controls in the $Z$-basis.
The idea of this proof is to leverage the completeness of $( \Sigma_{\CatCQC}, Q )$ with respect to $\sem{-}_C$.
We will first consider the simpler case, in which $( \Sigma_{\CatCQC}, Q )$ is a merely a prop presentation, as opposed to a controlled prop presentation.
\begin{enumerate}
\item Construct a functor $\encoder: \Core \to \CatCQC$ such that $\sem{\encoder( - )}_C = \sem{-}_H$.
      This means that if $\sem{U}_H = \sem{V}_H$, then $\sem{\encoder(U)}_C = \sem{\encoder(V)}_C$, which implies that $\encoder( U ) \approx_Q \encoder( V )$.
\item Construct a functor $\decoder: \CatCQC \to \Core$ which respects the relations in $Q$.
      This means that for each relation $( X, Y ) \in Q$ used to prove $\encoder( U ) \approx_Q \encoder( V )$, there is a corresponding sequence of relations $\decoder( X ) \approx_E \decoder( Y )$.
\item Ensure that $W \approx_E \decoder( \encoder( W ) )$ for all $W \in \Mor( \CatHQC )$.
      This is because the decoder can only prove that $\decoder( \encoder( U ) ) \approx_E \decoder( \encoder( V ) )$. 
\end{enumerate}
Unfortunately, $\CatCQC$ is a free controlled prop, so $\decoder( - )$ must map into the quotient $\Core / E$, as opposed to $\Core$.
This means that $\decoder( - )$ maps each circuit in $\CatCQC$ to an equivalence class of circuits in $\Core$, from which representative circuits must be chosen.

\begin{theorem}
    \label{Thm:Completeness1}
    Let $( \Gamma, Q )$ be a controlled prop presentation which is complete with respect to the interpretation $\sem{-}_\Gamma: P_C( \Gamma ) \to \mathcal{C}$, and $( \Sigma, E )$ be a prop presentation equipped with a control functor $C: P( \Sigma ) / E \to P( \Sigma ) / E$ and a prop functor $\sem{-}_\Sigma: P( \Sigma ) \to \mathcal{C}$.
    Assume that there exists a prop functor $\encoder: P( \Sigma ) \to P_C( \Gamma )$ and a controlled prop functor $\decoder: P_C( \Gamma ) \to ( P( \Sigma ) / E, C )$ such that the following properties hold.
    \begin{enumerate}
    \item If $x \in \Sigma$, then $\sem{\encoder(x)}_\Gamma = \sem{x}_\Sigma$.
    \item If $( X, Y ) \in Q$, then there exists $X^* \in \decoder( X )$ and $Y^* \in \decoder( Y )$ such that $X^* \approx_E Y^*$.
    \item If $x \in \Sigma$, then there exists $f \in \decoder( \encoder ( x ) )$ such that $x \approx_E f$.
    \end{enumerate}
    Then $( \Sigma, E )$ is complete with respect to the semantic interpretation $\sem{-}_\Sigma$.
\end{theorem}

It remains to show that the completeness results extend to all of $\CatHQC$.
The basic idea is to show that for each $U \in \Mor( \CatHQC )$, there exists a $V \in \Mor( \Core )$ such that $U \approx_E V$.
This means that every circuit in $\CatHQC$ can be rewritten to a circuit in $\Core$, in which $E$ is already known to be complete.

\begin{theorem}
    \label{Thm:Completeness2}
    Let $\Sigma \subseteq \Gamma$ be prop signatures with an equation theory $( \Sigma, E )$ that is sound and complete with respect to $\sem{-}: P( \Gamma ) \to \mathcal{C}$.
    If for each $g \in \Gamma$, there exists a $f\in P( \Sigma )$ such that $g \approx_E f$, then $( \Gamma, E )$ is a complete equational theory with respect to $\sem{-}$.
\end{theorem}

\vspace{1em}
\noindent
\textbf{Constructing the Sub-Language}.
The sub-language is constructed inductively, starting from the gate set
$\Sigma_{\Core}^0 =
    \{
        \scalebox{0.65}{\circimg{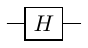}}
    \}
    \cup
    \{
        \scalebox{0.65}{\circimg{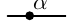}}
        ,
        \scalebox{0.65}{\circimg{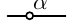}}
        ,
        \scalebox{0.65}{\circimg{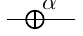}}
        ,
        \scalebox{0.65}{\circimg{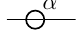}}
        ,
        \scalebox{0.65}{\circimg{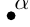}}
        :
        \alpha \in \mathbb{R}
    \}$.
The construction then proceeds as follows.
First, a free prop category $\Core[n] = P( \Sigma_{\Core}^n ) $ is constructed for each control depth $n \in \mathbb{N}$.
The category at control depth $n$ is then closed under a single application of the control functor, by setting $\Sigma_{\Core}^{n+1} = \Sigma_{\Core}^n \cup \{ \scalebox{0.65}{\circimg{tikz-cache/dagger__term1.pdf}} \odot U : U \in \Mor( \Core[n] ) \}$ for each $n \in \mathbb{N}$.
This yields an infinite diagram of chain of inclusions, as in \cref{Sect:Semantics}.
The categorical colimit of this sequence will be denoted $\Core$.
This means $\Core = P( \Sigma_{\Core} )$ where $\Sigma_{\Core} = \bigcup_{n=0}^{\infty} \Sigma_{\Core}^n$.
Clearly $C( - )$ restricts to $\Core / E$, since $\scalebox{0.65}{\circimg{tikz-cache/dagger__term1.pdf}} \odot U$ is a representative of $C( \pi_E( U ) )$.

\begin{lemma}
    \label{Lem:CRestrict}
    The functor $C: \CatHQC / E \to \CatHQC / E$ restricts to $\Core / E$.
\end{lemma}

Unfortunately, $( \dagger )$ does not restrict to $\Core$ since $U \odot V \in \Mor( \Core )$ does not imply that $( U \odot V ) \exparr (-1) \in \Mor( \Core )$.
However, there does exist a prop functor $\ddagger: \Core \to \Core^\text{op}$ such that $U^\ddagger \approx_E U^\dagger$ for all $U \in \Mor( \Core )$.
It follows from \cref{Lem:FuncDefFam} that the equations $d_0( g ) = g^\dagger$ and $d_{n+1}( \scalebox{0.65}{\circimg{tikz-cache/dagger__term1.pdf}} \odot U ) = \scalebox{0.65}{\circimg{tikz-cache/dagger__term1.pdf}} \odot d_n( U )$ define a prop functor $\ddagger: \Core \to \Core^{\text{op}}$.

\begin{lemma}
    \label{Lem:CoreDag}
    If $U \in \Mor( \Core )$, then $U^\ddagger \approx_E U^\dagger$.
\end{lemma}

\vspace{1em}
\noindent
\textbf{Constructing the Encoder}.
The encode will be defined by the following set of equations, together with the inductive rule $e_{n+1}( \scalebox{0.65}{\circimg{tikz-cache/dagger__term1.pdf}} \odot U ) = C( e_n( U ) )$ defined for each $n \in \mathbb{N}$.
\begin{align*}
    e_0\left( \scalebox{0.65}{\circimg{tikz-cache/encoder__eqn1_lhs.pdf}} \right)
    &=
    \scalebox{0.65}{\circimg{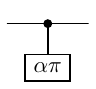}}
    & 
    e_0\left( \scalebox{0.65}{\circimg{tikz-cache/encoder__eqn2_lhs.pdf}} \right)
    &=
    \scalebox{0.65}{\circimg{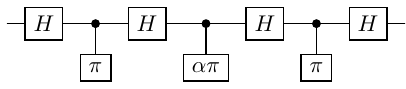}}
    & 
    e_0\left( \scalebox{0.65}{\circimg{tikz-cache/encoder__eqn3_lhs.pdf}} \right)
    &=
    \scalebox{0.65}{\circimg{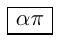}}
    \\ 
    e_0\left( \scalebox{0.65}{\circimg{tikz-cache/encoder__eqn4_lhs.pdf}} \right)
    &=
    \scalebox{0.65}{\circimg{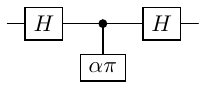}}
    & 
    e_0\left( \scalebox{0.65}{\circimg{tikz-cache/encoder__eqn5_lhs.pdf}} \right)
    &=
    \scalebox{0.65}{\circimg{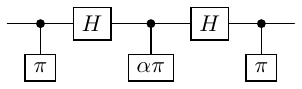}}
    & 
    e_0\left( \scalebox{0.65}{\circimg{tikz-cache/encoder__eqn6_lhs.pdf}} \right)
    &=
    \scalebox{0.65}{\circimg{tikz-cache/encoder__eqn6_lhs.pdf}}
\end{align*}
It follows immediately from \cref{Lem:FuncDefFam} that this defines a prop functor $\encoder: \Core \to \CatCQC$.
Moreover, this prop functor preserves the semantics of $\CatHQC$.
For gates in $\Sigma_{\Core}^0$, this follows by a direct matrix computation.
For gates in $\Sigma_{\Core}^{n+1} \setminus \Sigma_{\Core}^n$, this follows from \cref{Thm:GenCtrlNF}.

\begin{lemma}
    \label{Lem:Encoder}
    If $g \in \Sigma_{\Core}$, then $\sem{\encoder(g)}_C = \sem{g}_H$.
\end{lemma}

\vspace{1em}
\noindent
\textbf{Constructing the Decoder}.
Let $\decoder: \CatCQC \to ( \Core / E, C )$ be the unique controlled prop functor defined by the equations $\decoder( \scalebox{0.65}{\circimg{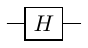}} ) = \pi_E( \scalebox{0.65}{\circimg{tikz-cache/decoder__eqn1.pdf}} )$ and $\decoder( \scalebox{0.65}{\circimg{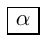}} ) = \pi_E\left( \scalebox{0.65}{\circimg{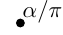}} \right)$.
The relations $Q( 1 )$, $Q( 2 )$, and $Q( 5 )$ appear in the set of relations for $\CatHQC$, and are therefore respected by $\decoder( - )$.
The relation $Q( 4 )$ is also respected by $\decoder( - )$ since it is an instance of unitarity.
The relations $Q( 6 )$ through to $Q( 9 )$ also hold since $C( - )$ is conjugated.
To show that relation $Q( 3 )$ is also respected, it suffices to prove \cref{Lemma:CNot}.

\begin{lemma}
    \label{Lemma:CNot}
    The following equations hold.
    \begin{align*}
        \scalebox{0.65}{\circimg{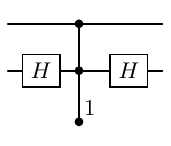}}
        &\in
        \decoder\left( \scalebox{0.65}{\circimg{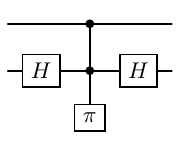}} \right)
        &
        \scalebox{0.65}{\circimg{tikz-cache/cnot__rep.pdf}}
        &\approx_E
        \scalebox{0.65}{\circimg{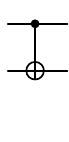}}
    \end{align*}
\end{lemma}

\begin{lemma}
    \label{Lem:Decoder}
    If $( X, Y ) \in Q$, then $X^* \approx_E Y^*$ for some $X^* \in \decoder( X )$ and $Y^* \in \decoder( Y )$.
\end{lemma}

\vspace{1em}
\noindent
\textbf{Round-Trip Translation}.
At this point, an encoder $\encoder: \Core \to \CatCQC$ and a decoder $\decoder: \CatCQC \to \Core / E$ have been established.
It remains to be shown that $\pi_E( g ) = \decoder( \encoder( g ) )$ for each $g \in \Sigma_{\Core}$.
Since $\decoder( - )$ is a free control functor, then it suffices to prove the condition for $g \in \mathcal{G}_0$.
This follows via routine derivations and consequently establishes completeness.

\begin{lemma}
    \label{Lem:RoundTrip}
    If $g \in \Sigma_{\Core}$, then there exists $U \in \decoder( \encoder( g ) )$ such that $g \approx_E U$.
\end{lemma}

\begin{theorem}
    \label{Thm:CoreComplete}
    $( \Sigma_{\Core}, E )$ is complete with respect to the semantic interpretation $\sem{-}_H$.
\end{theorem}

\subsection{Completeness for the Full Language}
\label{Sect:Completeness:Full}

It remains to be shown that every circuit in $\CatHQC$ can be reduced to a circuit in $\Core$.
The proof begins by showing that this is possible when a set of primitive gates is added to $\Core$.
This new language is denoted $\CoreExt[0]$.
The proof then proceeds by induction on $\CoreExt[n]$, in which each step adds increasingly more complex gates to $\CoreExt[n]$.
At each step of the proof, it is shown how a reduction procedure for $\CoreExt[n]$ can be used to obtain a reduction procedure for $\CoreExt[ n + 1 ]$.
Formally, let $\Sigma_{\CoreExt}^0 = \Sigma_{\Core} \cup \Sigma_{\mathbf{Prim}}$ and $\CoreExt[0] = P( \Sigma_{\CoreExt}^0 )$.
Then for each $n \in \mathbb{N}$, let $\Sigma_{\CoreExt}^{n+1} = \Sigma_{\CoreExt}^n \cup \Tower( \CoreExt[n] ) \cup \Power( \CoreExt[n] )$ with $\Core[n+1] = P( \Sigma_{\CoreExt}^n )$.
It is not hard to see that $\Sigma_{\CatHQC}^n \subseteq \Sigma_{\CoreExt}^n \subseteq \Sigma_{\CatHQC}$ for each $n \in \mathbb{N}$.
It follows that the categorical colimit of this sequence is $\CatHQC$.

\vspace{1em}
\noindent
\textbf{The Base Reduction}.
The only gates which appear in $\CoreExt[0]$ but not in $\Core$ are the primitive rotation gates free from exponents.
Of course, the relation $\HqcPowId$ in $E$ allows for these gate to be rewritten to the form $( g \exparr 1 )$.
This motivates the following reduction.
\begin{align*}
    r_0( \scalebox{0.65}{\circimg{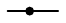}} )
    &=
    \scalebox{0.65}{\circimg{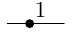}}
    &
    r_0( \scalebox{0.65}{\circimg{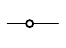}} )
    &=
    \scalebox{0.65}{\circimg{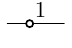}}
    &
    r_0( \scalebox{0.65}{\circimg{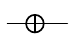}} )
    &=
    \scalebox{0.65}{\circimg{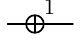}}
    &
    r_0( \scalebox{0.65}{\circimg{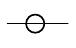}} )
    &=
    \scalebox{0.65}{\circimg{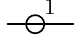}}
\end{align*}
\vspace{-2em}
\begin{align*}
    r_0( \scalebox{0.65}{\circimg{tikz-cache/dagger__term5.pdf}} )
    &=
    \scalebox{0.65}{\circimg{tikz-cache/dagger__term5.pdf}}
    &
    r_0( U \exparr \alpha )
    &=
    U \exparr \alpha
    &
    r_0( U \odot V )
    &=
    U \odot V
\end{align*}
Then by \cref{Lem:FuncDefPart}, these equations define a unique prop functor $\Reduce_0: \CoreExt[0] \to \Core$ with the property that $U \approx_E \Reduce_0( U )$ for each $U \in \Mor( \CoreExt[0] )$.

\vspace{1em}
\noindent
\textbf{The Inductive Construction}.
Assume that there exists a functor $\Reduce_n: \CoreExt[n] \to \Core$ with the property that $U \approx_E \Reduce_n( U )$ for each $U \in \Mor( \CoreExt[n] )$. 
The goal of this step is to construct a new functor $\Reduce_{n+1}: \CoreExt[n+1] \to \Core$ such that $U \approx_E \Reduce_{n+1}( U )$ for each $U \in \Mor( \CoreExt[n+1] )$.
Of course, the new gates in $\CoreExt[n+1]$ are gates of the form $U \odot V$ and $U \exparr \alpha$ where $U \in \CoreExt[n]$ and $V \in \CoreExt[n]$.
For simplicity, we will consider the case of exponentiation.
Since $U \in \CoreExt[n]$, then $( U \exparr \alpha )$ rewrites to $( V \exparr \alpha )$ where $V = \Reduce_n( U )$.
It is not hard to show that $\Core$ is universal for unitary qubit quantum computation.
Then means that there exists a circuit $P$ and a diagonal circuit $\Lambda$ such that $\sem{P^\ddagger \circ \Lambda \circ P}_H = \sem{V}_H$.
Since $E$ is complete for $\Core$ with respect to $\sem{-}_H$, then $V \approx_E P^\ddagger \circ \Lambda \circ P$.
It then follows from relations $\HqcPowDag$, $\HqcPowCong$, and \cref{Lem:CoreDag} that $U \exparr \alpha \approx_E P^\ddagger \circ ( \Lambda \exparr \alpha ) \circ P$.
Using relation $\HqcPowId$ and $\HqcPowExpand$, it is then possible to reduce $ ( \Lambda \exparr \alpha )$ to a diagonal circuit in $\Core$.
A similar procedure can be carried out for $U \odot V$, using relations $\HqcCtrlExpand$ and $\HqcCtrlMerge$.
These procedures are stated formally in~\cref{App:Routines} with proofs of termination, and are denoted $\Drop_n( U, \alpha )$ and $\Expand_n( U, V )$, respectively.
Using these procedures, the following reduction is obtained.
\begin{equation*}
    r_{n+1}( g ) = \begin{cases}
        r_n( g ) & \text{if } g \in \Sigma_{\CoreExt}^n
        \\
        \Expand_{n+1}( g ) & \text{if } g \in \Tower( \CoreExt[n] ) \cap \Mor( \CoreExt[n + 1] )
        \\
        \Drop_{n+1}( U, \alpha ) & \text{if } g = U \exparr \alpha \text{ for } U \in \Mor( \CoreExt[n] ) \text{ and } \alpha \in \mathbb{R}
    \end{cases}
\end{equation*}
This defines a unique prop functor $\Reduce_{n+1}: \CoreExt[n+1] \to \Core$ with the property that $U \approx_E \Reduce_{n+1}( U )$ for each $U \in \Mor( \CoreExt[n+1] )$.

\begin{theorem}
    \label{Thm:HQCComplete}
    $( \Sigma_{\CatHQC}, E )$ is complete with respect to the semantic interpretation $\sem{-}_H$.
\end{theorem}

\section{Conclusion}

In this paper, we introduced the notion of generalized controls.
We studied its algebraic properties and established its relation to circuit exponentiation and conjugated control functors.
Based on these operations, we then introduced the notion of a hierarchical circuit language, which extends a standard prop category with an abstract notion of generalized control and exponentiation.
It was then shown how the algebraic properties of generalized controls could be used to obtain a sound and complete equational theory for the category of hierarchical quantum circuits.
A key insight obtained from this proof is that reducing hierarchical quantum circuits to controlled quantum circuits can be seen as a form of diagrammatic circuit diagonalization.
There are several directions for future work.
\begin{enumerate}
\item Finding a purely categorical description for $( \odot )$ and $( \;\exparr\; )$, and their pointed extensions.
\item Exploring quantum circuit compilation from the perspective of hierarchical circuits.
\item Identifying sub-languages for which diagrammatic diagonalization is efficiently computable.
\end{enumerate}

\bibliographystyle{eptcs}
\bibliography{generic}

\appendix
\allowdisplaybreaks
\newpage
\section{Example: converting a QFT-based adder circuit to a ripple-carry adder circuit}
\label{App:Adders}

The following is a four-qubit example transpilation between two forms of adder circuits, the QFT-based adder circuit proposed by Draper \cite{Draper2000}, and the ripple-carry adder proposed by Takahashi et al. \cite{Takahashi2009} for performing addition modulo $2^n$. 
These circuits have the action $\ket{a}\ket{b} \mapsto \ket{a}\ket{a+b}$. 
The bit ordering of $b = b_1b_0$ has been reversed for the sake of visual clarity in the circuit diagram.

Before proceeding with the transpilation, we present some intermediate derivations.
First, we note from~\ref{App:involution_proofs} that $H^2 = Z^2 = X^2 = \mathbf{1}$ as usual. 
\begin{align}\label{eqn:gen_involutions}
    \scalebox{0.65}{\circimg{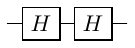}}
    &\approx
    \scalebox{0.65}{\circimg{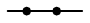}}
    \approx
    \scalebox{0.65}{\circimg{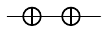}}
    \approx
    \scalebox{0.65}{\circimg{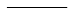}}
\end{align}
Second, since the QFT-based adder circuit involves some $\lambda$ generators, we note that the position of the exponent is irrelevant. 
In particular, whether the exponent is ``attached'' to the entire $\lambda$ generator or ``attached'' to just one of the constituent gates is irrelevant, as both result in the same gate.
\begin{align*}
    \scalebox{0.65}{\circimg{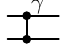}}
    &:=
    \scalebox{0.65}{\circimg{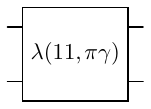}}
    \overset{\HqcCtrlMerge}{\approx}
    \scalebox{0.65}{\circimg{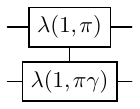}}
    =:
    \scalebox{0.65}{\circimg{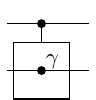}}
\end{align*}
We choose the path of least notation and place all exponents at the top right of their respective $\lambda$ generators, without drawing any boxes.

Third, pushing a Hadamard gate past a $\bullet$ or $\oplus$ gate flips one symbol to the other.
\begin{align}\label{eqn:type_flip}
    \scalebox{0.65}{\circimg{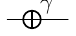}}
    &\overset{\HqcTargDef}{\approx}
    \scalebox{0.65}{\circimg{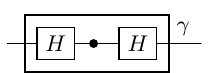}}
    \overset{\HqcPowDag}{\approx}
    \scalebox{0.65}{\circimg{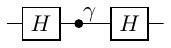}}
\end{align}
\begin{align}\label{eqn:HqcTargDef'}
    \scalebox{0.65}{\circimg{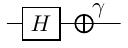}}
    &\overset{\ref{eqn:type_flip}}{\approx}
    \scalebox{0.65}{\circimg{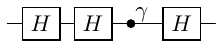}}
    \overset{\ref{eqn:gen_involutions}}{\approx}
    \scalebox{0.65}{\circimg{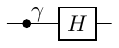}}
\end{align}
Fourth, by the same argument, pushing a $\oplus$ gate past a $\bullet$ or $\circ$ gate flips one symbol to the other.
\begin{align}\label{eqn:flip}
    \scalebox{0.65}{\circimg{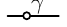}}
    &\overset{\HqcOPhaseDef}{\approx}
    \scalebox{0.65}{\circimg{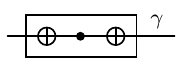}}
    \overset{\HqcPowDag}{\approx}
    \scalebox{0.65}{\circimg{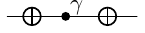}}
\end{align}
\begin{align}\label{eqn:HqcOPhaseDef'}
    \scalebox{0.65}{\circimg{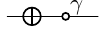}}
    &\overset{\ref{eqn:flip}}{\approx}
    \scalebox{0.65}{\circimg{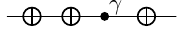}}
    \overset{\ref{eqn:gen_involutions}}{\approx}
    \scalebox{0.65}{\circimg{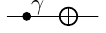}}
\end{align}
Fifth, CNOT is self-inverse as expected.
\begin{align}\label{eqn:CNOTId}
    \scalebox{0.65}{\circimg{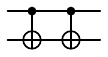}}
    &\overset{\HqcCtrlComp}{\approx}
    \scalebox{0.65}{\circimg{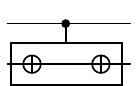}}
    \overunderset{\HQCCtrlCong}{\ref{eqn:gen_involutions}}{\approx}
    \scalebox{0.65}{\circimg{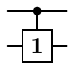}}
    \overset{\HqcCtrlId}{\approx}
    \scalebox{0.65}{\circimg{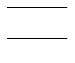}}
\end{align}
Sixth, relation \HqcCtrlComp holds in any basis. In particular, it holds in the $X$ basis.
\begin{align}\label{eqn:HqcCtrlComp'}
    \scalebox{0.65}{\circimg{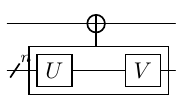}}
    &\overset{\HqcTargDef}{\approx}
    \scalebox{0.65}{\circimg{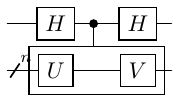}}
    \overset{\HqcCtrlComp}{\approx}
    \scalebox{0.65}{\circimg{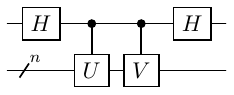}}
    \\
    &\overunderset{\HQCCtrlCong}{\ref{eqn:HqcTargDef'}}{\approx}
    \scalebox{0.65}{\circimg{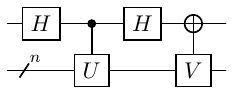}}
    \overunderset{\HQCCtrlCong}{\ref{eqn:HqcTargDef'}}{\approx}
    \scalebox{0.65}{\circimg{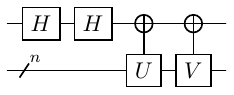}}
    \overset{\ref{eqn:gen_involutions}}{\approx}
    \scalebox{0.65}{\circimg{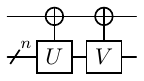}}
\end{align}
Seventh and last, relation \HqcCtrlPair holds in any basis. In particular, it holds in the $X$ basis.
\begin{align}\label{eqn:HqcCtrlPair'}
    \scalebox{0.65}{\circimg{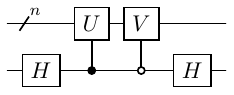}}
    &\overset{\ref{eqn:gen_involutions}}{\approx}
    \scalebox{0.65}{\circimg{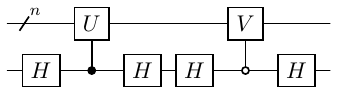}}
    \overset{\HqcCtrlDag}{\approx}
    \scalebox{0.65}{\circimg{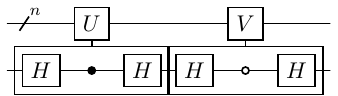}}
    \overset{\HqcTargDef,\HqcTargMDef}{\approx}
    \scalebox{0.65}{\circimg{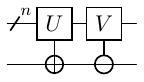}}
\end{align}

With those preparatory derivations out of the way, we proceed to the transpilation. We begin from a QFT-based adder circuit \cite{Draper2000} and show an explicit step-by-step transpilation to a ripple-carry adder circuit \cite{Takahashi2009}. All intermediate circuits are semantically equivalent by soundness of Fig. \ref{Fig:HQCEqnTheory}, meaning they all implement the same unitary transformation.
\begin{align*}
    \scalebox{0.65}{\circimg{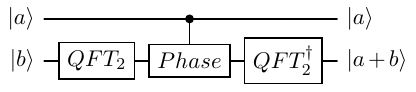}}
    &:=
    \scalebox{0.65}{\circimg{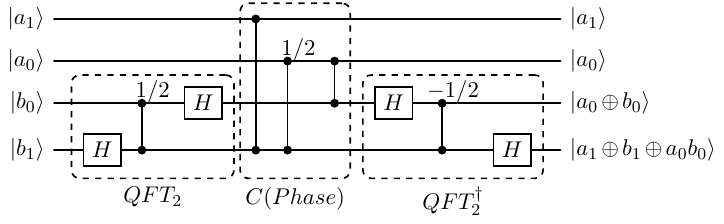}}
    \\
    \overunderset{\HQCCtrlCong}{\ref{eqn:HqcTargDef'}}{\approx}
    \scalebox{0.65}{\circimg{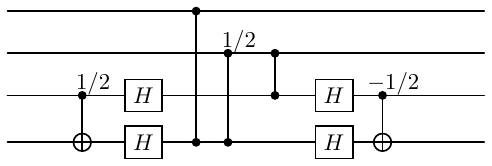}}
    &\overunderset{\HQCCtrlCong}{\ref{eqn:HqcTargDef'}}{\approx}
    \scalebox{0.65}{\circimg{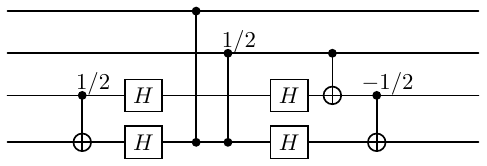}}
    \\
    \overunderset{\HQCCtrlCong}{\ref{eqn:HqcTargDef'}}{\approx}
    \scalebox{0.65}{\circimg{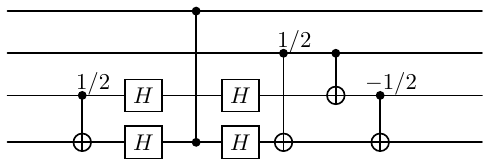}}
    &\overunderset{\HQCCtrlCong}{\ref{eqn:HqcTargDef'}}{\approx}
    \scalebox{0.65}{\circimg{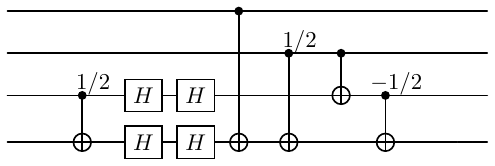}}
    \\
    \overset{\ref{eqn:gen_involutions}}{\approx}
    \scalebox{0.65}{\circimg{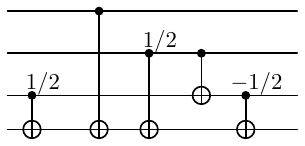}}
    &\overset{\HqcCtrlComp}{\approx}
    \scalebox{0.65}{\circimg{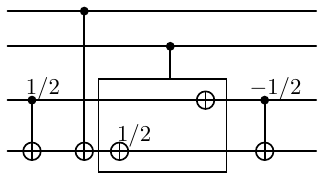}}
    \approx
    \scalebox{0.65}{\circimg{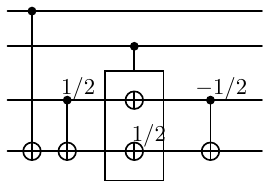}}
    \\
    \overset{\HqcCtrlDag}{\approx}
    \scalebox{0.65}{\circimg{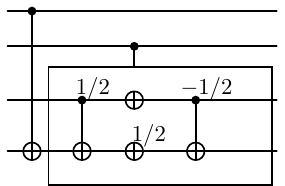}}
    &\overunderset{\HQCCtrlCong}{\ref{eqn:HqcOPhaseDef'}}{\approx}
    \scalebox{0.65}{\circimg{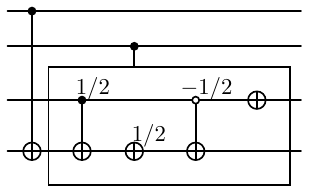}}
    \overunderset{\HQCCtrlCong}{\HqcCtrlPair}{\approx}
    \scalebox{0.65}{\circimg{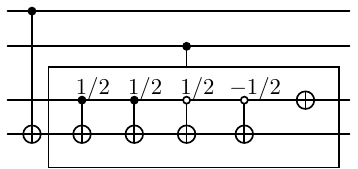}}
    \\
    \overunderset{\HQCCtrlCong}{\HqcPowAdd}{\approx}
    \scalebox{0.65}{\circimg{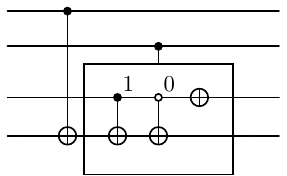}}
    &\overunderset{\HQCCtrlCong}{\HqcPowId,\HqcPowZero}{\approx}
    \scalebox{0.65}{\circimg{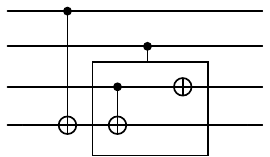}}
    \overset{\HqcCtrlComp}{\approx}
    \scalebox{0.65}{\circimg{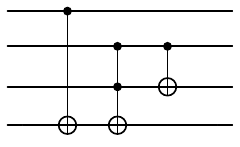}}
    \\
    \overset{\ref{eqn:CNOTId}}{\approx}
    \scalebox{0.65}{\circimg{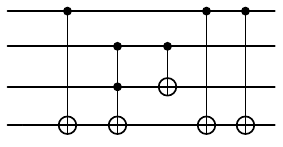}}
    &\approx
    \scalebox{0.65}{\circimg{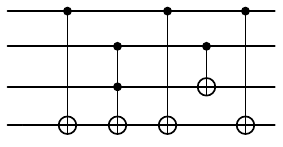}}
    \overset{\ref{eqn:HqcCtrlComp'}}{\approx}
    \scalebox{0.65}{\circimg{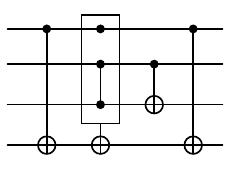}}
    \\
    \overunderset{\HQCCtrlCong}{\ref{eqn:HqcCtrlPair'}}{\approx}
    \scalebox{0.65}{\circimg{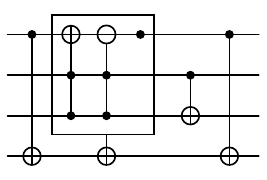}}
    &\overunderset{\HQCCtrlCong}{\HqcTargDef}{\approx}
    \scalebox{0.65}{\circimg{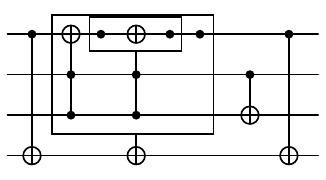}}
    \overunderset{\HQCCtrlCong}{\HqcCtrlDag}{\approx}
    \scalebox{0.65}{\circimg{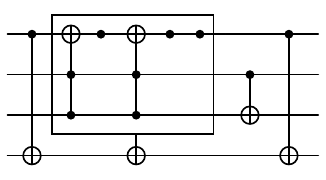}}
    \\
    \overunderset{\HQCCtrlCong}{\ref{eqn:gen_involutions}}{\approx}
    \scalebox{0.65}{\circimg{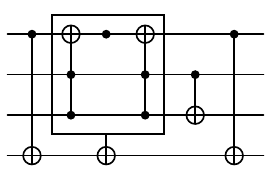}}
    &\overset{\HqcCtrlDag}{\approx}
    \scalebox{0.65}{\circimg{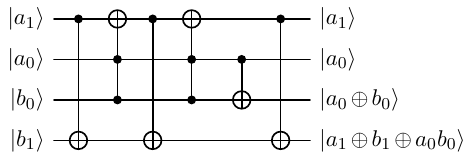}}
    =: 
    \begin{array}{c}
         \text{Takahashi} \\
         \text{ripple-carry} \\
         \text{adder } \cite{Takahashi2009}
    \end{array}
\end{align*}
\section{Proofs for Section~\ref{Sect:Semantics}}
%

This section contains all proofs for~\cref{Sect:Semantics}.

\subsection{Proof of Theorem~\ref{Thm:ControlAssoc}}
\begin{proof}
    First, it must be shown that $( \odot )$ is unital with unit $( -1 )$.
    Since $\Log( -1 ) = i\pi$, then the following equations hold for all $U \in \Mor( \CatUnitary )$.
    \begin{align*}
        (-1) \odot U
        &=
        \exp( \Log(-1) \otimes \Log(U) / (i\pi) )
        =
        \exp( \Log( U ) )
        =
        U
        \\
        U \odot (-1)
        &=
        \exp( \Log(U) \otimes \Log(-1) / (i\pi) )
        =
        \exp( \Log( U ) )
        =
        U
    \end{align*}
    Then $(-1) \odot U = U = U \odot (-1)$ for all $U \in \Mor( \CatUnitary )$.
    In other words, $( \odot )$ is unital with unit $( -1 )$.
    It remains to be shown that $( \odot )$ is associative.
    To this end, let $U, V, W \in \Mor( \CatUnitary )$.
    Let $\{ \alpha_j \}_{j=1}^n$ denote the eigenvalues of $\Log( U )$ with orthonormal eigenbasis $\{ b_j \}_{j=1}^n$ and $\{ \beta_j \}_{j=1}^m$ denote the eigenvalues of $\Log( V )$ with orthonormal eigenbasis $\{ c_j \}_{j=1}^m$.
    Then then following equation holds where $A = \Log( U ) \otimes \Log( V ) / (i\pi)$, since $( \otimes )$ is bilinear.
    \begin{equation*}
        A
        =
        \frac{1}{i\pi}
        \left( \sum_{j=1}^n \alpha_j \ket{b_j} \bra{b_j} \right)
        \otimes
        \left( \sum_{k=1}^m \beta_k \ket{c_k} \bra{c_k} \right)
        =
        \sum_{j=1}^n
        \sum_{k=1}^m
        \frac{\alpha_j\beta_k}{i\pi} ( \ket{b_j} \otimes \ket{c_k} ) ( \bra{b_j} \otimes \bra{c_k} )
    \end{equation*}
    Then $\{ \ket{b_j} \otimes \ket{c_k} : j \in \{ 1, 2, \ldots, n \}, k \in \{ 1, 2, \ldots, m \} \}$ is an orthonormal eigenbasis for $A$.
    Moreover, for each $j \in \{ 1, 2, \ldots, n \}$ and $k \in \{ 1, 2, \ldots, m \}$, the eigenvalue associated with $\ket{b_j} \otimes \ket{c_k}$ is $(\alpha_j\beta_k)/(i\pi)$.
    Since $\beta_k \in i(-\pi,\pi]$, then $\beta_k/(i\pi) \in (-1,1]$.
    Consequently, $(\alpha_j\beta_k)/(i\pi) \in i(-\pi,\pi]$.
    Since $j$ and $k$ were arbitrary, then all eigenvalues of $A$ fall within $i(-\pi,\pi]$.
    Consequently, $\Log( U \odot V ) = A$ and the following equation holds by the associativity and bilinearity of $( \otimes )$.
    \begin{equation*}
        ( U \odot V ) \odot W
        =
        \exp\left( \frac{\Log( U \odot V ) \otimes \Log(W)}{i\pi} \right)
        =
        \exp\left( \frac{ \Log(U) \otimes \Log(V) \otimes \Log(W)}{(i\pi)^2} \right)
    \end{equation*}
    By a similar argument, $\Log( V \odot W ) = B$ where $B = \Log( V ) \otimes \Log( W ) / (i\pi)$.
    Then the following equation also holds by the associativity and bilinearity of $( \otimes )$.
    \begin{equation*}
        U \odot ( V \odot W )
        =
        \exp\left( \frac{\Log(U) \otimes \Log( V \odot W )}{i\pi} \right)
        =
        \exp\left( \frac{ \Log(U) \otimes \Log(V) \otimes \Log(W)}{(i\pi)^2} \right)
    \end{equation*}
    In conclusion, $( U \odot V ) \odot W = U \odot ( V \odot W )$.
    Since $U$, $V$, and $W$ were arbitrary, then $( \odot )$ is associative.
    Next, the equation for $\bigodot_{j=1}^n U_j$ will be derived.
    The proof follows by induction.
    \begin{itemize}
    \item \textbf{Base Case}.
          $\bigodot_{j=1}^0 U_j = -1$ and $\Log( -1 ) = i\pi = i\pi \cdot 1 = i\pi \bigotimes_{j=1}^n \frac{\Log( U_j )}{i\pi}$.
    \item \textbf{Inductive Hypothesis}.
          For some $n \in \mathbb{N}$, if $\{ U_j \}_{j=1}^n$ is a sequence over $\Mor( \CatUnitary )$, then $\Log\left( \bigodot_{j=1}^n U_j \right) = i\pi \bigotimes_{j=1}^n \frac{\Log( U_j )}{i\pi}$.
    \item \textbf{Inductive Step}.
          Assume for some $n \in \mathbb{N}$, if $\{ U_j \}_{j=1}^n$ is a sequence over $\Mor( \CatUnitary )$, then $\Log\left( \bigodot_{j=1}^n U_j \right) = i\pi \bigotimes_{j=1}^n \frac{\Log( U_j )}{i\pi}$.
          Let $\{ U_j \}_{j=1}^{n+1}$ be a sequence over $\Mor( \CatUnitary )$.
          As shown previously, $\Log\left( \bigodot_{j=1}^{n+1} U_j \right) = \Log\left( \bigodot_{j=1}^n U_j \right) \otimes \Log( U_{n+1} ) / (i\pi)$.
          Then by the inductive hypothesis, $\Log\left( \odot_{j=1}^{n+1} U_j \right) = \left( i\pi \bigotimes_{j=1}^n \frac{\Log( U_j )}{i\pi} \right) \otimes \frac{\Log( U_{n+1} )}{i\pi} = i\pi \bigotimes_{j=1}^{n+1} \frac{\Log( U_j )}{i\pi}$.
          Then the inductive step holds.
    \end{itemize}
    Then by the principle of induction, $\Log\left( \bigodot_{j=1}^n \right) = i\pi \bigotimes_{j=1}^n \frac{\Log( U_j )}{i\pi}$ for each sequence $\{ U_j \}_{j=1}^n$ over $\Mor( \CatUnitary )$.
    In other words, $\bigodot_{j=1}^n U_j = \exp\left( i\pi \bigotimes_{j=1}^{n} \frac{\Log( U_j )}{i\pi} \right)$.
\end{proof}

\subsection{Proof of Theorem~\ref{Thm:GenCtrlNF}}
\begin{proof}
    Let $M$ be an $m \times m$ matrix.
    If $H$ were diagonal, then $H \otimes M$ would be block diagonal.
    This equation follows by the bilinearity of $( \otimes )$.
    \begin{equation*}
        \left( \sum_{j=1}^n \alpha_j \ket{e_j}\bra{e_j} \right) \otimes M
        =
        \sum_{j=1}^n \alpha_j \ket{e_j}\bra{e_j} \otimes M
        =
        \sum_{j=1}^n \ket{e_j}\bra{e_j} \otimes \alpha_j M
        =
        \begin{bsmallmatrix}
            \alpha_1 M & & \\
             & \ddots & \\
             &  & \alpha_n M
        \end{bsmallmatrix}
    \end{equation*}
    Since $H$ is Hermitian, then there exists a $P \in \CatUnitary( m, m )$ such that $P$ unitarily diagonalizes $N$.
    That is to say, $P\ket{j} = \ket{b_j}$ for each $j \in \{ 1, 2, \ldots, n \}$.
    It follows that $H \otimes M$ is block diagonal with respect to the basis $\{ b_j \}_{j=1}^n$ with blocks $\{ \alpha_j M \}_{j=1}^n$.
    \begin{equation*}
        H \otimes M
        =
        ( P \otimes \mathbb{I}_m)
        \left( \left( \sum_{j=1}^n \alpha_j \ket{e_j} \bra{e_j} \right) \otimes M \right)
        ( P^\dagger \otimes \mathbb{I}_m)
        =
        (P \otimes \mathbb{I}_m) 
        \begin{bsmallmatrix}
            \alpha_1 M & &  \\
             & \ddots & \\
             &  & \alpha_n M
        \end{bsmallmatrix}
        (P^\dagger \otimes \mathbb{I}_m)
    \end{equation*}
    Let $Q = P \otimes \mathbb{I}_m$.
    Then by \cref{Eqn:UnitaryConj}, the following equation also holds.
    \begin{equation*}
        \exp( H \otimes M )
        =
        \exp\left( 
        Q
        \begin{bsmallmatrix}
            \alpha_1 M & &  \\
             & \ddots & \\
             &  & \alpha_n M
        \end{bsmallmatrix}
        Q^\dagger
        \right)
        =
        Q
        \exp\left( 
        \begin{bsmallmatrix}
            \alpha_1 M & &  \\
             & \ddots & \\
             &  & \alpha_n M
        \end{bsmallmatrix}
        \right)
        Q^\dagger
    \end{equation*}
    Since $\exp( - )$ acts pointwise on block diagonal matrices, then we conclude that for each Hermitian matrix $M = \sum_{j=1}^n \beta_j \ket{b_j} \bra{b_j}$ and $n \times n$ matrix $M$, the matrix $\exp( N \otimes M )$ is block diagonal with respect to $\{ b_j \}_{j=1}^n$ with blocks $\{ \exp( \alpha_j M ) \}_{j=1}^n$.
    \begin{equation*}
        \exp( H \otimes M )
        =
        ( P \otimes \mathbb{I}_m )
        \begin{bsmallmatrix}
            \exp( \alpha_1 M ) & &  \\
             & \ddots & \\
             &  & \exp( \alpha_n M )
        \end{bsmallmatrix}
        ( P^\dagger \otimes \mathbb{I}_m )
    \end{equation*}
    Returning to the original basis, the following equataion holds.
    \begin{equation*}
        \exp( H \otimes M )
        =
        ( P \otimes \mathbb{I}_m )
        \left( \sum_{j=1}^n \ket{e_j} \bra{e_j} \otimes \exp( \alpha_j M ) \right)
        ( P^\dagger \otimes \mathbb{I}_m )
        =
        \sum_{j=1}^n \ket{b_j} \bra{b_j} \otimes \exp( \alpha_j M )
    \end{equation*}
    Letting $M = \Log( V )$, it follows that $\exp( H \otimes \Log( V ) ) = \sum_{j=1}^n \ket{b_j} \bra{b_j} \otimes V^{\alpha_j}$.
    Then the following equation holds.
    \begin{equation*}
        U \odot V
        =
        \exp( \Log(U) \otimes \Log(V) / (i\pi) ) 
        =
        \exp\left( H \otimes \Log(V) \right)
        =
        \sum_{j=1}^n \ket{b_j} \bra{b_j} \otimes V^{\alpha_j}
    \end{equation*}
    In particular, letting $U = Z = \ket{0}\bra{0} - \ket{1}\bra{1}$, we have $\Log(Z)/(i\pi) = 0\cdot\ket{0}\bra{0} + 1\cdot\ket{1}\bra{1}$, and hence $Z \odot V = \ket{0} \bra{0} \otimes V^0 + \ket{1} \bra{1} \otimes V^1 = \ket{0} \bra{0} \otimes \mathbb{I}_m + \ket{1} \bra{1} \otimes V$.
\end{proof}

\subsection{Proof of Corollary~\ref{Cor:GenCtrlFunc}}
\begin{proof}
    Let $U \in \CatUnitary( d, d )$ and define $F( - ) = U \odot ( - )$.
    Let $\{ \alpha_j \}_{j=1}^d$ denote the eigenvalues of $\Log( U ) / (i\pi)$ with orthonormal eigenbasis $\{ b_j \}_{j=1}^d$.
    There are four properties to verify.
    \begin{enumerate}
    \item $F( 1_n ) = \exp( \Log( U ) \otimes \Log( 1_n ) / ( i\pi ) ) = \exp( \Log( U ) \otimes \mathbb{O}_n / ( i\pi ) ) = \exp( \mathbb{O}_{dn} ) = \mathbb{I}_{dn}$ for $n \in \mathbb{N}$.
    \item Let $n \in \mathbb{N}$ and $V \in \Mor( \CatUnitary )$.
          Then the following equation holds.
          \begin{align*}
              U \odot ( V \otimes \mathbb{I}_n )
              &=
              \sum_{j=1}^d \ket{b_j} \bra{b_j} \otimes ( V \otimes \mathbb{I}_n )^{\alpha_j}
              &
              ( \text{by Cor.~\ref{Thm:GenCtrlNF}} )
              \\
              &=
              \sum_{j=1}^d \ket{b_j} \bra{b_j} \otimes \exp( \alpha_j \Log( V \otimes \mathbb{I}_n ) )
              \\
              &=
              \sum_{j=1}^d \ket{b_j} \bra{b_j} \otimes \exp( \alpha_j \Log( V ) \otimes \mathbb{I}_n )
              \\
              &=
              \sum_{j=1}^d \ket{b_j} \bra{b_j} \otimes \exp( \alpha_j \Log( V ) ) \otimes \mathbb{I}_n
              \\
              &=
              \left( \sum_{j=1}^d \ket{b_j} \bra{b_j} \otimes \exp( \alpha_j \Log( V ) ) \right) \otimes \mathbb{I}_n
              \\
              &=
              \left( \sum_{j=1}^d \ket{b_j} \bra{b_j} \otimes V^{\alpha_j} \right) \otimes \mathbb{I}_n
              =
              F( V ) \otimes \mathbb{I}_n
              &
              ( \text{by Cor.~\ref{Thm:GenCtrlNF}} )
          \end{align*}
    \item Let $V \in \CatUnitary( n, n )$.
          Since $\sigma_{d,d}$ is a monoidal symmetry, then $\sigma_{d,d}$ is self-adjoint with $\sigma_{d,d} \cdot ( \Log( U ) \otimes \Log( U ) ) \cdot \sigma_{d,d} = \Log( U ) \otimes \Log( U )$.
          Then the following equation holds where $H = \Log( V )$.
          \begin{align*}
              F( F( V ) )
              &=
              U \odot ( U \odot V )
              \\
              &=
              \exp( \Log( U ) \otimes \Log( U ) \otimes H / (i\pi)^2 )
              &
              ( \text{by Thm.~\ref{Thm:ControlAssoc}} )
              \\
              &=
              \exp( ( \sigma_{d,d} \cdot ( \Log( U ) \otimes \Log( U ) ) \cdot \sigma_{d,d} ) \otimes H / (i\pi)^2 )
              \\
              &=
              \exp( ( \sigma_{d,d} \otimes \mathbb{I}_n ) \cdot ( \Log( U ) \otimes \Log( U ) \otimes H / (i\pi)^2 ) \cdot ( \sigma_{d,d} \otimes \mathbb{I}_n ) )
              \\
              &=
              ( \sigma_{d,d} \otimes \mathbb{I}_n ) \cdot \exp( \Log( U ) \otimes \Log( U ) \otimes H / (i\pi)^2 ) \cdot ( \sigma_{d,d} \otimes \mathbb{I}_n )
              &
              ( \text{by~\cref{Eqn:UnitaryConj}} )
              \\
              &=
              ( \sigma_{d,d} \otimes \mathbb{I}_n ) \cdot ( U \odot ( U \odot V ) ) \cdot ( \sigma_{d,d} \otimes \mathbb{I}_n )
              &
              ( \text{by Thm.~\ref{Thm:ControlAssoc}} )
              \\
              &=
              ( \sigma_{d,d} \otimes \mathbb{I}_n ) \cdot F( F( V ) ) \cdot ( \sigma_{d,d} \otimes \mathbb{I}_n )
          \end{align*}
          Then $F( F( V ) ) = ( \sigma_{d,d} \otimes \mathbb{I}_n ) \cdot F( F( V ) ) \cdot ( \sigma_{d,d} \otimes \mathbb{I}_n )$.
          Since $\sigma_{d,d} \otimes \mathbb{I}_n$ is also self-adjoint, then $( \sigma_{d,d} \otimes \mathbb{I}_n ) \cdot F( F( V ) ) = F( F( V ) ) \cdot ( \sigma_{d,d} \otimes \mathbb{I}_n )$.
    \item If $V \in \CatUnitary( n, n )$ and $P \in \CatUnitary( n, n )$, then the following equation holds.
          \begin{align*}
              F( P^\dagger \circ V \circ P )
              &=
              \exp( \Log( U ) \otimes \Log( P^\dagger \circ V \circ P ) / (i\pi) )
              \\
              &=
              \exp( \Log( U ) \otimes ( P^\dagger \circ \Log( V ) \circ P ) / (i\pi) )
              &
              ( \text{by~\cref{Eqn:UnitaryConj}} )
              \\
              &=
              \exp( ( \mathbb{I}_d \otimes P^\dagger ) \circ ( \Log( U ) \otimes \Log( V ) / (i\pi) ) \circ ( \mathbb{I}_d \otimes P ) )
              \\
              &=
              ( \mathbb{I}_d \otimes P^\dagger ) \circ \exp( \Log( U ) \otimes \Log( V ) / (i\pi) ) \circ ( \mathbb{I}_d \otimes P )
              &
              ( \text{by~\cref{Eqn:UnitaryConj}} )
              \\
              &=
              ( \mathbb{I}_d \otimes P^\dagger ) \circ F( V ) \circ ( \mathbb{I}_d \otimes P )
          \end{align*}
    \end{enumerate}

    \noindent
    \textbf{Functoriality}.
    Next, it must be shown that $F( - )$ is an ordinary functor if and only if $U$ is Hermitian.
    As a preliminary result, a general equation will be computed for $F( V \circ W)$.
    Let $V, W \in \CatUnitary( n, n )$.
    Then by \cref{Thm:GenCtrlNF}, the following equations hold.
    \begin{align*}
        F( V ) &= \sum_{j=1}^{d} \ket{b_j} \bra{b_j} \otimes V^{\alpha_j}
        &
        F( W ) &= \sum_{j=1}^{d} \ket{b_j} \bra{b_j} \otimes W^{\alpha_j}
    \end{align*}
    Since $\{ b_j \}_{j=1}^d$ is orthonormal, then $\braket{b_j|b_k} = \delta_{j,k}$ where $\delta$ is the Kronecker delta function.
    Then the following equation holds by the bilinearity of $( \otimes )$.
    \begin{equation*}
        F( V ) \circ F( W )
        =
        \sum_{j=1}^{d} \sum_{k=1}^{d} \delta_{j,k} \ket{b_j} \bra{b_k} \otimes ( V^{\alpha_j} \circ W^{\alpha_k} )
        =
        \sum_{j=1}^{d} \ket{b_j} \bra{b_j} \otimes ( V^{\alpha_j} \circ W^{\alpha_j} )
    \end{equation*}
    Since $V$ and $W$ were arbitrary, then this equation holds in general.
    Now it can be shown that $F( - )$ is a functor if and only if $U$ is Hermitian.
    \begin{itemize}
    \item Assume that $F( - )$ is a functor.
          Let $j \in \{ 1, 2, \ldots, d \}$.
          Observe that the following equations hold where $M = \ket{b_j} \otimes \mathbb{I}_2$.
          \begin{align*}
              M^\dagger \circ F( -Z ) \circ M
              &=
              \sum_{k=1}^{d} \delta_{j,k} \delta_{k,j} ( -Z )^{\alpha_k}
              =
              ( -Z )^{\alpha_j}
              \\
              M^\dagger \circ F( -\mathbb{I}_2 ) \circ F( Z ) \circ M
              &=
              \sum_{k=1}^{d} \delta_{j,k} \delta_{k,j} ( -\mathbb{I}_2 )^{\alpha_k} Z^{\alpha_k}
              =
              ( -\mathbb{I}_2 )^{\alpha_j} \circ Z^{\alpha_j}
          \end{align*}
          Since $F( - )$ is a functor, then the following equation holds.
          \begin{equation*}
              ( -Z )^{\alpha_j}
              =
              M \circ F( -Z ) \circ M
              =
              M \circ F( -\mathbb{I}_2 ) \circ F( Z ) \circ M
              =
              ( -\mathbb{I}_2 )^{\alpha_j} \circ Z^{\alpha_j}
              =
              e^{i\alpha_j\pi} Z^{\alpha_j}
          \end{equation*}
          Since $\Log( -Z ) = i \ket{0}\bra{0} \pi$ and $\Log( Z ) = i \ket{1}\bra{1} \pi$, then the following equation holds.
          \begin{align*}
            ( -Z )^{\alpha_j}
            &=
            \exp( i\alpha_j\pi \ket{0} \bra{0} )
            =
            \begin{bsmallmatrix}
                e^{i\alpha_j\pi} & 0 \\
                0 & 1
            \end{bsmallmatrix}
            &
            Z^{\alpha_j}
            &=
            \exp( i\alpha_j\pi \ket{1} \bra{1} )
            =
            \begin{bsmallmatrix}
                1 & 0 \\
                0 & e^{i\alpha_j\pi}
            \end{bsmallmatrix}
          \end{align*}
          Then the following equation also holds.
          \begin{equation*}
              1
              =
              \braket{1|1}
              =
              \bra{1} ( -Z )^{\alpha_j} \ket{1}
              =
              e^{i\alpha_j\pi} \bra{1} Z^{\alpha_j} \ket{1}
              =
              e^{i\alpha_j\pi} \left( e^{i\alpha_j\pi} \braket{1|1} \right)
              =
              e^{i\alpha_j(2\pi)}
          \end{equation*}
          Then $\alpha_j(2\pi)$ is a multiple of $2\pi$.
          Then $\alpha_j$ is an integer.
          Then $e^{i\alpha_j\pi}$ is either $-1$ or $1$.
          Since $j$ was arbitrary, then all eigenvalues of $U$ are either $-1$ or $1$.
          Since $U$ is unitary, then this means $U$ is also Hermitian.
    \item Assume that $U$ is Hermitian.
          Since $U = \exp( iH\pi )$, then the eigenvalues of $U$ are $\{ e^{i\alpha_j\pi} \}_{j=1}^{d}$.
          Let $j \in \{ 1, 2, \ldots, d \}$.
          Since $U$ is Hermitian and unitary, then $e^{i\alpha_j\pi}$ is either $1$ or $-1$.
          Then $\alpha_j\pi$ is a multiple of $\pi$.
          Since $\alpha_j \in (-1, 1]$, then $\alpha_j \in \{ 0, 1 \}$.
          Then either $\alpha_j = 0$ and $V^{\alpha_j} \circ W^{\alpha_j} = \mathbb{I}_m = ( V \circ W )^{\alpha_j}$ or $\alpha_j = 1$ and $V^{\alpha_j} \circ W^{\alpha_j} = V \circ W = ( V \circ W )^{\alpha_j}$.
          Since $j$ was arbitrary, then $V^{\alpha_j} \circ W^{\alpha_j} = ( V \circ W )^{\alpha_j}$.
          Then the following equation holds.
          \begin{equation*}
              F( V ) \circ F( W )
              =
              \sum_{j=1}^{d} \ket{b_j} \bra{b_j} \otimes ( V^{\alpha_j} \circ W^{\alpha_j} )
              =
              \sum_{j=1}^{d} \ket{b_j} \bra{b_j} \otimes ( V \circ W )^{\alpha_j}
              =
              F( V \circ W )
          \end{equation*}
          Since $V$ and $W$ were arbitrary, then $F( - )$ preserves composition.
          It was already shown in (\textbf{C1}) that $F( - )$ preserves identities, so $F( - )$ is a functor.
    \end{itemize}
    Then $F( - )$ is a functor if and only if $U$ is Hermitian.

    \vspace{1em}
    \noindent
    \textbf{Control Functoriality}.
    By definition, $( \CatUnitary_d )_\mathbf{endo} = \CatUnitary_d$.
    Notice that for each $k \in \mathbb{N}$, $\Ob( F )( d^k ) = d d^k = d^{k+1}$.
    Then there exists a functor $C: \CatUnitary_d \to \CatUnitary_d$ such that $C( V: k \to k ) = F( V ) \in \CatUnitary( d^{k+1}, d^{k+1} ) = \CatUnitary_d( k + 1, k + 1 )$.
    It can then be shown that $C( - )$ is a conjugated control functor.
    \begin{itemize}
    \item If $V \in \CatUnitary_d( k, k )$, then $C( V \otimes 1 ) = F( V \otimes \mathbb{I}_d ) = F( V ) \otimes \mathbb{I}_d = C( V ) \otimes 1$ by (\textbf{C2}).
    \item Let $V \in \CatUnitary_d( k, k )$ and write $n = d^k$.
          It follows from (\textbf{C3}) that $( \sigma \otimes 1_k ) \circ C( C( V ) ) = ( \sigma_{d,d} \otimes \mathbb{I}_n ) \circ F( F( V ) ) = F( F( V ) )\circ ( \sigma_{d,d} \otimes \mathbb{I}_n ) = C( C( V ) ) \circ ( \sigma \otimes 1_k )$.
          Since $V$ was arbitrary, then $C( - )$ satisfies~\cref{Fig:ControlDefn:ContrlSym}.
    \item Let $V \in \CatUnitary_d( k, k )$ and $j \in \{ 1, 2, \ldots, d \}$.
          Since $\gamma_{k,j}$ is self-adjoint, then it follows by (\textbf{C4}) that $C( \gamma_{k,j} \circ V \circ \gamma_{k,j} ) = ( 1 \otimes \gamma_{k,j} ) \circ C( V ) \circ ( 1 \otimes \gamma_{k,j} )$.
          Since $V$ and $j$ were arbitrary, then $C( - )$ satisfies~\cref{Fig:ControlDefn:DataSym}.
    \end{itemize}
    This means that $C( - )$ is a control functor.
    Since every morphism in $\CatUnitary_d$ is unitary, then $C( - )$ is a dagger control functor by~\cref{Lem:UnitaryCtrl}.
    Finally, if $V, P \in \CatUnitary_d( k, k )$, then by (\textbf{C4}), $C( P^\dagger \circ V \circ P ) = F( P^\dagger \circ V \circ P ) = ( \mathbb{I}_d \otimes P^\dagger ) \circ F( V ) \circ ( \mathbb{I}_d \otimes P ) = ( 1 \otimes P^\dagger ) \circ C( V ) \circ ( 1 \otimes P )$.
    Since $V$ and $P$ were arbitrary, then $C( - )$ is a conjugated control functor.

    \vspace{1em}
    \noindent
    \textbf{Pointed Control Functors}.
    Assume that $\ket{u} \in \mathbb{C}^d$ and $\ket{v} \in \mathbb{C}^d$.
    Since $U$ is Hermitian, then $\mathbb{C}^d$ can be partitioned into the $(+1)$ and $(-1)$ eigenspaces of $U$, where $\alpha_j = 0$ and $\alpha_j = 1$ respectively.
    To this end, define the following two sets where $[n] = \{ 1, 2, \ldots, n \}$.
    \begin{align*}
        X &= \{ b_j : j \in [d] \text{ and } \alpha_j = 0 \}
        &
        Y &= \{ b_j : j \in [d] \text{ and } \alpha_j = 1 \}
    \end{align*}
    It follows that $\{ b_j : j \in [d] \} = X \cup Y$.
    This partitioning will be used to determine the conditions under which $\ket{u}$ and $\ket{v}$ are points of $C( - )$.
    \begin{itemize}
    \item Assume that $C( - )$ is $( u, v )$-pointed.
          It will be shown that $U\ket{u} = -\ket{u}$.
          It then follows by a symmetric argument that $U\ket{v} = \ket{v}$.
          Construct a matrix $M = \sum_{j=0}^{d-1} \ket{d-j} \bra{j}$, so that $M$ is a permutation of the standard basis vectors within $\mathbb{C}^d$.
          Since $M$ is a permutation matrix, then $M$ is unitary, and hence $M \in \CatUnitary_d( 1, 1 )$.
          Since $C( - )$ is a $( u, v )$-pointed control functor, then the following equation holds where $\gamma_x = \braket{x|u}$.
          \begin{align*}
              \ket{u} \otimes M
              &=
              C( M ) \circ \left( \ket{u} \otimes \mathbb{I}_d \right)
              \\
              &=
              \sum_{j=1}^n \ket{b_j} \braket{b_j|u} \otimes M^{\alpha_j}
              &
              ( \text{by \cref{Thm:GenCtrlNF}} )
              \\
              &=
              \sum_{b_j \in X} \gamma_{b_j} \ket{b_j} \otimes M^0
              +
              \sum_{b_j \in Y} \gamma_{b_j} \ket{b_j} \otimes M^1
              \\
              &=
              \sum_{b_j \in X} \gamma_{b_j} \ket{b_j} \otimes \mathbb{I}_d
              +
              \sum_{b_j \in Y} \gamma_{b_j} \ket{b_j} \otimes M
          \end{align*}
          Since $\otimes$ is bilinear, then the following equation holds.
          \begin{equation*}
              \left( \ket{u} - \sum_{b_j \in Y} \gamma_{b_j} \ket{b_j} \right) \otimes M
              =
              \left( \sum_{b_j \in X} \gamma_{b_j} \ket{b_j} \right) \otimes \mathbb{I}_d
          \end{equation*}
          Since $\{ \ket{\ell} \bra{k} \}_{j=1,\ell=1}^{d}$ is a basis for $\CatUnitary_d( 1, 1 )$, then $\mathbb{I}_d$ and $M$ are not linear combinations of one-another.
          Then the left-hand side and right-hand side also not linear combinations of one-another.
          Then by equating components, $\ket{u} = \sum_{b_j \in Y} \gamma_{b_j} \ket{b_j}$.
          Then $\ket{u}$ is in the $(-1)$-eigenspace of $U$.
          In conclusion, $U\ket{u} = -\ket{u}$.
    \item Assume that $U\ket{u} = -\ket{u}$ and $U\ket{v} = \ket{v}$.
          Since $U\ket{u} = -\ket{u}$, then by definition $\ket{u}$ is in the $(-1)$-eigenspace of $U$.
          This means that $\ket{u} = \sum_{b_j \in Y} \gamma_j \ket{b_j}$ where $\gamma_j = \braket{b_j|u}$.
          Recall that $\braket{b_j|b_k} = \delta_{j,k}$ where $\delta$ is the Kronecker delta function.
          Let $V \in \CatUnitary_d( k ,k )$ and $n = d^k$.
          Then the following equation holds.
          \begin{align*}
              C( V ) \circ ( \ket{u} \otimes \mathbb{I}_n )
              &=
              \sum_{j=1}^n \ket{b_j} \braket{b_j|u} \otimes V^{\alpha_j}
              &
              ( \text{by \cref{Thm:GenCtrlNF}} )
              \\
              &=
              \sum_{j=1}^n \ket{b_j} \left( \sum_{b_k \in X} \gamma_k \braket{b_j|b_k} \right) \otimes V^{\alpha_j}
              &
              ( \text{by the bilinearity of } \braket{-|-} )
              \\
              &=
              \sum_{j=1}^n \ket{b_j} \left( \sum_{b_k \in X} \gamma_x \delta_{j,k} \right) \otimes V^{\alpha_j}
              \\
              &=
              \sum_{j=1}^n \, \sum_{b_k \in X} \gamma_k \delta_{j,k} \ket{b_j} \otimes V^{\alpha_j}
              &
              ( \text{by the bilinearity of } \otimes )
              \\
              &=
              \sum_{b_j \in X} \gamma_j \ket{b_j} \otimes V^1
              \\
              &=
              \sum_{b_j \in X} \gamma_j \ket{b_j} \otimes V
              \\
              &=
              \left( \sum_{b_j \in X} \gamma_j \ket{b_j} \right) \otimes V
              =
              \ket{u} \otimes V
              &
              ( \text{by the bilinearity of } \otimes )
          \end{align*}
          Since $V$ was arbitrary, then $C( V ) \circ ( \ket{u} \otimes \mathbb{I}_n ) = \ket{u} \otimes V$ for all $V \in \CatUnitary_d( k, k )$ with $n = d^k$.
          By a symmetric argument, starting from the fact that $\ket{v}$ in the $(+1)$-eigenspace of $U$, it follows that $C( V ) \circ ( \ket{v} \otimes \mathbb{I}_n ) = \ket{v} \otimes \mathbb{I}_n$ for all $V \in \CatUnitary_d( k, k )$ with $n = d^k$.
          Then $C( - )$ is $( u, v )$-pointed.
    \end{itemize}
    In conclusion, $C( - )$ is $( u, v )$-pointed if and only if $U\ket{u} = -\ket{u}$ and $U\ket{v} = \ket{v}$.
\end{proof}

\subsection{Proof of Theorem~\ref{Thm:PowAsCtrl}}
\begin{proof}
    Let $\alpha \in \mathbb{R}$.
    First, it must be shown that $U \odot e^{i\alpha\pi} = U^\alpha$ for all $U \in \Mor( \CatUnitary )$ if and only if $\alpha \in (-1, 1]$.
    \begin{itemize}
    \item Assume that $\alpha \in ( -1, 1 ]$.
          Then $i\alpha\pi \in i( -\pi, \pi ]$.
          Then $\Log( e^{i\alpha\pi} ) = i\alpha\pi$ since $\Log( - )$ is the principle branch of $\log( - )$.
          Let $U \in \Mor( \CatUnitary )$.
          Then the following equation holds.
          \begin{equation*}
              U \odot e^{i\alpha\pi}
              =
              \exp\left( \Log( U ) \otimes ( i\alpha\pi ) / (i\pi) \right)
              =
              \exp( \alpha \Log( U ) )
              =
              U^\alpha
          \end{equation*}
          Since $U$ was arbitrary, then $U \odot e^{i\alpha\pi} = U^\alpha$ for all $U \in \Mor( \CatUnitary )$.
    \item Assume that $\alpha \not \in ( -1, 1 ]$.
          Then there exists a unique $k \in \mathbb{Z} \setminus \{ 0 \}$ such that $\alpha - 2k \in ( -1, 1 ]$.
          Then $i\alpha\pi - i2k\pi \in i( -\pi, \pi ]$, which implies that $\Log( e^{i\alpha\pi} ) = i\alpha\pi - i2k\pi$.
          Since $k \ne 0$, then let $U = Z^{1/|4k|}$, so that $\Log( U ) = i \frac{\pi}{|4k|} \ket{1} \bra{1}$.
          Then the following equations hold.
          \begin{align*}
              U^\alpha
              &=
              \exp\left( \alpha \left( i\frac{\pi}{|4k|} \ket{1} \bra{1} \right) \right)
              =
              \exp\left( i \frac{\alpha\pi}{|4k|} \ket{1} \bra{1} \right)
              \\
              U \odot e^{i\alpha\pi}
              &=
              \exp\left( \left( i\frac{\pi}{|4k|} \ket{1} \bra{1} \right) \otimes ( i\alpha\pi - i2k\pi ) / (i\pi) \right)
              =
              \exp\left( i \frac{(\alpha-2k)\pi}{|4k|} \ket{1} \bra{1} \right)
          \end{align*}
          Since $-1 < \alpha - 2k \le 1$, then $2k - 1 < \alpha < 2k + 1$.
          The value of $\alpha / |4k|$ then depends on the value of $k$.
          If $k > 0$, then the following equations hold.
          \begin{align*}
              (2k - 1)/|4k| &= 2k/|4k| - 1/|4k| = 1/2 - 1/|4k| \ge 1/2 - 1/4 > -1
              \\
              (2k + 1)/|4k| &= 2k/|4k| + 1/|4k| = 1/2 + 1/|4k| \le 1/2 + 1/4 < 1
          \end{align*}
          This means that $\alpha / |4k| \in ( -1, 1 )$.
          If $k < 0$, then the following equations hold.
          \begin{align*}
              (2k - 1)/|4k| &= 2k/|4k| - 1/|4k| = -1/2 - 1/|4k| \ge -1/2 - 1/4 > -1
              \\
              (2k + 1)/|4k| &= 2k/|4k| + 1/|4k| = -1/2 + 1/|4k| \le -1/2 + 1/4 < 1
          \end{align*}
          This means that $\alpha / |4k| \in ( -1, 1 )$.
          Then in either case, $\alpha / |4k| \in ( -1, 1 )$.
          It follows that $\Log( U^\alpha ) = i \frac{\alpha\pi}{|4k|} \ket{1} \bra{1}$.
          Since $i( \alpha - 2k )\pi \in i( -\pi, \pi ]$, then $i \frac{(\alpha - 2k)\pi}{|4k|} \in i( -\pi, \pi ]$.
          It follows that $\Log( U \odot e^{i\alpha\pi} ) = i \frac{(\alpha - 2k)\pi}{|4k|} \ket{1} \bra{1}$.
          Clearly $\Log( U^\alpha ) \ne \Log( U \odot e^{i\alpha\pi} )$.
          Then $U^\alpha \ne U \odot e^{i\alpha\pi}$.
          Then there exists a $U \in \Mor( \CatUnitary )$ such that $U \odot e^{i\alpha\pi} \ne U^\alpha$.
    \end{itemize}
    It remains to be shown that if $U \in \Mor( \CatUnitary )$ is Hermitian, then $U \odot e^{i\alpha\pi} = U^\alpha$.
    Assume that $U \in \Mor( \CatUnitary )$ with $U$ Hermitian.
    Let $\{ \beta_j \}_{j=1}^n$ denote the eigenvalues of $\Log( U ) / (i\pi)$ with orthonormal eigenbasis $\{ b_j \}_{j=1}^n$.
    Then the following equation holds where $z = e^{i\alpha\pi}$.
    \begin{equation*}
        U \odot z
        =
        \exp\left( \Log( U ) \otimes \Log( z ) / (i\pi) \right)
        =
        \exp\left( \sum_{j=1}^n \Log( z ) \beta_j \ket{b_j} \bra{b_j} \right)
        =
        \sum_{j=1}^n e^{\Log( z ) \beta_j} \ket{b_j} \bra{b_j}
    \end{equation*}
    Let $j \in \{ 1, 2, \ldots, n \}$.
    Then $\exp( i\beta_j\pi )$ is an eigenvalue of $U$.
    Since $U$ is Hermitian, then $e^{i\beta_j\pi}$ is either $-1$ or $1$.
    Then $\beta_j\pi$ is a multiple of $\pi$.
    Since $\beta_j\pi \in ( -\pi, \pi ]$, then $\beta_j$ is either $0$ or $1$.
    If $\beta_j = 0$, then the following equation holds.
    \begin{equation*}
        \exp( \Log(z)\beta_j )
        =
        \exp( 0 \Log(z) )
        =
        \exp( 0 )
        =
        \exp( 0 (i\alpha\beta_j) )
        =
        \exp( i\alpha\beta_j\pi )
    \end{equation*}
    If $\beta_j = 1$, then the following equation holds.
    \begin{equation*}
        \exp( \Log(z)\beta_j )
        =
        \exp( 1 \Log(z) )
        =
        \exp( \Log(z) )
        =
        \exp( i\alpha\pi )
        =
        \exp( 1 (i\alpha\pi) )
        =
        \exp( i\alpha\beta_j\pi )
    \end{equation*}
    In either case, $e^{\Log(z)\beta_j} = e^{i\alpha\beta_j\pi}$.
    Since $j$ was arbitrary, then the following equation holds.
    \begin{align*}
        \sum_{j=1}^n e^{\Log(z)\beta_j} \ket{b_j} \bra{b_j}
        =
        \sum_{j=1}^n e^{i\alpha\beta_j\pi} \ket{b_j} \bra{b_j}
        =
        \exp\left( \sum_{j=1}^n i\alpha\beta_j\pi \ket{b_j} \bra{b_j} \right)
        =
        \exp( \alpha \Log( U ) )
        =
        U^\alpha
    \end{align*}
    In conclusion, $U \odot e^{i\alpha\pi} = U^\alpha$.
    Since $U$ was arbitrary, then $U \odot e^{i\alpha\pi} = U^\alpha$ for each Hermitian $U \in \Mor( \CatUnitary )$.
    Since $\alpha$ was arbitrary, then this completes the proof.
\end{proof}

\subsection{Proof of Theorem~\ref{Thm:AddPow}}
\begin{proof}
    Let $\{ b_j \}_{j=1}^n$ denote the orthonormal eigenbasis associated with $\{ \lambda_j \}_{j=1}^n$.
    Then let $\alpha \in \mathbb{R}$ and $\beta \in \mathbb{R}$.
    First, equations will be found for $( U^\alpha )^\beta$ and $U^{\alpha\beta}$.
    Observe that the following equation holds.
    \begin{equation*}
        U^\alpha
        =
        \exp( \alpha\Log( U ) )
        =
        \exp\left( \alpha \sum_{j=1}^n i\lambda_j\pi \ket{b_j} \bra{b_j} \right)
        =
        \exp\left( \sum_{j=1}^n i\alpha\lambda_j\pi \ket{b_j} \bra{b_j} \right)
        =
        \sum_{j=1}^n e^{i\alpha\lambda_j\pi} \ket{b_j} \bra{b_j}
    \end{equation*}
    This can then be used to compute $\Log( U^\alpha )$ where $\zeta_j = \Log\left( e^{\alpha\lambda_j} \right)$.
    \begin{equation*}
        \Log( U^\alpha )
        =
        \Log\left( \sum_{j=1}^n e^{i\alpha\lambda_j\pi} \ket{b_j} \bra{b_j} \right)
        =
        \sum_{j=1}^n \Log\left( e^{i\alpha\lambda_j\pi} \right) \ket{b_j} \bra{b_j}
        =
        \sum_{j=1}^n \zeta_j \ket{b_j} \bra{b_j}
    \end{equation*}
    This can then be used to compute $( U^\alpha )^\beta$.
    \begin{equation*}
        ( U^\alpha )^\beta
        =
        \exp( \beta\Log( U^\alpha ) )
        =
        \exp\left( \beta \sum_{j=1}^n \zeta_j \ket{b_j} \bra{b_j} \right)
        =
        \exp\left( \sum_{j=1}^n \beta \zeta_j \ket{b_j} \bra{b_j} \right)
        =
        \sum_{j=1}^n e^{\beta\zeta_j} \ket{b_j} \bra{b_j}
    \end{equation*}
    Likewise, $U^{\alpha\beta} = \sum_{j=1}^n e^{i\alpha\beta\lambda_j\pi} \ket{b_j} \bra{b_j}$.
    Next an explicit equation will be calculated for each $\zeta_j$ in terms of $\alpha$ and $\lambda_j$.
    Let $j \in \{ 1, 2, \ldots, n \}$.
    Since $\zeta_j = \Log( \exp( i\alpha\lambda_j\pi ) )$, then there exists a $k \in \mathbb{Z}$ such that $i\alpha\lambda_j\pi = \zeta_j + k(i2\pi)$.
    Then the following sequence of statements hold.
    \begin{align*}
        \zeta_j
        &\in
        i( -\pi, \pi ]
        &
        ( \text{since } \zeta_j = \Log( \exp( \alpha\lambda_j ) ) )
        \\
        i\alpha\lambda_j\pi
        &\in
        i( -\pi + k(2\pi), \pi + k(2\pi) ]
        &
        ( \text{since } \alpha\lambda_j = \zeta_j + k(i2\pi) )
        \\
        \alpha\lambda_j / 2
        &\in
        ( k - 1/2, k + 1/2 ]
        \\
        \alpha\lambda_j / 2 - 1/2
        &\in
        ( k - 1, k ]
    \end{align*}
    Since $k - 1 < ( \alpha\lambda_j - 1 ) / 2 \le k$, then $\ceil*{( \alpha \lambda_j - 1 ) / 2} = k$.
    Since $j$ was arbitrary, then it follows that $\alpha\lambda_j = \zeta_j + \ceil*{( \alpha \lambda_j - 1 ) / 2} (i2\pi)$ for each $j \in \{ 1, 2, \ldots, n \}$.
    Next, it must be shown that $( U^\alpha )^\beta = U^{\alpha\beta}$ if and only if for each $j \in \{ 1, 2, \ldots, n \}$, either $\alpha \lambda_j \in ( -1, 1 ]$ or $\beta \in \mathbb{Q}$ with reduced denominator dividing $\ceil*{( \alpha \lambda_j - 1 ) / 2}$.
    \begin{itemize}
    \item Assume that $U^{\alpha\beta} = ( U^\alpha )^\beta$.
          Let $j \in \{ 1, 2, \ldots, n \}$.
          Since $\{ b_j \}_{j=1}^n$ is an orthonormal basis, then $e^{\beta\zeta_j} = e^{i\alpha\beta\lambda_j\pi}$ by equating components in $( U^\alpha )^\beta = U^{\alpha\beta}$.
          Then $\beta\zeta_j - i\alpha\beta\lambda_j\pi$ is a multiple of $i2\pi$.
          Then there exists some $m \in \mathbb{Z}$ such that $m (i2\pi) = \beta(i\alpha\lambda_j\pi - \zeta_j) = \beta k (i2\pi)$ where $k = \ceil*{( \alpha \lambda_j - 1 ) / 2}$.
          In other words, $\beta k \in \mathbb{Z}$.
          If $k = 0$, then clearly $\beta k \in \mathbb{Z}$.
          This happens precisely when $\zeta_j = i\alpha\lambda_j\pi$, which implies that $\alpha\lambda_j \in (-1,1]$.
          Assume instead that $k \ne 0$.
          Then $\beta \in \mathbb{Q}$, otherwise $\beta k$ is irrational.
          Then there exists some $d \in \mathbb{Z}$ and $q \in \mathbb{N}$ such that $\gcd( d, q ) = 1$ and $\beta = d / q$.
          Then $dk / q \in \mathbb{Z}$.
          Then $q$ divides $dk$.
          Since $\gcd( q, d ) = q$, then $q$ divides $k$.
          This means that $\lambda_j$ is $( \alpha, \beta )$-admissible.
          Since $j$ was arbitrary, then $\lambda_j$ is $( \alpha, \beta )$-admissible for each $j \in \{ 1, 2, \ldots, n \}$.
    \item Assume that for each $j \in \{ 1, 2, \ldots, n \}$, $\lambda_j$ is $( \alpha, \beta )$-admissible.
          Let $j \in \{ 1, 2, \ldots, n \}$.
          There are two cases to consider.
          If $\alpha \lambda_j \in (-1,1]$, then $\zeta_j = i\alpha \lambda_j\pi$, and consequently $\beta\zeta_j = i\alpha\beta\lambda_j\pi = i\alpha\beta\pi + 0(i2\pi)$.
          Assume instead that $\beta \in \mathbb{Q}$ with reduced denominator dividing $k = \ceil*{( \alpha \lambda_j - 1 ) / 2}$.
          Then there exists $q \in \mathbb{Z}$ and $d \in \mathbb{N}$ such that $\beta = q / d$, $\gcd( q, d ) = 1$ and $d$ divides $k$.
          It follows that $\beta\zeta_j = \beta( i\alpha\lambda_j\pi + k ) = i\alpha\beta\lambda_j\pi + q k (i2\pi) / d$.
          Since $d$ divides $k$, then there exists a $m \in \mathbb{Z}$ such that $\beta\zeta_j = i\alpha\beta\lambda_j\pi + m(i2\pi)$.
          In either case, there exists an $m \in \mathbb{N}$ such that $\beta\zeta_j = i\alpha\beta\lambda_j\pi + m(i2\pi)$.
          Then $\exp( \beta\zeta_j ) = \exp( i\alpha\beta\lambda_j\pi )$.
          Since $j$ was arbitrary, then $\exp( \beta\zeta_j ) = \exp( i\alpha\beta\lambda_j\pi )$ for each $j \in \{ 1, 2, \ldots, n \}$.
          Then the following equation holds.
          \begin{equation*}
              ( U^\alpha )^\beta
              =
              \sum_{j=1}^n e^{\beta\zeta_j} \ket{b_j} \bra{b_j}
              =
              \sum_{j=1}^n e^{i\alpha\beta\lambda_j\pi} \ket{b_j} \bra{b_j}
              =
              U^{\alpha\beta}
          \end{equation*}
    \end{itemize}
    Then $( U^\alpha )^\beta = U^{\alpha\beta}$ if and only if for each $j \in \{ 1, 2, \ldots, n \}$, either $\alpha \lambda_j \in ( -1, 1 ]$ or $\beta \in \mathbb{Q}$ with reduced denominator dividing $\ceil*{( \alpha \lambda_j - 1 ) / 2}$.
    It remains to be shown that $\mathbb{Z}$ and $( -1, - 1]$ are maximal submonoids of $\mathbb{R}$ for which for which $( U, \alpha ) \mapsto U^\alpha$ is a monoid action on $\Mor( \CatUnitary )$.
    \begin{itemize}
    \item Let $\alpha \in \mathbb{Z}$, $\beta \in \mathbb{Z}$, and $U \in \Mor( \CatUnitary )$.
          It must be shown that $( U^\alpha )^\beta = U^{\alpha\beta}$.
          To this end, let $\{ \lambda_j \}_{j=1}^n$ denote the eigenvalues of $\Log( U ) / (i\pi)$.
          Let $j \in \{ 1, 2, \ldots, n \}$.
          Assume that $\alpha\lambda_j \not \in ( -1, 1 ]$.
          Since $\beta \in \mathbb{Z}$, then its reduced denominator is $1$.
          Then clearly the reduced denominator of $\beta$ divides $\ceil*{( \alpha \lambda_j - 1 ) / 2}$.
          Then either $\alpha\lambda_j \in ( -1, 1 ]$ or $\beta \in \mathbb{Q}$ with reduced denominator dividing $\ceil{( \alpha \lambda_j - 1) / 2 }$.
          Since $j$ was arbitrary, then $( U^\alpha )^\beta = U^{\alpha\beta}$.
          Since $\alpha$, $\beta$, and $U$ were arbitrary, then $( U, \alpha ) \mapsto U^\alpha$ is a monoid action on $\Mor( \CatUnitary )$ with respect to $\mathbb{Z}$.
          It remains to be shown that $\mathbb{Z}$ is maximal.
          Assume that $M$ is a submonoid of $\mathbb{R}$ for which $( U, \alpha ) \mapsto U^\alpha$ is a monoid action on $\Mor( \CatUnitary )$ and $\mathbb{Z}$ is a submonoid of $M$.
          Let $\beta \in M$.
          Then $( Z^2 )^\beta = Z^{2\beta}$ by definition of $M$.
          Since $\lambda = 1$ is an eigenvalue of $\Log( Z ) / (i\pi)$ and $2\lambda \not \in (-1,1]$, then it follows by the first result of this proof that $\beta \in \mathbb{Q}$.
          Then there exists some $q \in \mathbb{Z}$ and $d \in \mathbb{N}$ such that $\beta = q / d$ and $\gcd( q, d ) = 1$.
          Since $2d + 3 \in \mathbb{Z} \subseteq M$, then $( Z^{2d + 3} )^\beta = Z^{\beta(2d+3)}$ by definition of $M$.
          Since $\lambda = 1$ is an eigenvalue of $\Log( Z ) / (i\pi)$ and $(2d+3) \not \in (-1,1]$, then by the first result of this proof $d$ divides $\ceil{((2d+3)-1)/2} = d + 1$.
          This is only possible when $d = 1$.
          Then $\beta = q / d = q \in \mathbb{Z}$.
          Since $\beta$ was arbitrary, then $M \subseteq \mathbb{Z}$.
          In conclusion, $\mathbb{Z}$ is maximal.
    \item Let $\alpha \in ( -1, 1 ]$, $\beta \in ( -1, 1 ]$, and $U \in \Mor( \CatUnitary )$.
          It must be shown that $( U^\alpha )^\beta = U^{\alpha\beta}$.
          To this end, let $\{ \lambda_j \}_{j=1}^n$ denote the eigenvalues of $\Log( U ) / (i\pi)$.
          Since $\alpha \in ( -1, 1 ]$ and $i\lambda_j\pi \in ( -\pi, \pi ]$ for each $j \in \{ 1, 2, \ldots, n \}$, then $\alpha\lambda_j \in ( -1, 1 ]$ for each $j \in \{ 1, 2, \ldots, n \}$.
          Then by the first result of this proof, $( U^\alpha )^\beta = U^{\alpha\beta}$ regardless of the value of $\beta$.
          It remains to be shown that $( -1, 1 ]$ is maximal.
          Assume that $M$ is a submonoid of $\mathbb{R}$ for which $( U, \alpha ) \mapsto U^\alpha$ is a monoid action on $\Mor( \CatUnitary )$ and $( -1, 1 ]$ is a submonoid of $M$.
          Let $\alpha \in \mathbb{R}$.
          There are three cases to consider.
          \begin{enumerate}
          \item If $\alpha \in ( -1, 1 ]$, then $\alpha \in M$ by definition.
          \item If $\alpha = -1$, then $( Z^{\alpha} )^{1/2} = ( Z^{-1} )^{1/2} = Z^{1/2} = S \ne S^\dagger = Z^{-1/2} = Z^{\alpha/2}$.
                Since $1/2 \in M$, then $\alpha \not \in M$, since $M$ acts on $\Mor( \CatUnitary )$ by exponentiation.
          \item Assume that $\alpha \not \in [ -1, 1 ]$.
                Then $\alpha^2 > 1$ and there exists some $m \in \mathbb{N}$ such that $\alpha^{2m} > 2$.
                Let $\hat{\alpha} = \alpha^{2m}$.
                Then $( \hat{\alpha} - 1 ) / 2 > 1 / 2$.
                Then $\ceil*{( \hat{\alpha} / 2} \ge 1$.
                Let $k = \ceil*{( \hat{\alpha} - 1 ) / 2}$ and $\beta = 1 / ( k + 1 )$.
                Since $\lambda = 1$ is an eigenvalue of $\Log( Z ) / (i\pi)$ and $\hat{\alpha}\lambda = \hat{\alpha} \not \in ( -1, 1 ]$, then it follows from the first result of this proof that $( U{\hat{\alpha}} )^\beta = U^{\hat{\alpha}\beta}$ if and only if $k + 1$ divides $k$.
                Since $k \ne 0$ and $k + 1 > k$, then clearly $k + 1$ does not divide $k$.
                This means that $( U^{\hat{\alpha}} )^\beta \ne U^{\hat{\alpha}\beta}$.
                Since $M$ acts on $\Mor( \CatUnitary )$ by exponentiation and $\beta \in M$, then $\hat{\alpha} \not \in M$.
                Then $\alpha \not \in M$, since otherwise $\hat{\alpha} = \alpha^{2m} \in M$.
          \end{enumerate}
          This means that if $\alpha \in M$, then $\alpha \in ( -1, 1 ]$.
          In other words, $M \cap \mathbb{R} = ( -1, 1 ]$.
          However, $M$ was a submonoid of $\mathbb{R}$, so $M = ( -1, 1 ]$.
          In conclusion, $( -1, 1 ]$ is maximal.
    \end{itemize}
    Then $\mathbb{Z}$ and $( -1, 1 ]$ are maximal submonoids of $\mathbb{R}$ for which for which $( U, \alpha ) \mapsto U^\alpha$ is a monoid action on $\Mor( \CatUnitary )$.
\end{proof}

\section{Proofs for Section~\ref{Sect:Hierarchical}}
%

This section contains all proofs for~\cref{Sect:Hierarchical}, except for those which appear in \cref{Sect:Background:Theory}.

\subsection{Proof of Lemma~\ref{Lem:FuncDefFam}}
\begin{proof}
    First, it must be shown that $\Sigma_{\CatHQC} = \bigsqcup_{n=0}^\infty \Sigma_n^*$.
    This will require showing that if $j \ne k$, then $\Sigma_j^*$ is disjoint from $\Sigma_k^*$.
    To this end, let $j \in \mathbb{N}$ and $k \in \mathbb{N}$ such that $j \ne k$.
    Assume, without loss of generality, that $j < k$.
    There are two cases to consider.
    \begin{enumerate}
    \item If $j = 0$, then $\Sigma_j^* = \Sigma_{\CatHQC}^j \subseteq \Sigma_{\CatHQC}^j$.
    \item If $j > 0$, then $\Sigma_j^* = \Sigma_{\CatHQC}^j \setminus \Sigma_{\CatHQC}^{j-1} \subseteq \Sigma_{\CatHQC}^j$.
    \end{enumerate}
    In either case, $\Sigma_j^* \subseteq \Sigma_{\CatHQC}^j$.
    Since $j \le k - 1$, then by the inductive construction $\Sigma_j^* \subseteq \Sigma_{\CatHQC}^{k-1}$.
    Since $\Sigma_k^* = \Sigma_{\CatHQC}^k \setminus \Sigma_{\CatHQC}^{k-1}$, then $\Sigma_j^* \cap \Sigma_k^* = \varnothing$.
    Since $j$ and $k$ were arbitrary, then $\Sigma_j^*$ and $\Sigma_k^*$ are disjoint for each $j \in \mathbb{N}$ and $k \in \mathbb{N}$.
    Then $\bigsqcup_{n=0}^\infty \Sigma_n^* \subseteq \bigcup_{n=0}^\infty \Sigma_{\CatHQC}^n = \Sigma_{\CatHQC}$.
    Then let $g \in \Sigma_{\CatHQC}$.
    Then there exists some least $n \in \mathbb{N}$ such that $g \in \Sigma_{\CatHQC}^n$.
    There are two cases to consider.
    \begin{enumerate}
    \item If $n = 0$, then $g \in \Sigma_{\CatHQC}^n = \Sigma_n^*$.
    \item If $n > 0$, then $g \not \in \Sigma_{\CatHQC}^{n-1}$.
          Then $g \in \Sigma_{\CatHQC}^n \setminus \Sigma_{\CatHQC}^{n-1} = \Sigma_n^*$.
    \end{enumerate}
    In either case, $g \in \Sigma_n^*$.
    Since $g$ was arbitrary, then $\Sigma_{\CatHQC} = \bigsqcup_{n=0}^\infty \Sigma_n^*$.
    Then for each $g \in \Sigma_{\CatHQC}$, there exists a unique $\nu(g) \in \mathbb{N}$ such that $g \in \Sigma_{\nu(g)}^*$.
    This defines an assignment $\nu: \Sigma_{\CatHQC} \to \mathbb{N}$.
    This define a prop signature morphism $\rho: \Sigma_{\CatHQC} \to U( \mathcal{C} )$ such that $\rho( g ) = \tau_{\nu(g)}( g )$.
    This is well-defined since $g \in \Sigma_{\nu(g)}^*$ and $\tau_{\nu(g)}: \Sigma_{\nu(g)}^* \to U( \mathcal{C} )$.
    Then there exists a unique prop functor $F: \CatHQC \to \mathcal{C}$ such that $F( g ) = \rho( g )$ for each $g \in \Sigma_{\CatHQC}$.
    Moreover, if $n \in \mathbb{N}$ and $g \in \Sigma_n^*$, then $\nu( g ) = n$ and consequently $F( g ) = \rho( g ) = \tau_{\nu(g)}( g ) = \tau_n( g )$.
    It remains to be shown that $F$ is unique.
    Assume that $G: \CatHQC \to \mathcal{C}$ is a prop functor such that $G( g ) = \tau_n( g )$ for each $n \in \mathbb{N}$ and $g \in \Sigma_n^*$.
    Let $g \in \Sigma_{\CatHQC}$.
    Then $g \in \Sigma_{\nu(g)}^*$.
    Then $G( g ) = \tau_{\nu(g)}( g ) = F( g )$.
    Since $g$ was arbitrary, then $G \circ i = F \circ i$.
    Then by the universal property of free prop categories, $G = F$.
    Since $G$ was arbitrary, then $F$ is unique.
\end{proof}

\subsection{Proof of Lemma~\ref{Lem:FuncDefPart}}
\begin{proof}
    Clearly $\Sigma_{\textbf{Prim}}$, $\Sigma_{\textbf{Ctrl}}$, and $\Sigma_{\textbf{Pow}}$ are pairwise disjoint.
    This is because $( \odot )$ and $( \exparr )$ are distinct term constructors and $\Sigma_{\textbf{Prim}}$ consists only of base terms.
    This means that the set $\Sigma_{\textbf{Prim}} \sqcup \Sigma_{\textbf{Ctrl}} \sqcup \Sigma_{\textbf{Pow}}$ is well-defined.
    Moreover, $\Sigma_{\textbf{Prim}}$, $\Sigma_{\textbf{Ctrl}}$, and $\Sigma_{\textbf{Pow}}$ are all subsets of $\Sigma_{\CatHQC}$ by construction.
    Then it suffices to show that $\Sigma_{\CatHQC}$ is a subset of $\Sigma_{\textbf{Prim}} \sqcup \Sigma_{\textbf{Ctrl}} \sqcup \Sigma_{\textbf{Pow}}$.
    Let $g \in \Sigma_{\CatHQC}$.
    Then there exists some least $n \in \mathbb{N}$ such that $g \in \Sigma_{\CatHQC}^n$.
    If $n = 0$, then it follows that $g \in \Sigma_{\CatHQC}^0 = \Sigma_{\textbf{Prim}}$.
    Otherwise, $n > 0$ and $g \in \Sigma_{\CatHQC}^n \setminus \Sigma_{\CatHQC}^{n-1}$.
    By definition, the following equation holds.
    \begin{align*}
        \Sigma_{\CatHQC}^n \setminus_{\CatHQC}^{n-1}
        &=
        \left(
            \Sigma_{\CatHQC}^{n-1}
            \cup
            \Tower( \CatHQC[n-1] )
            \cup
            \Power( \CatHQC[n-1] )
        \right)
        \setminus
        \Sigma_{\CatHQC}^{n-1}
        \\
        &\subseteq
        \Tower( \CatHQC[n-1] )
        \cup
        \Power( \CatHQC[n-1] )
    \end{align*}
    Then either $g \in \Tower( \CatHQC[n-1] )$ or $g \in \Power( \CatHQC[n-1] )$.
    There are three cases to consider.
    \begin{enumerate}
    \item Assume that $g \in \Tower( \CatHQC[n-1] )$ and $|g| = 2$
          Then $g_1 \in \Mor( \CatHQC[n-1] ) \subseteq \CatHQC$ and $g_2 \in \Mor( \CatHQC[n-1] ) \subseteq \CatHQC$.
          Moreover, $g = g_1 \odot g_2$.
          In conclusion, $g \in \Sigma_{\mathbf{Ctrl}}$.
    \item Assume that $g \in \Tower( \CatHQC[n-1] )$ and $|g| > 2$
          Then $g_1 \in \Mor( \CatHQC[n-1] ) \subseteq \CatHQC$.
          Define $V = g_2 \odot g_3 \odot \cdots \odot g_k$ where $k = |g|$.
          Then the following sequence of inclusions holds.
          \begin{equation*}
              V \in \Tower( \CatHQC[n-1] )
              \subseteq
              \Sigma_{\CatHQC}^n
              \subseteq
              \CatHQC[n]
              \subseteq
              \CatHQC
          \end{equation*}
          Moreover, $g = g_1 \odot V$.
          In conclusion, $g \in \Sigma_{\mathbf{Ctrl}}$.
    \item Assume that $g \in \Power( \CatHQC[n-1] )$.
          Then there exists a $U \in \CatHQC[n-1]$ and $\alpha \in \mathbb{R}$ such that $g = U \exparr \alpha$.
          Since $\CatHQC[n-1] \subseteq \CatHQC$, then $g \in \Sigma_{\mathbf{Pow}}$.
    \end{enumerate}
    In either case, $g \in \Sigma_{\textbf{Ctrl}} \sqcup \Sigma_{\textbf{Pow}}$.
    Then $g \in \Sigma_{\textbf{Prim}} \sqcup \Sigma_{\textbf{Ctrl}} \sqcup \Sigma_{\textbf{Pow}}$ regardless of the value of $n$.
    Since $g$ was arbitrary, then $\Sigma_{\CatHQC} \subseteq \Sigma_{\textbf{Prim}} \sqcup \Sigma_{\textbf{Ctrl}} \sqcup \Sigma_{\textbf{Pow}}$.
    This completes the proof.
\end{proof}

\subsection{Proof of Therem~\ref{Thm:DagSem}}
\begin{proof}
    Clearly $\sem{-}_H \circ \dagger$ and $\dagger \circ \sem{-}_H$ are both contravariant prop functors from $\CatHQC$ to a subcategory of $\CatHQC$.
    Then $\sem{-}_H \circ \dagger = \dagger \circ \sem{-}_H$ if and only if $\sem{-} \circ \dagger \circ i = \dagger \circ \sem{-}_H \circ i$.
    In other words, $\sem{U^\dagger}_H = \sem{U}_H^\dagger$ for all $U \in \Mor( \CatHQC )$ if and only if $\sem{g^\dagger}_H = \sem{g}_H^\dagger$ for all $g \in \Sigma_{\CatHQC}$.
    Let $g \in \Sigma_{\CatHQC}$.
    There are three cases to consider.
    \begin{enumerate}
    \item Assume $g \in \Sigma_{\textbf{Prim}}$.
          By definition, $g^\dagger = g$ and $\sem{g}_H$ is Hermitian and unitary.
          Since $\sem{g}_H$ is both Hermitian and unitary, then $\sem{g}_H = \sem{g}_H^\dagger$.
          In conclusion, $\sem{g^\dagger}_H = \sem{g}_H = \sem{g}_H^\dagger$.
    \item Assume $g \in \Sigma_{\textbf{Ctrl}}$ with $\dom( g ) = n$.
          Then by definition, $\sem{g^\dagger}_H = \sem{g \exparr (-1)}_H = \sem{g}_H^{-1}$.
          Then the following equation holds.
          \begin{align*}
              \sem{g}_H \sem{g^\dagger}_H
              &=
              \sem{g}_H^1 \sem{g}_H^{-1}
              \\
              &=
              \exp( \Log\sem{g}_H ) \cdot \exp( \Log\sem{g}_H (-1) )
              \\
              &=
              \exp( \Log\sem{g}_H + \Log\sem{g}_H (-1) )
              &
              ( \text{by~\cref{Eqn:SimulDiag}} )
              \\
              &=
              \exp( \mathbb{O}_{2^n} )
              =
              \mathbb{I}_{2^n}
          \end{align*}
          In other words, $\sem{g^\dagger}_H$ is the inverse to $\sem{g}_H$.
          Since $\sem{g}_H$ is unitary, then $\sem{g^\dagger}_H = \sem{g}_H^\dagger$.
    \item Assume $g \in \Sigma_{\textbf{Pow}}$.
          Then there exists a $U \in \Mor( \CatHQC )$ and an $\alpha \in \mathbb{R}$ such that $g = U \exparr \alpha$.
          Then $g^\dagger = U \exparr ( -\alpha )$.
          Then by definition, $\sem{g}_H = \sem{U}_H^\alpha$ and $\sem{g^\dagger} = \sem{U}_H^{-\alpha}$.
          Then the following equation holds.
          \begin{align*}
              \sem{g}_H \sem{g^\dagger}_H
              &=
              \sem{U}_H^\alpha \sem{U}_H^{-\alpha}
              \\
              &=
              \exp( \Log\sem{g}_H \alpha ) \cdot \exp( \Log\sem{g}_H (-\alpha) )
              \\
              &=
              \exp( \Log\sem{g}_H \alpha + \Log\sem{g}_H (-\alpha) )
              &
              ( \text{by~\cref{Eqn:SimulDiag}} )
              \\
              &=
              \exp( \mathbb{O}_{2^n} )
              =
              \mathbb{I}_{2^n}
          \end{align*}
          In other words, $\sem{g^\dagger}_H$ is the inverse to $\sem{g}_H$.
          Since $\sem{g}_H$ is unitary, then $\sem{g^\dagger}_H = \sem{g}_H^\dagger$.
    \end{enumerate}
    In each case, $\sem{g^\dagger}_H = \sem{g}_H^\dagger$.
    Since $g$ was arbitrary, then $\sem{g^\dagger}_H = \sem{g}_H^\dagger$ for all $g \in \Sigma_{\CatHQC}$.
    In conclusion, $\sem{U^\dagger}_H = \sem{U}_H^\dagger$ for all $U \in \Mor( \CatHQC )$.
\end{proof}

\section{Lambda Generators for Diagonal Circuits}
\label{App:DiagGate}

\newcommand{\One}{\mathsf{One}}

This section shows how to construct the $\lambda( x, \alpha )$ circuits as described in~\cref{Sect:Hierarchical}.
It is also shown that these circuits belong to the sub-language $\Core$ as defined in in~\cref{Sect:Completeness}.
Up to a change of basis, every $\lambda( x, \alpha )$ generator can be thought of as an $n$-fold controlled global phase by an angle of $\alpha\pi$.
The aforementioned change of basis should apply $X$-gates to the qubits, so that the state $\ket{x} := \bigotimes_{j=0}^{n-1} \ket{x_j}$ is sent to the state $\ket{\One( n )} := \bigotimes_{j=1}^n \ket{1}$.

\subsection{Constructing the Change of Basis}

\newcommand{\Nots}{\mathsf{Nots}}

Let $x \in \{ 0, 1 \}^n$.
Then define $\Nots( x ) = \bigboxtimes_{j=0}^{n-1} \left( \scalebox{0.65}{\circimg{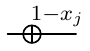}} \right)$.
Note that in $\Nots( x )$, each $X$-gate has been exponentiated by either $0$ or $1$.
This ensures that $\Nots( x ) \in \Core( n, n )$.
It follows that $\sem{\Nots( x )}_H = \bigotimes_{j=0}^{n-1} X^{1-x_j}$.
To see how this operation acts on $\ket{x}$, there are two cases to consider.
\begin{itemize}
\item If $x_j = 0$, then $X^{1-x_j} \ket{x_j} = X^1 \ket{0} = \ket{1}$.
\item If $x_j = 1$, then $X^{1-x_j} \ket{x_j} = X^0 \ket{1} = \ket{1}$.
\end{itemize}
In either case, $X^{1-x_j} \ket{x_j} = \ket{1}$.
This means that $\sem{\Nots( x )}_H \ket{x} = \bigotimes_{j=0}^{n-1} X^{1-x_j} \ket{x_j} = \ket{\One( n )}$, as desired.
Since $\sem{\Nots( x )}_H$ is unitary, then $\sem{\Nots( x )}_H \ket{\psi} = \ket{\One( n )}$ if and only if $\ket{\psi} = \ket{x}$.
Moreover, $\sem{\Nots( x )}_H$ is self-inverse.

\subsection{Constructing the Controlled Phases}

\newcommand{\Phase}{\mathsf{CPh}}

Let $n \in \mathbb{N}$ and $\alpha \in \mathbb{R}$.
Then the $n$-fold controlled phase gate can be defined inductively as follows.
\begin{align*}
    \Phase( \alpha, 0 )
    &=
    \scalebox{0.65}{\circimg{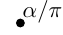}}
    &
    \Phase( \alpha, k + 1 )
    &=
    \scalebox{0.65}{\circimg{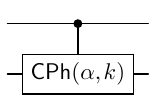}}
\end{align*}
Since $\Phase( \alpha, 0 )$ is a generator for $\Core$ which is closed under the application of $Z$-controls, then $\Phase( \alpha, n ) \in \Core( n, n )$ for each $n \in \mathbb{N}$.
It follows by induction that the following equation holds.
\begin{equation}
    \label{Eqn:DiagHyp}
    \sem{\Phase( \alpha, n )}_H \ket{y} = \begin{cases}
        e^{i\alpha} \ket{y} & \text{if } \ket{y} = \ket{\One( n )} \\
        \ket{y} & \text{otherwise}
    \end{cases}
\end{equation}
The proof proceeds as follows.
\begin{itemize}
\item \textbf{Base Case}.
      Since $\ket{\One( 0 )}$ is the only basis vector for $\mathbb{C}^1$ and $\Phase( \alpha, 0 ) = e^{i\alpha}$, then \cref{Eqn:DiagHyp} holds when $n = 0$.
\item \textbf{Inductive Hypothesis}.
      The equation \cref{Eqn:DiagHyp} holds for some $k \in \mathbb{N}$.
\item \textbf{Inductive Step}.
      Assume that the inductive hypothesis holds for some $k \in \mathbb{N}$.
      This means that \cref{Eqn:DiagHyp} holds for $k$.
      To show that \cref{Eqn:DiagHyp} holds for $k + 1$, let $x \in \{ 0, 1 \}^{k+1}$.
      Then there exists some $a \in \{ 0, 1 \}$ and $y \in \{ 0, 1 \}^k$ such that $x = ay$.
      This means that $\ket{x} = \ket{a} \otimes \ket{y}$.
      Now recall from \cref{Thm:GenCtrlNF} that $\sem{\Phase( \alpha, k + 1 )}_H = \ket{0}\bra{0} \otimes \mathbb{I}_{2^k} + \ket{1}\bra{1} \otimes \sem{\Phase( \alpha, k )}_H$.
      There are three cases to consider.
      \begin{enumerate}
      \item Assume that $a = 0$.
            Then $\sem{\Phase( \alpha, k + 1 )}_H \ket{x} = \ket{a} \otimes \ket{y} = \ket{x}$, since $\braket{0|a} = 1$ and $\braket{1|a} = 0$.
            Then $\ket{x} \ne \ket{\One( k + 1 )}$ and $\sem{\Phase( \alpha, k + 1 )} \ket{x} = \ket{x}$.
      \item Assume $a = 1$ and $\ket{y} \ne \ket{\One( k )}$.
            Then $\sem{\Phase( \alpha, k + 1 )}_H \ket{x} = \ket{a} \otimes \sem{\Phase( \alpha, k )} \ket{y}$, since $\braket{0|a} = 1$ and $\braket{1|a} = 0$.
            Since \cref{Eqn:DiagHyp} holds for $k$ and $\ket{y} \ne \ket{\One( k )}$, then $\sem{\Phase( \alpha, k )} \ket{y} = \ket{y}$.
            It follows that $\sem{\Phase( \alpha, k + 1 )} \ket{x} = \ket{a} \otimes \ket{y} = \ket{x}$.
            In conclusion, $\ket{x} \ne \ket{\One( k + 1 )}$ and $\sem{\Phase( \alpha, k + 1 )} \ket{x} = \ket{x}$.
      \item Assume $a = 1$ and $\ket{y} = \ket{\One( k )}$.
            Then $\sem{\Phase( \alpha, k + 1 )}_H \ket{x} = \ket{a} \otimes \sem{\Phase( \alpha, k )} \ket{y}$, since $\braket{0|a} = 1$ and $\braket{1|a} = 0$.
            Since \cref{Eqn:DiagHyp} holds for $k$ and $\ket{y} = \ket{\One( k )}$, then $\sem{\Phase( \alpha, k )} \ket{y} = e^{i\alpha} \ket{y}$.
            It follows that $\sem{\Phase( \alpha, k + 1 )} \ket{x} = \ket{a} \otimes e^{i\alpha} \ket{y} = e^{i\alpha} \ket{x}$.
            In conclusion, $\ket{x} \ne \ket{\One( k + 1 )}$ and $\sem{\Phase( \alpha, k + 1 )} \ket{x} = e^{i\alpha} \ket{x}$.
      \end{enumerate}
      Then \cref{Eqn:DiagHyp} holds for $k + 1$ by case distinction.
\end{itemize}
Then by the principle of induction, \cref{Eqn:DiagHyp} holds for each $n \in \mathbb{N}$.

\subsection{Constructing the Lambda Generators}

Let $x \in \{ 0, 1 \}^*$ and $\alpha \in (-\pi, \pi]$.
Since $x \in \{ 0, 1 \}^*$, then $x$ is a possibly empty word over the alphabet $\{ 0, 1 \}$.
The \emph{$(x, \alpha)$-lambda generator} is the circuit $\lambda( x, \alpha ) = \Nots( x ) \circ \Phase( \alpha, n ) \circ \Nots( x )$ where $n = |x|$.
Since $\Nots( x ) \in \Core( n, n )$ and $\Phase( \alpha, n ) \in \Core( n, n )$, then $\lambda( x, \alpha ) \in \Core( n, n )$ as well.
It must be shown that $\lambda( x, \alpha )$ has the intended semantics.
Clearly $\{ \ket{y} : y \in \{ 0, 1 \}^n \}$ is a basis for $\mathbb{C}^{2^n}$.
Let $y \in \{ 0, 1 \}^n$.
There are two cases to consider.
\begin{enumerate}
\item Assume that $\ket{y} = \ket{x}$.
      Then the following equation holds.
      \begin{align*}
          \sem{\lambda( x, \alpha )}_H \ket{y}
          &=
          \sem{\Nots( x )}_H \circ \sem{\Phase( \alpha, n )}_H \circ \sem{\Nots( x )}_H \ket{y}
          \\
          &=
          \sem{\Nots( x )}_H \circ \sem{\Phase( \alpha, n )}_H \ket{\One( n )}
          \\
          &=
          e^{i\alpha\pi} \sem{\Nots( x )}_H \ket{\One( n )}
          =
          e^{i\alpha} \ket{x}
      \end{align*}
      Then $\ket{y}$ is an eigenvector of $\sem{\lambda( x, \alpha)}_H$ with eigenvalue $e^{i\alpha}$.
\item Assume that $\ket{y} = \ket{x}$.
      Then there exists some $z \in \{ 0, 1 \}^n$ such that $\ket{z} = \sem{\Nots( x )}_H \ket{y}$ with $z \ne x$.
      Then the following equation holds.
      \begin{align*}
          \sem{\lambda( x, \alpha )}_H \ket{y}
          &=
          \sem{\Nots( x )}_H \circ \sem{\Phase( \alpha, n )}_H \circ \sem{\Nots( x )}_H \ket{y}
          \\
          &=
          \sem{\Nots( x )}_H \circ \sem{\Phase( \alpha, n )}_H \ket{z}
          \\
          &=
          \sem{\Nots( x )}_H \ket{z}
          =
          \ket{y}
      \end{align*}
      Then $\ket{y}$ is an eigenvector of $\sem{\lambda( x, \alpha)}_H$ with eigenvalue $1$.
\end{enumerate}
Then $\sem{\lambda( x, \alpha )}_H$ has eigenvalue $e^{i\alpha}$ with eigenspace $\spn\{ \ket{x} \}$ and eigenvalue $1$ with eigenspace $\spn\{ \ket{y} : y \in \{ 0, 1 \}^n \setminus \{ x \} \}$.
Since $\alpha \in (-\pi, \pi]$, then $\Log\sem{\lambda( x, \alpha )}_H = i \alpha \ket{x} \bra{x} = i \alpha \bigotimes_{j=0}^{n-1} \ket{x_j} \bra{x_j}$.
In other words, $\sem{\lambda( x, \alpha )}_H = \exp( iH )$ where $H = \alpha \bigotimes_{j=0}^{n-1} \ket{x_j} \bra{x_j}$.

\begin{theorem}
    \label{Thm:LambdaSem}
    If $x \in \{ 0, 1 \}^n$ and $\alpha \in \mathbb{R}$, then $\lambda( x, \alpha ) \in \Core( n,n )$ and $\sem{\lambda( x, \alpha )}_H = \exp( iH )$ where $H = \alpha \bigotimes_{j=0}^{n-1} \ket{x_j} \bra{x_j}$.
    If $\alpha \in ( -\pi, \pi ]$, then $\Log\sem{\lambda( x, \alpha )}_H = iH$ as well.
\end{theorem}

\begin{corollary}
    \label{Cor:Diag}
    Let $n \in \mathbb{N}$ with a bijection $f: \{ 1, 2, \ldots, 2^n \} \to \{ 0, 1 \}^n$.
    If $M \in \CatUnitary( 2^n, 2^n )$ is a diagonal matrix, then there exists a unique function $\alpha: \{ 1, 2, \ldots, 2^n \} \to ( -\pi, \pi ]$ such that $M = \sem{C}_H$ where $C = \lambda( f( 1 ), \alpha( 1 ) ) \circ \lambda( f( 2 ), \alpha( 2 ) ) \circ \cdots \circ \lambda( f( 2^n ), \alpha( 2^n ) )$.
\end{corollary}

\begin{proof}
    Since $M$ is a diagonal matrix, then $M$ has eigenbasis $\{ \ket{y} : y \in \{ 0, 1 \}^n \}$.
    For each $j \in \mathbb{N}$, let $\ket{b_j}$ denote $\ket{f( j )}$ and $\beta_j$ denote the eigenvalue associated to $\ket{b_j}$.
    Then define $\alpha( j ) = \Log( \beta_j ) / i$.
    It remains to be shown that $\sem{C}_H = M$.
    By \cref{Thm:LambdaSem}, $\Log\sem{\lambda( f( j ), \alpha( j ) )}_H = i \alpha( j ) \ket{b_j} \bra{b_j}$ for each $j \in \{ 1, 2, \ldots, 2^n \}$.
    Then it suffices to show that each $\Log\sem{\lambda( f(j), \alpha(j) )}_H$ commutes with each $\Log\sem{\lambda( f(k), \alpha(k) )}_H$.
    To this end, let $j \in \{ 1, 2, \ldots, 2^n \}$ and $k \in \{ 1, 2, \ldots, 2^n \}$ such that $j \ne k$.
    Since $f$ is a bijection, then $f(j) \ne f(k)$.
    Then $\braket{b_j|b_k} = \delta_{j,k} = 0 = \delta_{k,j} = \braket{b_k|b_j}$ where $\delta$ is the Kronecker delta function.
    Then the following equation holds.
    \begin{equation*}
        \Log\sem{\lambda( f( j ), \alpha( j ) )}_H\Log\sem{\lambda( f( k ), \alpha( k ) )}_H
        =
        \mathbb{O}_{2^n}
        =
        \Log\sem{\lambda( f( k ), \alpha( k ) )}_H\Log\sem{\lambda( f( j ), \alpha( j ) )}_H
    \end{equation*}
    Since $j$ and $k$ were arbitrary, then $\Log\sem{\lambda( f(j), \alpha(j) )}_H$ commutes $\Log\sem{\lambda( f(k), \alpha(k) )}_H$ for each $j \in \{ 1, 2, \ldots, 2^n \}$ and $k \in \{ 1, 2, \ldots, 2^n \}$.
    Then the following equation holds.
    \begin{align*}
        \sem{C}_H
        =
        \prod_{j=1}^{2^n} \sem{\lambda( f( j ), \alpha( j )}_H
        &=
        \prod_{j=1}^{2^n} \exp( i \alpha(j) \ket{b_j} \bra{b_j} )
        \\
        &=
        \exp\left( \sum_{j=1}^{2^n} i \alpha(j) \ket{b_j} \bra{b_j} \right)
        &
        ( \text{by~\cref{Eqn:SimulDiag}} )
        \\
        &=
        \sum_{j=1}^{2^n} \exp( i \alpha(j) ) \ket{b_j} \bra{b_j}
        \\
        &=
        \sum_{j=1}^{2^n} \exp( i \Log( \beta_j ) / i ) \ket{b_j} \bra{b_j}
        \\
        &=
        \sum_{j=1}^{2^n} \beta_j \ket{f(j)} \bra{f(j)}
        =
        M
    \end{align*}
    It remains to be shown that $\alpha$ is unique.
    Assume that there exists some $\gamma: \{ 1, 2, \ldots, 2^n \} \to (-\pi, \pi]$ such that $M = \sem{D}_H$ where $D = \lambda( f( 1 ), \gamma( 1 ) ) \circ \lambda( f( 2 ), \gamma( 2 ) ) \circ \cdots \circ \lambda( f( 2^n ), \gamma( 2^n ) )$.
    It suffices to show that $\alpha$ and $\gamma$ agree pointwise.
    Let $j \in \{ 1, 2, \ldots, 2^n \}$.
    Then the following equation holds.
    \begin{equation*}
        e^{i\alpha(j)}
        =
        \sum_{k=1}^{2^n} e^{i\alpha(j)} \delta_{j,k} \delta_{k,j}
        =
        \bra{b_j} \sem{C}_H \ket{b_j}
        =
        \bra{b_j} \sem{D}_H \ket{b_j}
        =
        \sum_{k=1}^{2^n} e^{i\gamma(j)} \delta_{j,k} \delta_{k,j}
        =
        e^{i\gamma(j)}
    \end{equation*}
    Since $\alpha(j) \in (-\pi,\pi]$ and $\beta(j) \in (-\pi,\pi]$, then the following equation holds.
    \begin{equation*}
        \alpha(j)
        =
        i\alpha(j)/i
        =
        \Log( \exp( i\alpha( j ) ) ) / i
        =
        \Log( \exp( i\gamma( j ) ) ) / i
        =
        i\gamma(j)/i
        =
        \gamma(j)
    \end{equation*}
    Since $j$ was arbitrary, then $\alpha = \gamma$.
    Since $\gamma$ was arbitrary, then $\alpha$ is unique.
\end{proof}

\section{Proofs for Section~\ref{Sect:Hierarchical}}
%

This section contains all proofs for~\cref{Sect:Background:Theory}.

\subsection{Proof of Theorem~\ref{Thm:Soundness}}
\begin{proof}
    First it must be shown that each relation in $E_0$ is sound.
    Let $( X, Y ) \in E_0$.
    Start by noting the following.
    \begin{align*}
        \sem{\scalebox{0.65}{\circimg{tikz-cache/encoder__eqn1_lhs.pdf}}}_H
        &=
        e^{\Log( Z ) \alpha}
        =
        \begin{bsmallmatrix}
            1 & 0 \\
            0 & e^{i\alpha\pi}
        \end{bsmallmatrix}
        =
        Z( \alpha\pi )
        &
        \sem{\scalebox{0.65}{\circimg{tikz-cache/encoder__eqn4_lhs.pdf}}}_H
        &=
        e^{\Log( X ) \alpha} 
        =
        \tfrac{1}{2}
        \begin{bsmallmatrix}
            1 + e^{i\alpha\pi} & 1 - e^{i\alpha\pi} \\
            1 - e^{i\alpha\pi} & 1 + e^{i\alpha\pi}
        \end{bsmallmatrix}
        =
        X( \alpha\pi )
    \end{align*}
    There are several cases to consider.
    \begin{itemize}
    \item Assume that $( X, Y )$ is \HqcHDef. Then the following equation holds.
          \begin{align*}
              \sem{X}_H
              &=
              \sem*{\scalebox{0.65}{\circimg{tikz-cache/hqc__eqn8_lhs.pdf}}}_H
              =
              \tfrac{1}{\sqrt{2}}
              \begin{bsmallmatrix}
                1 & 1 \\
                1 & -1
              \end{bsmallmatrix}
              \\
              &=
              \overbrace{
                \underbrace{
                    \tfrac{1}{2}
                    \begin{bsmallmatrix}
                        1 - i & 1 + i \\
                        1 + i & 1 - i
                    \end{bsmallmatrix}
                }_{X(-\pi/2)}
                \underbrace{
                    \begin{bsmallmatrix}
                        1 & 0 \\
                        0 & e^{i\pi/4}
                    \end{bsmallmatrix}
                }_{Z(\pi/4)}
                \underbrace{
                    \tfrac{1}{2}
                    \begin{bsmallmatrix}
                        1 + i & 1 - i \\
                        1 - i & 1 + i
                    \end{bsmallmatrix}
                }_{X(\pi/2)}
              }^{Y(\pi/4)}
              \underbrace{
                \begin{bsmallmatrix}
                    1 & 0 \\
                    0 & -1
                \end{bsmallmatrix}
              }_{Z}
              \overbrace{
                \underbrace{
                    \tfrac{1}{2}
                    \begin{bsmallmatrix}
                        1 - i & 1 + i \\
                        1 + i & 1 - i
                    \end{bsmallmatrix}
                }_{X(-\pi/2)}
                \underbrace{
                    \begin{bsmallmatrix}
                        1 & 0 \\
                        0 & e^{-i\pi/4}
                    \end{bsmallmatrix}
                }_{Z(-\pi/4)}
                \underbrace{
                    \tfrac{1}{2}
                    \begin{bsmallmatrix}
                        1 + i & 1 - i \\
                        1 - i & 1 + i
                    \end{bsmallmatrix}
                }_{X(\pi/2)}
              }^{Y(-\pi/4)}
              \\
              &=
              \sem*{\scalebox{0.65}{\circimg{tikz-cache/hqc__eqn8_rhs.pdf}}}_H
              =
              \sem{Y}_H
          \end{align*}

    \item Assume that $( X, Y )$ is \HqcTargDef.
          Then the following equation holds.
          \begin{align*}
            \sem{X}_H
            =
            \sem*{\scalebox{0.65}{\circimg{tikz-cache/hqc__eqn10_lhs.pdf}}}_H
            =
            \begin{bsmallmatrix}
                0 & 1 \\
                1 & 0
            \end{bsmallmatrix}
            =
            \tfrac{1}{\sqrt{2}}\begin{bsmallmatrix}
                1 & 1 \\
                1 & -1
            \end{bsmallmatrix}
            \begin{bsmallmatrix}
                1 & 0 \\
                0 & -1
            \end{bsmallmatrix}
            \tfrac{1}{\sqrt{2}}\begin{bsmallmatrix}
                1 & 1 \\
                1 & -1
            \end{bsmallmatrix}
            =
            \sem*{\scalebox{0.65}{\circimg{tikz-cache/hqc__eqn10_rhs.pdf}}}_H
            =
            \sem{Y}_H
          \end{align*}

    \item Assume that $( X, Y )$ is \HqcOPhaseDef.
          Then the following equation holds.
          \begin{align*}
            \sem{X}_H
            =
            \sem*{\scalebox{0.65}{\circimg{tikz-cache/hqc__eqn11_lhs.pdf}}}_H
            =
            \begin{bsmallmatrix}
                -1 & 0 \\
                0 & 1
            \end{bsmallmatrix}
            =
            \begin{bsmallmatrix}
                0 & 1 \\
                1 & 0
            \end{bsmallmatrix}
            \begin{bsmallmatrix}
                1 & 0 \\
                0 & -1
            \end{bsmallmatrix}
            \begin{bsmallmatrix}
                0 & 1 \\
                1 & 0
            \end{bsmallmatrix}
            =
            \sem*{\scalebox{0.65}{\circimg{tikz-cache/hqc__eqn11_rhs.pdf}}}_H
            =
            \sem{Y}_H
          \end{align*}

    \item Assume that $( X, Y )$ is \HqcTargMDef.
          Then the following equation holds.
          \begin{align*}
            \sem{X}_H
            =
            \sem*{\scalebox{0.65}{\circimg{tikz-cache/hqc__eqn12_lhs.pdf}}}_H
            =
            \begin{bsmallmatrix}
                0 & -1 \\
                -1 & 0
            \end{bsmallmatrix}
            =
            \begin{bsmallmatrix}
                1 & 0 \\
                0 & -1
            \end{bsmallmatrix}
            \begin{bsmallmatrix}
                0 & 1 \\
                1 & 0
            \end{bsmallmatrix}
            \begin{bsmallmatrix}
                1 & 0 \\
                0 & -1
            \end{bsmallmatrix}
            =
            \sem*{\scalebox{0.65}{\circimg{tikz-cache/hqc__eqn12_rhs.pdf}}}_H
            =
            \sem{Y}_H
          \end{align*}

    \item Assume that $( X, Y )$ is \HqcSwapDef.
          Then the following equation holds.
          \begin{align*}
            \sem{X}_H
            =
            \sem*{\scalebox{0.65}{\circimg{tikz-cache/proofs__decode_q3_lhs.pdf}}}_H
            &=
            \begin{bsmallmatrix}
                1 & 0 & 0 & 0 \\
                0 & 0 & 1 & 0 \\
                0 & 1 & 0 & 0 \\
                0 & 0 & 0 & 1
            \end{bsmallmatrix}
            \\
            &=
            \begin{bsmallmatrix}
                1 & 0 & 0 & 0 \\
                0 & 1 & 0 & 0 \\
                0 & 0 & 0 & 1 \\
                0 & 0 & 1 & 0
            \end{bsmallmatrix}
            \begin{bsmallmatrix}
                1 & 0 & 0 & 0 \\
                0 & 0 & 0 & 1 \\
                0 & 0 & 1 & 0 \\
                0 & 1 & 0 & 0
            \end{bsmallmatrix}
            \begin{bsmallmatrix}
                1 & 0 & 0 & 0 \\
                0 & 1 & 0 & 0 \\
                0 & 0 & 0 & 1 \\
                0 & 0 & 1 & 0
            \end{bsmallmatrix}
            \\
            &=
            \sem{Z \odot X}_H \sem{X \odot Z}_H \sem{Z \odot X}_H
            \\
            &=
            \sem*{\scalebox{0.65}{\circimg{tikz-cache/proofs__decode_q3_mid.pdf}}}_H
            =
            \sem{Y}_H
          \end{align*}

    \item Assume that $( X, Y )$ is an instance of the relation \HqcEuler.
          Then let $\theta_1, \theta_2 \in \mathbb{R}$ and define $( \phi_0\pi, \phi_1\pi, \phi_2\pi, \phi_3\pi ) = \mathsf{Euler}( \theta_1\pi, \theta_2\pi )$ as in an arbitrary instance of \HqcEuler.
          Then by the soundness of Q(5), the equation $H Z( \theta_1\pi ) H Z( \theta_2\pi ) H = e^{i\phi_0\pi} Z( \phi_1\pi ) H Z( \phi_2\pi ) H Z( \phi_3\pi )$ holds.
          Then the following equation holds.
          \begin{align*}
            \sem{X}_H
            =
            \sem*{\scalebox{0.65}{\circimg{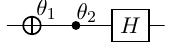}}}_H
            &=
            H Z( \theta_2\pi ) X( \theta_1\pi )
            \\
            &=
            H Z( \theta_2\pi ) H Z( \theta_1\pi ) H
            \\
            &=
            e^{i\phi_0\pi} Z( \phi_1\pi ) H Z( \phi_2\pi ) H Z( \phi_3\pi )
            \\
            &=
            e^{i\phi_0\pi} Z( \phi_1\pi ) X( \phi_2\pi ) Z( \phi_3\pi )
            \\
            &=
            \sem*{\scalebox{0.65}{\circimg{tikz-cache/hqc__euler_rhs.pdf}}}_H
            =
            \sem{Y}_H
          \end{align*}

    \item Assume that $( X, Y )$ is \HqcPOrd.
          Then the following equation holds.
          \begin{equation*}
            \sem{X}_H
            =
            \sem*{\scalebox{0.65}{\circimg{tikz-cache/hqc__eqn7_lhs.pdf}}}_H
            =
            \exp( 2 \Log\sem{\scalebox{0.65}{\circimg{tikz-cache/dagger__term8.pdf}}}_H )
            =
            \exp( 2 \Log(-1) )
            =
            \exp( 2 (i\pi) )
            =
            \mathbb{I}_1
            =
            \sem{1_0}_H
            =
            \sem{Y}_H
          \end{equation*}

    \item Assume that $( X, Y )$ is an instance of \HqcCtrlId.
          Then the following holds by \cref{Cor:GenCtrlFunc}.
          \begin{equation*}
            \sem{X}_H
            =
            \sem*{\scalebox{0.65}{\circimg{tikz-cache/hqc__eqn1_lhs.pdf}}}_H
            =
            \sem{U}_H \odot \sem{1_m}_H
            =
            \mathbb{I}_{2^{n+m}}
            =
            \mathbb{I}_{2^n} \otimes \mathbb{I}_{2^m}
            =
            \sem*{\scalebox{0.65}{\circimg{tikz-cache/hqc__eqn1_rhs.pdf}}}_H
            =
            \sem{Y}_H
          \end{equation*}
   
    \item Assume that $( X, Y )$ is an instance of \HqcCtrlComp.
          Since $Z$ is unitary with eigenvalues $\{ -1, 1 \}$, then $Z$ is Hermitian.
          Then the following equation holds.
          \begin{align*}
            \sem{X}_H
            =
            \sem*{\scalebox{0.65}{\circimg{tikz-cache/hqc__eqn2_lhs.pdf}}}_H
            &=
            Z \odot \sem{V \circ U}_H
            \\
            &=
            Z \odot ( \sem{V}_H \circ \sem{U}_H )
            \\
            &=
            ( Z \odot \sem{V}_H ) \circ ( Z \odot \sem{U} )_H
            &
            ( \text{by Cor.~\cref{Cor:GenCtrlFunc}} )
            \\
            &=
            \sem*{\scalebox{0.65}{\circimg{tikz-cache/hqc__eqn2_rhs.pdf}}}_H
            =
            \sem{Y}_H
          \end{align*}
          
    \item Assume that $( X, Y )$ is an instance of \HqcCtrlSwap.
          Since $\sem{-}_H: \CatHQC \to \CatUnitary$ is symmetric monoidal, then $\sem*{\scalebox{0.5}{\circimg{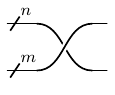}}}_H$ is the monoidal symmetry $\sigma_{2^n,2^m}: \mathbb{C}^{2^n} \otimes \mathbb{C}^{2^m} \cong \mathbb{C}^{2^m} \otimes \mathbb{C}^{2^n}$.
          Then $\sigma_{2^n,2^m}$ is the self-adjoint unitary operator whose action is $\sigma_{2^n,2^m} \cdot ( N \otimes M ) \cdot \sigma_{2^n,2^m} = M \otimes N$ for all $M \in \CatUnitary( 2^n, 2^n )$ and $N \in \CatUnitary( 2^m, 2^m )$.
          Then the following equation holds.
          \begin{align*}
            \sem{X}
            =
            \sem*{\scalebox{0.65}{\circimg{tikz-cache/hqc__eqn3_lhs.pdf}}}_H
            &=
            \exp( \Log\sem{U}_H \otimes \Log\sem{V}_H / (i\pi) )
            \\
            &=
            \exp( \sigma_{2^n,2^m} \cdot ( \Log\sem{V}_H \otimes \Log\sem{U}_H / (i\pi) ) \cdot \sigma_{2^n,2^m} )
            \\
            &=
            \sigma_{2^n,2^m} \cdot \exp( \Log\sem{V}_H \otimes \Log\sem{U}_H / (i\pi) ) \cdot \sigma_{2^n,2^m}
            &
            ( \text{by~\cref{Eqn:UnitaryConj}} )
            \\
            &=
            \sem*{\scalebox{0.65}{\circimg{tikz-cache/hqc__eqn3_rhs.pdf}}}_H
            =
            \sem{Y}_H
          \end{align*}

    \item Assume that $( X, Y )$ is an instance of \HqcFreePhase.
          Then the following holds by \cref{Thm:PowAsCtrl}.
          \begin{align*}
            \sem{X}_H
            =
            \sem*{\scalebox{0.65}{\circimg{tikz-cache/hqc__eqn18_lhs.pdf}}}_H
            =
            Z \odot \sem*{\scalebox{0.65}{\circimg{tikz-cache/encoder__eqn3_lhs.pdf}}}
            =
            Z \odot e^{i\alpha\pi}
            =
            Z^\alpha
            =
            \sem*{\scalebox{0.65}{\circimg{tikz-cache/encoder__eqn1_lhs.pdf}}}_H
            =
            \sem{Y}_H
          \end{align*}

    \item Assume that $( X, Y )$ is an instance of \HqcCtrlDag.
          Then the following equation holds.
          \begin{align*}
            \sem{X}_H
            =
            \sem*{\scalebox{0.65}{\circimg{tikz-cache/hqc__eqn14_lhs.pdf}}}_H
            &=
            \sem{V}_H \odot \sem{P^\dagger \circ U \circ P}
            \\
            &=
            \sem{V}_H \odot ( \sem{P^\dagger}_H \circ \sem{U}_H \circ \sem{P}_H )
            \\
            &=
            \sem{V}_H \odot ( \sem{P}_H^\dagger \circ \sem{U}_H \circ \sem{P}_H )
            &
            ( \text{by Thm.~\ref{Thm:DagSem}} )
            \\
            &=
            ( \mathbb{I}_{2^m} \otimes \sem{P}_H^\dagger ) ( \sem{V} \odot \sem{U} ) ( \mathbb{I}_{2^m} \otimes \sem{P}_H )
            &
            ( \text{by Cor.~\ref{Cor:GenCtrlFunc}} )
            \\
            &=
            ( \mathbb{I}_{2^m} \otimes \sem{P^\dagger}_H ) ( \sem{V} \odot \sem{U} ) ( \mathbb{I}_{2^m} \otimes \sem{P}_H )
            &
            ( \text{by Thm.~\ref{Thm:DagSem}} )
            \\
            &=
            \sem*{\scalebox{0.65}{\circimg{tikz-cache/hqc__eqn14_rhs.pdf}}}_H
            =
            \sem{Y}_H
          \end{align*}

    \item Assume that $( X, Y )$ is an instance of \HqcCtrlPair.
          First, note that for each $M \in \CatUnitary( 2^n, 2^n )$, the matrix $M \otimes \ket{0} \bra{0}$ commutes with the matrix $M \otimes \ket{1} \bra{1}$ since $\braket{0|1} = 0 = \braket{1|0}$.
          Then the following equation holds where $A = \Log( -Z ) = i \ket{0} \bra{0} \pi$ and $B = \Log( Z ) = i \ket{1} \bra{1} \pi$.
          \begin{align*}
            \sem{X}_H
            =
            \sem*{\scalebox{0.65}{\circimg{tikz-cache/hqc__eqn15_lhs.pdf}}}_H
            &=
            \exp( \Log\sem{U}_H \otimes A / (i\pi) ) \exp( \Log\sem{U}_H \otimes B / (i\pi) )
            \\
            &=
            \exp( \Log\sem{U}_H \otimes \ket{0}\bra{0} ) \exp( \Log\sem{U}_H \otimes \ket{1}\bra{1} )
            &
            ( \text{by~\cref{Eqn:SimulDiag}} )
            \\
            &=
            \exp\big( ( \Log\sem{U}_H \otimes \ket{0}\bra{0} ) + ( \Log\sem{U}_H \otimes \ket{1}\bra{1} ) \big) 
            \\
            &=
            \exp( \Log\sem{U}_H \otimes (\ket{0}\bra{0} + \ket{1}\bra{1}) )
            \\
            &=
            \exp( \Log\sem{U}_H \otimes \mathbb{I}_2 )
            \\
            &=
            \exp( \Log\sem{U}_H ) \otimes \mathbb{I}_2
            \\
            &=
            \sem{U}_H \otimes \mathbb{I}_2
            \\
            &=
            \sem*{\scalebox{0.65}{\circimg{tikz-cache/hqc__eqn15_rhs.pdf}}}_H
            =
            \sem{Y}_H
          \end{align*}

    \item Assume that $( X, Y )$ is an instance of \HqcCtrlExpand.
          As in \cref{App:DiagGate}, let $\ket{x} = \bigotimes_{j=0}^{n-1} \ket{x_j}$.
          Moreover, let $\Lambda$ be the prefix of a diagonal form for which the term $\lambda( x, \rho )$ does not appear.
          This means that the eigenvalue of $\ket{x} \in \sem{\Lambda}_H$ is $1$, and consequently $\sem{\Lambda}_H$ has the form $\sem{\Lambda}_H = \exp\left( \sum_{y \in \mathcal{B}} i \theta_y \ket{y} \bra{y} \right)$ where $\mathcal{B} = \{ 0, 1 \}^n \setminus \{ x \}$ and $\theta_y \in ( -\pi, \pi ]$ for each $y \in \mathcal{B}$ (see~\cref{Cor:Diag}).
          Since $x \not \in \mathcal{B}$, then for each $y \in \mathcal{B}$, $\braket{x|y} = 0 = \braket{y|x}$ and consequently the matrix $\ket{x}\bra{x}$ commutes with $\ket{y}\bra{y}$.
          Then the following equation holds for $M = \Log\sem{\Gamma}_H / (i\pi)$.
          \begin{align*}
            \sem{X}_H
            &=
            \sem*{\scalebox{0.65}{\circimg{tikz-cache/hqc__eqn22_lhs.pdf}}}_H
            \\
            &=
            \exp\left( \Log\sem{\Lambda \circ \lambda(x,\rho)}_H \otimes M \right)
            \\
            &=
            \exp\left( \Log\left( \sem{\Lambda}_H \sem{\lambda(x,\rho)}_H  \right) \otimes M \right)
            \\
            &=
            \exp\left( \Log\left( \exp\left( \sum_{y \in \mathcal{B}} i \theta_y \ket{y}\bra{y} \right) \exp( i\rho\ket{x}\bra{x} ) \right) \otimes M \right)
            &
            ( \text{by Thm.~\ref{Thm:LambdaSem}} )
            \\
            &=
            \exp\left( \Log\left( \exp\left( \sum_{y \in \mathcal{B}} \left( i \theta_y \ket{y}\bra{y} \right) + i\rho \ket{x}\bra{x} \right) \right) \otimes M \right)
            &
            ( \text{by \cref{Eqn:SimulDiag}} )
            \\
            &=
            \exp\left( \Log\left( \sum_{y \in \mathcal{B}} \left( e^{i \theta_y} \ket{y}\bra{y} \right) + e^{i \rho} \ket{x}\bra{x} \right) \otimes M \right)
            \\
            &=
            \exp\left( \left( \sum_{y \in \mathcal{B}} \left( \Log\left( e^{i \theta_y} \right) \ket{y}\bra{y} \right) + \Log\left( e^{i \rho} \right) \ket{x}\bra{x} \right) \otimes M \right)
            \\
            &=
            \exp\left( \left( \left( \sum_{y \in \mathcal{B}} i \theta_y \ket{y}\bra{y} \right) + i \rho \ket{x}\bra{x} \right) \otimes M \right)
            \\
            &=
            \exp\left( \left( \Log\sem{\Lambda}_H + \Log\sem{\lambda(x,\rho)}_H \right) \otimes M \right)
            \\
            &=
            \exp \left( \left( \Log\sem{\Lambda}_H \otimes M \right) + \left( \Log\sem{\lambda(x,\rho)}_H \otimes M \right) \right)
            \\
            &=
            \exp \left( \Log\sem{\Lambda}_H \otimes M \right) \exp \left( \Log\sem{\lambda(x,\rho)}_H \otimes M \right)
            &
            ( \text{by \cref{Eqn:SimulDiag}} )
            \\
            &=
            \sem*{\scalebox{0.65}{\circimg{tikz-cache/hqc__eqn22_rhs.pdf}}}_H
            =
            \sem{Y}_H
          \end{align*}
          
    \item Assume that $( X, Y )$ is an instance of \HqcCtrlMerge.
          Then let $\rho \in (-\pi, \pi]$ and $\gamma \in (-\pi, \pi]$ as in an arbitrary instance of \HqcCtrlMerge.
          Then the following equation holds.
          \begin{align*}
            \sem{X}_H
            =
            \sem*{\scalebox{0.65}{\circimg{tikz-cache/hqc__eqn23_lhs.pdf}}}_H
            &=
            \exp( \Log\sem{\lambda( x, \rho )}_H \otimes \Log\sem{\lambda( y, \gamma )}_H / ( i\pi ) )
            \\
            &=
            \exp\left( (i \rho\ket{x}\bra{x}) \otimes (i\gamma\ket{y}\bra{y}) / ( i\pi ) \right)
            &
            ( \text{by Thm.~\ref{Thm:LambdaSem}} )
            \\
            &=
            \exp\left( \left( i^2 \rho\gamma \ket{xy} \bra{xy} \right) / ( i\pi ) \right)
            \\
            &=
            \exp( i ( \rho \gamma / \pi ) \ket{xy}\bra{xy} )
            \\
            &=
            \sem*{\scalebox{0.65}{\circimg{tikz-cache/hqc__eqn23_rhs.pdf}}}_H
            =
            \sem{Y}_H
          \end{align*}

    \item Assume that $( X, Y )$ is an instance of \HqcPowZero.
          Then the following equation holds.
          \begin{align*}
            \sem{X}_H
            =
            \sem*{\scalebox{0.65}{\circimg{tikz-cache/hqc__eqn9_lhs.pdf}}}_H
            =
            \exp( 0 \Log\sem{U}_H )
            =
            \exp( \mathbb{O}_{2^n} )
            =
            \mathbb{I}_{2^n}
            =
            \sem*{\scalebox{0.65}{\circimg{tikz-cache/hqc__eqn9_rhs.pdf}}}_H
            =
            \sem{Y}_H
          \end{align*}
    
    \item Assume that $( X, Y )$ is an instance of \HqcPowId.
          Then the following equation holds.
          \begin{align*}
            \sem{X}_H
            =
            \sem*{\scalebox{0.65}{\circimg{tikz-cache/hqc__eqn4_lhs.pdf}}}_H
            =
            \exp( 1 \Log\sem{U}_H )
            =
            \exp( \Log\sem{U}_H )
            =
            U
            =
            \sem*{\scalebox{0.65}{\circimg{tikz-cache/hqc__eqn4_rhs.pdf}}}_H
            =
            \sem{Y}_H
          \end{align*}

    \item Assume that $( X, Y )$ is an instance of \HqcPowAdd.
          Then the following equation holds.
          \begin{align*}
            \sem{X}_H
            =
            \sem*{\scalebox{0.65}{\circimg{tikz-cache/hqc__eqn5_lhs.pdf}}}_H
            &=
            \exp( \beta \Log\sem{U}_H ) \exp( \alpha \Log\sem{U}_H )
            \\
            &=
            \exp( (\alpha + \beta) \Log\sem{U}_H )
            &
            ( \text{by~\cref{Eqn:SimulDiag}} )
            \\
            &=
            \sem*{\scalebox{0.65}{\circimg{tikz-cache/hqc__eqn5_rhs.pdf}}}_H
            =
            \sem{Y}_H
          \end{align*}

    \item Assume that $( X, Y )$ is an instance of \HqcPowIdem.
          Then the following equation holds.
          \begin{align*}
            \sem{X}_H
            =
            \sem*{\scalebox{0.65}{\circimg{tikz-cache/hqc__eqn24_lhs.pdf}}}_H
            =
            \exp( \alpha \Log( \mathbb{I}_{2^n} ) )
            =
            \exp( \alpha \mathbb{O}_{2^n} )
            =
            \exp( \mathbb{O}_{2^n} )
            =
            \mathbb{I}_{2^n}
            =
            \sem*{\scalebox{0.65}{\circimg{tikz-cache/hqc__eqn24_rhs.pdf}}}_H
            =
            \sem{Y}_H
          \end{align*}

    \item Assume that $( X, Y )$ is an instance of \HqcPowDag.
          Then the following equation holds.
          \begin{align*}
            \sem{X}_H
            =
            \sem*{\scalebox{0.65}{\circimg{tikz-cache/hqc__eqn19_lhs.pdf}}}_H
            &=
            \sem{P^\dagger}_H \exp( \Log\sem{U}_H \alpha ) \sem{P}_H
            \\
            &=
            \sem{P}_H^\dagger \exp( \Log\sem{U}_H \alpha ) \sem{P}_H
            &
            ( \text{by Thm.~\ref{Thm:DagSem}} )
            \\
            &=
            \exp\left( \sem{P}_H^\dagger \Log\sem{U}_H \sem{P}_H \alpha \right) 
            &
            ( \text{by~\cref{Eqn:UnitaryConj}} )
            \\
            &=
            \exp\left( \Log \left( \sem{P}_H^\dagger \sem{U}_H \sem{P}_H \right) \alpha \right)
            &
            ( \text{by~\cref{Eqn:UnitaryConj}} )
            \\
            &=
            \sem*{\scalebox{0.65}{\circimg{tikz-cache/hqc__eqn19_rhs.pdf}}}_H
            =
            \sem{Y}_H
          \end{align*}

    \item Assume that $( X, Y )$ is an instance of \HqcPowExpand.
          As in \cref{App:DiagGate}, let $\ket{x} = \bigotimes_{j=0}^{n-1} \ket{x_j}$.
          Moreover, let $\Lambda$ be the prefix of a diagonal form for which the term $\lambda( x, \rho )$ does not appear.
          This means that the eigenvalue of $\ket{x} \in \sem{\Lambda}_H$ is $1$, and consequently $\sem{\Lambda}_H$ has the form $\sem{\Lambda}_H = \exp\left( \sum_{y \in \mathcal{B}} i \theta_y \ket{y} \bra{y} \right)$ where $\mathcal{B} = \{ 0, 1 \}^n \setminus \{ x \}$ and $\theta_y \in ( -\pi, \pi ]$ for each $y \in \mathcal{B}$ (see~\cref{Cor:Diag}).
          Since $x \not \in \mathcal{B}$, then for each $y \in \mathcal{B}$, $\braket{x|y} = 0 = \braket{y|x}$ and consequently the matrix $\ket{x}\bra{x}$ commutes with $\ket{y}\bra{y}$.
          Then the following equation holds.
          \begin{align*}
            \sem{X}_H
            &=
            \sem*{\scalebox{0.65}{\circimg{tikz-cache/hqc__eqn21_lhs.pdf}}}_H
            \\
            &=
            \exp\left( \Log\sem{\Lambda \circ \lambda(x,\rho)}_H \alpha \right)
            \\
            &=
            \exp\left( \Log\left( \sem{\Lambda}_H \sem{\lambda(x,\rho)}_H  \right) \alpha \right)
            \\
            &=
            \exp\left( \Log\left( \exp\left( \sum_{y \in \mathcal{B}} i \theta_y \ket{y}\bra{y} \right) \exp( i\rho\ket{x}\bra{x} ) \right) \alpha \right)
            &
            ( \text{by Thm.~\ref{Thm:LambdaSem}} )
            \\
            &=
            \exp\left( \Log\left( \exp\left( \sum_{y \in \mathcal{B}} \left( i \theta_y \ket{y}\bra{y} \right) + i\rho \ket{x}\bra{x} \right) \right) \alpha \right)
            &
            ( \text{by \cref{Eqn:SimulDiag}} )
            \\
            &=
            \exp\left( \Log\left( \sum_{y \in \mathcal{B}} \left( e^{i \theta_y} \ket{y}\bra{y} \right) + e^{i \rho} \ket{x}\bra{x} \right) \alpha \right)
            \\
            &=
            \exp\left( \left( \sum_{y \in \mathcal{B}} \left( \Log\left( e^{i \theta_y} \right) \ket{y}\bra{y} \right) + \Log\left( e^{i \rho} \right) \ket{x}\bra{x} \right) \alpha \right)
            \\
            &=
            \exp\left( \sum_{y \in \mathcal{B}} \left( i \theta_y \alpha \ket{y}\bra{y} \right) + i \alpha \rho \ket{x}\bra{x} \right)
            \\
            &=
            \exp \left( \sum_{y \in \mathcal{B}} i \theta_y \alpha \ket{y}\bra{y} \right) \exp \Big( i \alpha \rho \ket{x}\bra{x} \Big)
            &
            ( \text{by \cref{Eqn:SimulDiag}} )
            \\
            &=
            \exp\left( \Log \sem{\Lambda}_H \alpha \right) \cdot \sem{\lambda(x,\alpha\rho)}_H
            \\
            &=
            \sem*{\scalebox{0.65}{\circimg{tikz-cache/hqc__eqn21_rhs.pdf}}}_H
            =
            \sem{Y}_H
          \end{align*}
    \end{itemize}
    Since $( X, Y )$ was arbitrary, then $\sem{-}_H$ is sound with respect to $E_0$.
    It then follows by induction on $E_n$ that $E$ is sound with respect to $\sem{-}_H$.
    \begin{itemize}
    \item \textbf{Base Case}.
             It follows by the proceeding analysis that $E_0$ is sound with respect to $\sem{-}_H$.
    \item \textbf{Inductive Hypothesis}.
             For some $n \in \mathbb{N}$, $E_n$ is sound with respect to $\sem{-}_H$.
    \item \textbf{Inductive Step}.
          Assume that for some $n \in \mathbb{N}$, that $E_n$ is sound with respect to $\sem{-}_H$.
          Then let $( X, Y ) \in E_{n+1}$.
          There are three cases to consider.
          \begin{enumerate}
          \item Assume that $( X, Y ) \in E_n$.
                Then $\sem{X}_H = \sem{Y}_H$ by the inductive hypothesis.
          \item Assume that there exists some $U \in \Mor( \Core )$ and $V \approx_{E_n} W$ such that $X = U \odot V$ and $Y = U \odot W$.
                Then by the inductive hypothesis, $\sem{V}_H = \sem{W}_H$.
                It follows that $\sem{X}_H = \exp( \Log\sem{U}_H \otimes \sem{V}_H / (i\pi) ) = \exp( \Log\sem{U}_H \otimes \sem{W}_H / (i\pi) ) = \sem{Y}_H$.
          \item Assume that there exists some exponent $\alpha \in \mathbb{R}$ and $V \approx_{E_n} W$ such that $X = V \exparr \alpha$ and $Y = W \exparr \alpha$.
                Then by the inductive hypothesis, $\sem{V}_H = \sem{W}_H$.
                It follows that $\sem{X}_H = \exp( \alpha \Log\sem{U}_H ) = \exp( \alpha \Log\sem{W}_H ) = \sem{Y}_H$.
          \end{enumerate}
          In each case, $\sem{X}_H = \sem{Y}_H$.
          Since $( X, Y )$ was arbitrary, then $E_{n+1}$ is sound with respect to $\sem{-}_H$.
          Then the inductive step holds.
    \end{itemize}
    Then by the principle of inductive, $E_n$ is sound with respect to $\sem{-}_H$ for each $n \in \mathbb{N}$.
    Assume that $( X, Y ) \in E$.
    Then there exists some $n \in \mathbb{N}$ such that $( X, Y ) \in E_n$.
    Then $\sem{X}_H = \sem{Y}_H$.
    Since $( X, Y )$ was arbitrary, then $E$ is sound with respect to $\sem{-}_H$.
\end{proof}

\subsection{Proof of Theorem~\ref{Thm:PropInv}}
\begin{proof}
    First, it must be shown that $F( f ) \circ f \approx_E 1_n$ and $f \circ F( f ) \approx_E 1_m$ for each $f \in P( \Sigma )( n, m )$.
    This follows by structural induction on the morphisms in $P( \Sigma )$.
    \begin{itemize}
    \item \textbf{Identity}.
          Assume that $f = 1_n$.
          Since $F$ is a prop functor, then $F( f ) = F( 1_n ) = 1_n$.
          Then $F( f ) \circ f = 1_n \circ 1_n = 1_n$ and $f \circ F( f ) = 1_n \circ 1_n = 1_n$.
          Since $\approx_E$ is reflexive, then $F( f ) \circ f \approx_E 1_n$ and $f \circ F( f ) \approx_E 1_n$.
    \item \textbf{Symmetries}.
          Assume that $f = \sigma$.
          Since $F$ is a strict symmetric monoidal functor, then $F( \sigma ) = \sigma$.
          By definition, $F( f ) \circ f = \sigma \circ \sigma = 1_2$ and $f \circ F( f ) = \sigma \circ \sigma = 1_2$.
          Since $\approx_E$ is reflexive, then $F( f ) \circ f \approx_E 1_2$ and $f \circ F( f ) \approx_E 1_2$.
    \item \textbf{Generators}.
          Assume that $f = x$ for some $x \in \Sigma$.
          Then $n = \dom( x )$ and $m = \cod( x )$.
          Moreover, $F( f ) \circ f = F( x ) \circ x \approx_E 1_n$ and $f \circ F( f ) = x \circ F( x ) \approx_E 1_m$.
    \item \textbf{Sequential Composition}.
          Assume that $f = b \circ a$ for $a: n \to s$ and $b: s \to m$.
          Then by the inductive hypothesis, the following equations hold.
          \begin{align*}
                F( a ) \circ a &\approx_E 1_n
                &
                a \circ F( a ) &\approx_E 1_x
                &
                F( b ) \circ b &\approx_E 1_x
                &
                b \circ F( b ) &\approx_E 1_m
          \end{align*}
          Since $F$ is a contravariant functor, then the following equation holds.
          \begin{align*}
              F( f ) \circ f
                &=
                ( F( a ) \circ F( b ) ) \circ ( b \circ a )
                \approx_E
                F( a ) \circ 1_x \circ a
                =
                F( a ) \circ a
                \approx_E
                1_n
              \\
              f \circ F( f )
                &=
                ( b \circ a ) \circ ( F( a ) \circ F( b ) )
                \approx_E
                b \circ 1_x \circ F( b )
                =
                b \circ F( b )
                \approx_E
                1_m
          \end{align*}
          Then $F( f ) \circ f \approx_E 1_n$ and $f \circ F( f ) \approx_E 1_m$ by the transitivity of $\approx_E$.
    \item \textbf{Parallel Composition}.
          Assume that $f = a \boxtimes b$ for $a: s \to t$ and $b: x \to y$.
          Clearly $n = s + x$ and $m = t + y$.
          Then by the inductive hypothesis, the following equations hold.
          \begin{align*}
                F( a ) \circ a &\approx_E 1_n
                &
                a \circ F( a ) &\approx_E 1_x
                &
                F( b ) \circ b &\approx_E 1_x
                &
                b \circ F( b ) &\approx_E 1_m
          \end{align*}
          Since $F$ is a strict monoidal functor, then the following equation holds.
          \begin{align*}
              F( f ) \circ f
                &=
                ( F( a ) \boxtimes F( b ) ) \circ ( a \boxtimes b )
                =
                ( F( a ) \circ a ) \boxtimes ( F( b ) \circ b )
                \approx_E
                1_s \boxtimes ( F( b ) \circ b )
                \approx_E
                1_s \boxtimes 1_x
            \\
            f \circ F( f )
                &=
                ( a \boxtimes b ) \circ ( F( a ) \boxtimes F( b ) )
                =
                ( a \circ F( a ) ) \boxtimes ( b \circ F( b ) )
                \approx_E
                1_t \boxtimes ( b \circ F( b ) )
                \approx_E
                1_t \boxtimes 1_y
          \end{align*}
          Then $F( f ) \circ f \approx_E 1_s \boxtimes 1_x = 1_n$ and $f \circ F( f ) \approx_E 1_t \boxtimes 1_y = 1_m$ by the transitivity of $\approx_E$.
    \end{itemize}
    Then by the principle of structural induction, $F( f ) \circ f \approx_E 1_n$ and $f \circ F( f ) \approx_E 1_m$ for each $f \in P( \Sigma )( n, m )$.
    Next, let $( f, g ) \in E$ where $f: n \to m$.
    Then $f \approx_E g$, $F( f ) \circ f \approx_E 1_n$, and $g \circ F( g ) \approx_E 1_m$ by the previous result.
    Then the following equation holds.
    \begin{equation*}
        F( f )
        =
        F( f ) \circ 1_m
        \approx_E
        F( f ) \circ g \circ F( g )
        \approx_E
        F( f ) \circ f \circ F( g )
        \approx_E
        1_n \circ F( g )
        =
        F( g )
    \end{equation*}
    Then $F( f ) \approx_E F( g )$ by the transitivity of $( \approx_E )$.
    Then $\pi_E^\text{op}( F( f ) ) = \pi_E^{\text{op}}( F( g ) )$.
    Then by the universal property of prop quotients, there exists a unique prop functor $H: P( \Sigma ) / E \to ( P( \Sigma ) / E )^\text{op}$ such that $H \circ \pi_E = \pi_E^\text{op} \circ F$.
    It remains to be shown that $H$ is a dagger functor.
    To this end, let $f \in P( \Sigma )( n, m )$.
    Then $f \circ F( f ) \approx_E 1_m$.
    Likewise, $F( f ) \circ F^\text{op}( F( f ) ) \approx 1_n$.
    This means that $f = f \circ 1_n \approx_E f \circ F( f ) \circ F^\text{op}( F( f ) ) \approx_E 1_m \circ F^\text{op}( F( f ) ) F( F^\text{op}( f ) )$.
    Then it follows that $\pi_E( f ) = \pi_E( F^\text{op}( F( f ) ) ) = H^\text{op}( \pi_E^\text{op}( F( f ) ) ) = H^\text{op}( H( \pi_E( f ) ) )$.
    Since $f$ was an arbitrary, then $\pi_E = H^\text{op} \circ H \circ \pi_E$.
    Then by uniqueness, $H^\text{op} \circ H = 1_{P( \Sigma ) / E}$.
    Then $H$ is a dagger functor.
    Finally, let $f: n \to n$.
    Then $F( f ) \circ f \approx_E 1_n$ and $f \circ F( f ) \approx_E 1_m$.
    This means that $H( \pi_E( f ) ) \circ \pi_E( f ) = 1_n$ and $\pi_E( f ) \circ H( \pi_E( f ) ) = 1_n$.
    Since $f$ was arbitrary and $\pi_E$ is surjective, then $H( f ) \circ f = 1_n$ and $f \circ H( f ) = 1_m$ for each $f \in P( \Sigma ) / E$.
    In other words, every morphism in $P( \Sigma ) / E$ is unitary.
\end{proof}

\subsection{Proof of Theorem~\ref{Thm:HQCInv}}
\begin{proof}
    First it must be shown that $g^\dagger \circ g \approx_E 1_{\dom(g)}$ and $g \circ g^\dagger \approx_E 1_{\cod(g)}$ for each $g \in \Sigma_{\CatHQC}$.
    Let $g \in \Sigma_{\CatHQC}$.
    There are three cases to consider.
    \begin{enumerate}
    \item Assume that $g \in \Sigma_{\mathbf{Prim}}$.
          Then $g^\dagger = g$ and $n = \dom( g ) = \cod( g )$.
          Then it suffices to show that $g \circ g \approx_E 1_n$.
          There are six cases to consider.
          \begin{enumerate}
          \item Assume that $g = \scalebox{0.65}{\circimg{tikz-cache/dagger__term1.pdf}}$.
                Then the following equation holds.
                \begin{equation*}
                    \scalebox{0.65}{\circimg{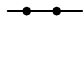}}
                    \overset{\HqcPowId}{\approx}
                    \scalebox{0.65}{\circimg{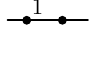}}
                    \overset{\HqcPowId}{\approx}
                    \scalebox{0.65}{\circimg{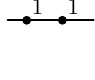}}
                    \overset{\HqcPowAdd}{\approx}
                    \scalebox{0.65}{\circimg{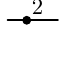}}
                    \overset{\HqcFreePhase}{\approx}
                    \scalebox{0.65}{\circimg{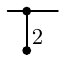}}
                    \overunderset{\HQCCtrlCong}{\HqcPOrd}{\approx}
                    \scalebox{0.65}{\circimg{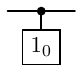}}
                    \overset{\HqcCtrlId}{\approx}
                    \scalebox{0.65}{\circimg{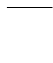}}
                \end{equation*}
                Then $g \circ g \approx_E 1_n$ by transitivity.
          \item Assume that $g = \scalebox{0.65}{\circimg{tikz-cache/dagger__term8.pdf}}$.
                Then the following equation holds.
                \begin{equation*}
                   \scalebox{0.65}{\circimg{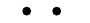}}
                    \overset{\HqcPowId}{\approx}
                    \scalebox{0.65}{\circimg{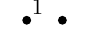}}
                    \overset{\HqcPowId}{\approx}
                    \scalebox{0.65}{\circimg{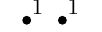}}
                    \overset{\HqcPowAdd}{\approx}
                    \scalebox{0.65}{\circimg{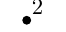}}
                    \overset{\HqcPOrd}{\approx}
                    1_0
                \end{equation*}
                Then $g \circ g \approx_E 1_n$ by transitivity.
          \item Assume that $g = \scalebox{0.65}{\circimg{tikz-cache/dagger__term5.pdf}}$.
                Then the following equation holds.
                \begin{align*}
                    \scalebox{0.65}{\circimg{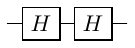}}
                    \;\;\;\overset{\HqcHDef}{\approx}\;\;\;&
                    \scalebox{0.65}{\circimg{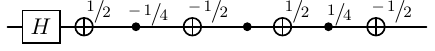}}
                    \\
                    \;\;\;\overset{\HqcHDef}{\approx}\;\;\;&
                    \scalebox{0.65}{\circimg{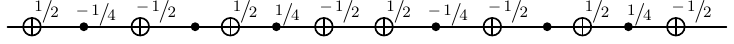}}
                    \\
                    \overset{\text{Lem(\ref{Lem:InvPow})}}{\approx}&
                    \scalebox{0.65}{\circimg{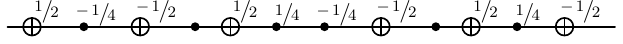}}
                    \\
                    \overset{\text{Lem(\ref{Lem:InvPow})}}{\approx}&
                    \scalebox{0.65}{\circimg{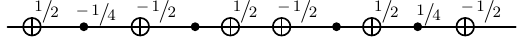}}
                    \\
                    \overset{\text{Lem(\ref{Lem:InvPow})}}{\approx}&
                    \scalebox{0.65}{\circimg{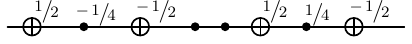}}
                    \\
                    \quad\overset{\text{(a)}}{\approx}\quad&
                    \scalebox{0.65}{\circimg{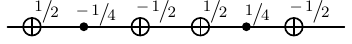}}
                    \\
                    \overset{\text{Lem(\ref{Lem:InvPow})}}{\approx}&
                    \scalebox{0.65}{\circimg{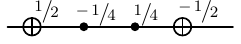}}
                    \\
                    \overset{\text{Lem(\ref{Lem:InvPow})}}{\approx}&
                    \scalebox{0.65}{\circimg{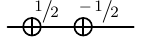}}
                    \\
                    \overset{\text{Lem(\ref{Lem:InvPow})}}{\approx}&
                    \scalebox{0.65}{\circimg{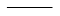}}
                \end{align*}
                Then $g \circ g \approx_E 1_n$ by transitivity.
          \item Assume that $g = \scalebox{0.65}{\circimg{tikz-cache/dagger__term3.pdf}}$.
                Then the following equation holds.
                \begin{align*}
                    \scalebox{0.62}{\circimg{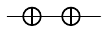}}
                    \overset{\HqcTargDef}{\approx}
                    \scalebox{0.62}{\circimg{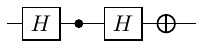}}
                    \overset{\HqcTargDef}{\approx}
                    \scalebox{0.62}{\circimg{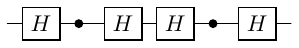}}
                    \overset{\text{(c)}}{\approx}
                    \scalebox{0.62}{\circimg{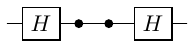}}
                    \overset{\text{(a)}}{\approx}
                    \scalebox{0.62}{\circimg{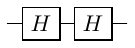}}
                    \overset{\text{(c)}}{\approx}
                    \scalebox{0.62}{\circimg{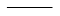}}
                \end{align*}
          \item Assume that $g = \scalebox{0.65}{\circimg{tikz-cache/dagger__term2.pdf}}$.
                Then the following equation holds.
                \begin{equation*}
                    \scalebox{0.65}{\circimg{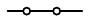}}
                    \overset{\HqcOPhaseDef}{\approx}
                    \scalebox{0.65}{\circimg{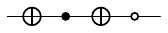}}
                    \overset{\HqcOPhaseDef}{\approx}
                    \scalebox{0.65}{\circimg{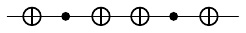}}
                    \overset{\text{(d)}}{\approx}
                    \scalebox{0.65}{\circimg{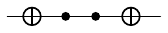}}
                    \overset{\text{(a)}}{\approx}
                    \scalebox{0.65}{\circimg{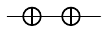}}
                    \overset{\text{(d)}}{\approx}
                    \scalebox{0.65}{\circimg{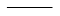}}
                \end{equation*}
                Then $g \circ g \approx_E 1_n$ by transitivity.
          \item Assume that $g = \scalebox{0.65}{\circimg{tikz-cache/dagger__term4.pdf}}$.
                Then the following equation holds.
                \begin{align*}
                    \scalebox{0.65}{\circimg{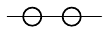}}
                    \overset{\HqcTargMDef}{\approx}
                    \scalebox{0.65}{\circimg{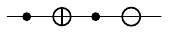}}
                    \overset{\HqcTargMDef}{\approx}
                    \scalebox{0.65}{\circimg{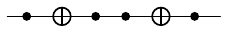}}
                    \overset{\text{(a)}}{\approx}
                    \scalebox{0.65}{\circimg{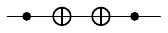}}
                    \overset{\text{(d)}}{\approx}
                    \scalebox{0.65}{\circimg{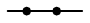}}
                    \overset{\text{(a)}}{\approx}
                    \scalebox{0.65}{\circimg{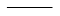}}
                \end{align*}
          \end{enumerate}
          In each case $g \circ g \approx_E 1_n$.
          Then  $g^\dagger \circ g \approx_E 1_{\dom(g)}$ and $g \circ g^\dagger \approx_E 1_{\cod(g)}$.
    \item Assume $g \in \Sigma_{\mathbf{Ctrl}}$.
          Then there exists $U \in \Mor(\CatHQC)$ and $V \in \Mor(\CatHQC)$ such that $g = U \odot V$.
          Then the following equations hold.
          \begin{align*}
              g \circ g^\dagger
              &=
              \scalebox{0.65}{\circimg{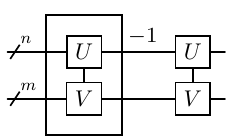}}
              \overset{\HqcPowId}{\approx}
              \scalebox{0.65}{\circimg{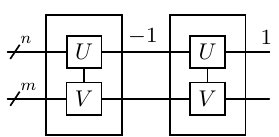}}
              \overset{\text{Lem(\ref{Lem:InvPow})}}{\approx}
              \scalebox{0.65}{\circimg{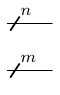}}
              \\
              g^\dagger \circ g
              &=
              \scalebox{0.65}{\circimg{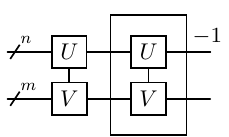}}
              \overset{\HqcPowId}{\approx}
              \scalebox{0.65}{\circimg{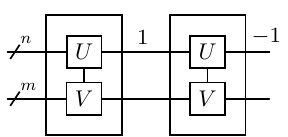}}
              \overset{\text{Lem(\ref{Lem:InvPow})}}{\approx}
              \scalebox{0.65}{\circimg{tikz-cache/proofs__ctrl_dag_3.pdf}}
          \end{align*}
          Then $g^\dagger \circ g \approx_E 1_{\dom(g)}$ and $g \circ g^\dagger \approx_E 1_{\cod(g)}$ by transitivity.
    \item Assume $g \in \Sigma_{\mathbf{Pow}}$.
          Then there exists $U \in \Mor( \CatHQC )$ and $\alpha \in \mathbb{R}$ such that $g = U \exparr \alpha$.
          It follows that $g^\dagger = U \exparr (-\alpha)$.
          Then $g^\dagger \circ g \approx_E 1_{\dom(g)}$ and $g \circ g^\dagger \approx_E 1_{\cod(g)}$ by \cref{Lem:InvPow}.
    \end{enumerate}
    In each case, $g^\dagger \circ g \approx_E 1_{\dom(g)}$ and $g \circ g^\dagger \approx_E 1_{\cod(g)}$.
    Since $g$ was an arbitrary generator, then $g^\dagger \circ g \approx_E 1_{\dom(g)}$ and $g \circ g^\dagger \approx_E 1_{\cod(g)}$ for each $g \in \Sigma_{\CatHQC}$.
    Then by \cref{Thm:PropInv}, there exists a unique functor $\overline{\dagger}: \CatHQC / E \to \CatHQC / E$ such that $\overline{\dagger} \circ \pi_E = \pi_E^{\text{op}} \circ \dagger$ with every morphism in $\CatHQC / E$ unitary with respect to $( \overline{\dagger} )$.
\end{proof}
\label{App:involution_proofs}

\subsection{Proof of Lemma~\ref{Lem:UnitaryCtrl}}
\begin{proof}
    Let $f \in \mathcal{C}( n, n )$.
    By assumption, $f$ and $C( f )$ are unitary.
    This means that $f^\dagger \circ f = 1_n$ and $C( f ) \circ C( f )^\dagger = 1_{n+1}$.
    Since $C$ is a functor, then $C( f^\dagger ) \circ C( f ) = C( f^\dagger \circ f ) = C( 1_n ) = 1_{n+1}$.
    Then $C( f^\dagger ) = C( f^\dagger) \circ C( f ) \circ C( f )^\dagger = C( f )^\dagger$.
    Since $f$ was arbitrary, then $C$ is a dagger functor.
\end{proof}

\subsection{Proof of Theorem~\ref{Thm:DefCtrl}}
\begin{proof}
    Assume that conditions (1) and (2) hold and define an assignment $F: P( \Sigma ) \to P( \Sigma ) / E$ such that $F_0( n ) = n + 1$ and $F( f : n \to n ) = \pi_E( \tau_n( f ) )$.
    Then $F$ is fully determined, and is therefore unique.
    It remains to be shown that $F$ is a functor.
    \begin{itemize}
    \item \textbf{Identities}.
          Let $n \in \mathbb{N}$.
          Since $\tau_n( 1_n ) \approx_E 1_{n+1}$ by condition (1), then the following holds.
          \begin{equation*}
              F( 1_n )
              =
              \pi_E( \tau_n( 1_n ) )
              =
              \pi_E( 1_{n+1} )
              =
              1_{n+1}
          \end{equation*}
          Since $n$ was arbitrary, then $F$ respects identities.
    \item \textbf{Composition}.
          Let $f: n \to n$ and $g: n \to n$.
          Since $\tau_n( g \circ f ) \approx_E \tau_n( g ) \circ \tau_n( f )$ by condition (2), then the following holds.
          \begin{equation*}
              F( g \circ f )
              =
              \pi_E( \tau_n( g \circ f ) )
              =
              \pi_E( \tau_n( g ) \circ \tau_n( f ) )
              =
              \pi_E( \tau_n( g ) ) \circ \pi_E( \tau_n( f ) )
              =
              F( g ) \circ F( f )
          \end{equation*}
          Since $f$ and $g$ were arbitrary, then $F$ respects composition.
    \end{itemize}
    Since $F$ respects identities and composition, then $F$ is a functor.
    Next, assume that condition (3) holds.
    It must be shown that there exists a unique functor $H: P( \Sigma ) / E \to P( \Sigma ) / E$ such that $H \circ \pi_E = F$.
    Let $( f, g ) \in E$.
    Since $\tau_E( f ) \approx_E \tau_E( g )$, then $F( f ) = \pi_E( \tau_n( f ) ) = \pi_E( \tau_n( g ) ) = F( g )$.
    Since $( f, g )$ was arbitrary, then there exists a unique functor $H: P( \Sigma ) / E \to P( \Sigma ) / E$ such that $H \circ \pi_E = F$.
    Finally, assume that conditions (4) through to (6) hold.
    It remains to be shown that $H$ is a control functor.
    \begin{enumerate}
    \item Let $f \in ( P( \Sigma ) / E )( n, n )$.
          Then there exists a $g \in P( \Sigma )( n, n )$ such that $\pi_E( g ) = f$.
          Then $H( f \boxtimes 1 ) = H( \pi_E( g ) \boxtimes 1 ) = H( \pi_E( g \boxtimes 1 ) ) = F( g \boxtimes 1)$.
          Since $\tau_{n+1}( g \boxtimes 1 ) \approx_E \tau_n( g ) \boxtimes 1$ by condition (4), then the following holds.
          \begin{equation*}
            H( f \boxtimes 1 )
            =
            F( g \boxtimes 1 )
            =
            \pi_E( \tau_n( g \boxtimes 1 ) )
            =
            \pi_E( \tau_n( g ) \boxtimes 1  )
            =
            \pi_E( \tau_n( g ) ) \boxtimes 1
            =
            F( g ) \boxtimes 1
            =
            H( f ) \boxtimes 1
          \end{equation*}
          Since $f$ was arbitrary, then $H( f \boxtimes 1 ) = H( f ) \boxtimes 1$ for each $f \in ( P( \Sigma ) / E )( n, n )$.
    \item Let $f \in ( P( \Sigma ) / E )( n, n )$.
          Then there exists a $g \in P( \Sigma )( n, n )$ such that $\pi_E( g ) = f$.
          Then the following equation holds.
          \begin{equation*}
              H( H( f ) )
              =
              H( H( \pi_E( g ) ) )
              =
              H( F( g ) )
              =
              H( \pi_E( \tau_n( g ) ) )
              =
              F( \tau_n( g ) )
              =
              \pi_E( \tau_{n+1}( \tau_n( g ) ) )
          \end{equation*}
          Since $\tau_{n+1}( \tau_n( g ) ) \approx_E ( \sigma \boxtimes 1_n ) \circ \tau_{n+1}( \tau_n( g ) ) \circ ( \sigma \boxtimes 1_n )$ by condition (5), then the following equation holds where $\sigma = \sigma \boxtimes 1_n$ and $\overline{\sigma} = \pi_E( \sigma )$.
          \begin{equation*}
              \pi_E( \tau_{n+1}( \tau_n( g ) ) )
              =
              \pi_E( \sigma \circ \tau_{n+1}( \tau_n( g ) ) \circ \sigma )
              \\
              =
              \overline{\sigma} \circ \pi_E( \tau_{n+1}( \tau_n( g ) ) ) \circ \overline{\sigma}
          \end{equation*}
          In conclusion, the following equation holds.
          \begin{align*}
              H( H( f ) )
              =
              \pi_E( \tau_{n+1}( \tau_n( g ) ) )
              =
              \overline{\sigma} \circ \pi_E( \tau_{n+1}( \tau_n( g ) ) ) \circ \overline{\sigma}
              =
              \overline{\sigma} \circ H( H( f ) ) \circ \overline{\sigma}
          \end{align*}
          Since $f$ was arbitrary, then $H( H( f ) ) = \overline{\sigma} \circ H( H( f ) ) \circ \overline{\sigma}$ for each $f \in ( P( \Sigma ) / E )( n, n )$.
    \item Let $f \in ( P( \Sigma ) / E )( n + 2, n + 2 )$ and $0 \le k \le n$.
          Then there exists a $g \in P( \Sigma )( n + 2, n + 2 )$ such that $\pi_E( g ) = f$.
          Then $H( f ) = H( \pi_E( g ) ) = F( g )$ the following equation holds.
          \begin{equation*}
              H( \gamma_{n,k} \circ f \circ \gamma_{n,k} )
              =
              H( \gamma_{n,k} \circ \pi_E( g ) \circ \gamma_{n,k} )
              =
              H( \pi_E( \gamma_{n,k} \circ g \circ \gamma_{n,k} ) )
              =
              F( \gamma_{n,k} \circ g \circ \gamma_{n,k} )
          \end{equation*}
          Since $\tau_{n+2}( \gamma_{n,k} \circ g \circ \gamma_{n,k} ) = ( 1 \boxtimes \gamma_{n,k} ) \circ \tau_{n+2}( g ) \circ ( 1 \boxtimes \gamma_{n,k} )$, then the following equation holds.
          \begin{align*}
              F( \gamma_{n,k} \circ g \circ \gamma_{n,k} )
              &=
              \pi_E( \tau_{n+2}( \gamma_{n,k} \circ g \circ \gamma_{n,k} ) )
              \\
              &=
              \pi_E( ( 1 \boxtimes \gamma_{n,k} ) \circ \tau_{n+2}( g ) \circ ( 1 \boxtimes \gamma_{n,k} ) )
              \\
              &= ( 1 \boxtimes \gamma_{n,k} ) \circ \pi_E( \tau_{n+2}( g ) ) \circ ( 1 \boxtimes \gamma_{n,k} )
              \\
              &= ( 1 \boxtimes \gamma_{n,k} ) \circ F( g ) \circ ( 1 \boxtimes \gamma_{n,k} )
              \\
              &= ( 1 \boxtimes \gamma_{n,k} ) \circ H( f ) \circ ( 1 \boxtimes \gamma_{n,k} )
          \end{align*}
          Since $f$ was an arbitrary morphism, then $H( \gamma_{n,k} \circ f \circ \gamma_{n,k} ) = ( 1 \boxtimes \gamma_{n,k} ) \circ H( f ) \circ ( 1 \boxtimes \gamma_{n,k} )$ for each $f \in ( P( \Sigma ) / E )( n + 2, n + 2 )$ and $0 \le k \le n$.
    \end{enumerate}
    Then $H$ is a control functor.
\end{proof}

\subsection{Proof of Theorem~\ref{Thm:HQCCtrl}}
\begin{proof}
    Since all generators in $\Sigma_{\CatHQC}$ are endomorphic, then \cref{Thm:DefCtrl} is applicable to $\CatHQC$.
    Let $\{ \tau_n: \CatHQC( n, n ) \to \CatHQC( n + 1, n + 1 ) \}_{n \in \mathbb{N}}$ be the family of functions defined by the equation $\tau_n( U ) = \scalebox{0.65}{\circimg{tikz-cache/dagger__term1.pdf}} \odot U$.
    The six conditions hold as follows.
    \begin{enumerate}
    \item If $n \in \mathbb{N}$, $\tau_n( 1_n ) \approx_E 1 \boxtimes 1_n = 1_{n+1}$ by \HqcCtrlId.
    \item If $U: n \to n$ and $V: n \to n$, then $\tau_n( V \circ U ) = \tau_n( V ) \circ \tau_n( U )$ by \HqcCtrlComp.
    \item If $U: n \to n$ and $( U, V ) \in E$, then $\tau_n( U ) \approx_E \tau_n( V )$ by \HQCCtrlCong.
    \item Let $U: n \to n$.
          Then the following derivation holds. 
          \begin{align*}
              \scalebox{0.65}{\circimg{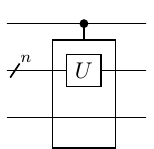}}
              \overunderset{\HQCCtrlCong}{\HqcCtrlPair}{\approx}
              \scalebox{0.65}{\circimg{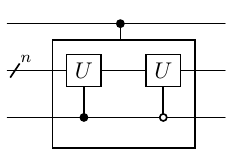}}
              \overset{\HqcCtrlComp}{\approx}
              \scalebox{0.65}{\circimg{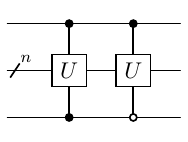}}
              \overset{\HqcCtrlPair}{\approx}
              \scalebox{0.65}{\circimg{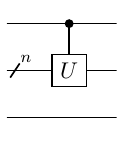}}
          \end{align*}
          Then $\tau_{n+1}( U \boxtimes 1 ) \approx_E \tau_n( U ) \boxtimes 1$.
    \item Let $U: n \to n$. Then the following implication holds.
          Since $\sigma^\dagger = \sigma$, then the following derivation also holds. 
          \begin{align*}
            \scalebox{0.63}{\circimg{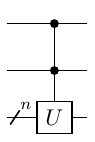}}
            \overset{\HqcCtrlSwap}{\approx}&
            \scalebox{0.63}{\circimg{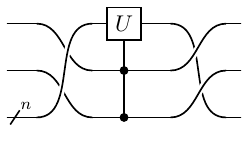}}
            \overunderset{\HQCCtrlCong}{\HqcCtrlSwap}{\approx}
            \scalebox{0.63}{\circimg{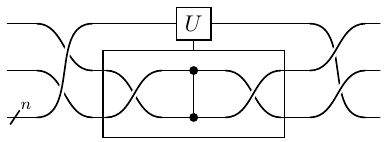}}
            \overset{\HqcCtrlDag}{\approx}
            \scalebox{0.63}{\circimg{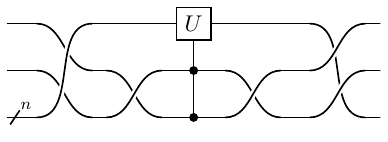}}
            \\
            =\;\;\;&
            \scalebox{0.63}{\circimg{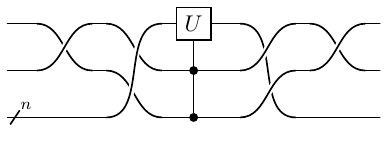}}
            \overset{\HqcCtrlSwap}{\approx}
            \scalebox{0.63}{\circimg{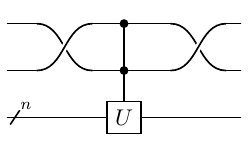}}
          \end{align*}
          Then $\tau_{n+1}( \tau_n( U ) ) \approx_E ( \sigma \boxtimes 1_n ) \circ \tau_{n+1}( \tau_n( U ) ) \circ ( \sigma \boxtimes 1_n )$.
    \item Let $U: n + 2 \to n + 2$ and $0 \le k \le n$.
          Since $1_k^\dagger = 1_k$, $1_{n-k}^\dagger = 1_{n-k}$, and $\sigma^\dagger = \sigma$, then it follows that $\gamma_{n,k}^\dagger = ( 1_k \boxtimes \sigma \boxtimes 1_{n-k} )^\dagger = 1_k^\dagger \boxtimes \sigma^\dagger \boxtimes 1_{n-k}^\dagger = \gamma_{n,k}$.
          Since $\gamma_{n,k}^\dagger = \gamma_{n,k}$, then it follows by \HqcCtrlDag that $\tau_{n+2}( \gamma_{n,k} \circ U \circ \gamma_{n,k} ) \approx_E (1 \boxtimes \gamma_{n,k} ) \circ \tau_{n+2}( U ) \circ ( 1 \boxtimes \gamma_{n,k} )$.
    \end{enumerate}
    Since conditions (1) and (2) hold, then by \cref{Thm:DefCtrl} there exists a unique ordinary functor $F: \CatHQC \to \CatHQC / E$ such that $F_0( n ) = n + 1$ and $F( U: n \to n ) = \pi_E( \tau_n( U ) )$.
    Since condition (3) also holds, then by \cref{Thm:DefCtrl} exists a unique ordinary functor $C: \CatHQC / E \to \CatHQC / E$ such that $H \circ \pi_E = F$.
    Then $C( \pi_E( U ) ) = F( U ) = \pi_E( \scalebox{0.65}{\circimg{tikz-cache/dagger__term1.pdf}} \odot U )$ for each $U \in \CatHQC( n, n )$.
    Since conditions (4) through to (6) hold, then $C$ is a control functor.
    Since every morphism in $\CatHQC / E$ is unitary with respect to $( \overline{\dagger} )$ by \cref{Thm:HQCInv}, then $C$ is a dagger control functor with respect to $( \overline{\dagger} )$ by \cref{Lem:UnitaryCtrl}.
    
    It remains to be shown that $C$ is a conjugated control functor.
    Let $X \in ( \CatHQC / E )( n, n )$ and $Y \in ( \CatHQC / E )( n, n )$.
    Then there exists $U \in \CatHQC( n, n )$ and $V \in \CatHQC( n, n )$ such that $\pi_E( U ) = X$ and $\pi_E( V ) = Y$.
    Then $( 1 \boxtimes U^\dagger ) \circ ( \scalebox{0.65}{\circimg{tikz-cache/dagger__term1.pdf}} \odot V ) \circ ( 1 \boxtimes U ) \approx_E ( \scalebox{0.65}{\circimg{tikz-cache/dagger__term1.pdf}} \odot ( U^\dagger \circ V \circ U ) )$ by $\HqcCtrlDag$.
    Then the following equation holds.
    \begin{align*}
        \left( 1 \boxtimes X^{\overline{\dagger}} \right) \circ C( \pi_E( V ) ) \circ ( 1 \boxtimes X )
        &=
        \left( 1 \boxtimes \pi_E( U )^{\overline{\dagger}} \right) \circ C( \pi_E( V ) ) \circ ( 1 \boxtimes \pi_E( U ) )
        \\
        &=
        \left( 1 \boxtimes \pi_E\left( U^{\dagger} \right) \right) \circ C( \pi_E( V ) ) \circ ( 1 \boxtimes \pi_E( U ) )
        &\text{(by \cref{Thm:HQCInv})}
        \\
        &=
        \pi_E\left( \scalebox{0.65}{\circimg{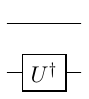}} \right) \circ C( \pi_E( V ) ) \circ \pi_E\left( \scalebox{0.65}{\circimg{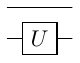}} \right)
        \\
        &=
        \pi_E\left( \scalebox{0.65}{\circimg{tikz-cache/proofs__ctrl_conj_eqn_term2.pdf}} \right) \circ \pi_E\left( \scalebox{0.65}{\circimg{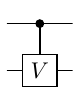}} \right) \circ \pi_E\left( \scalebox{0.65}{\circimg{tikz-cache/proofs__ctrl_conj_eqn_term1.pdf}} \right)
        &\text{(by \cref{Thm:HQCCtrl})}
        \\
        &=
        \pi_E\left( \scalebox{0.65}{\circimg{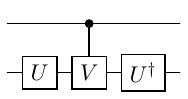}} \right)
        \\
        &=
        \pi_E\left( \scalebox{0.65}{\circimg{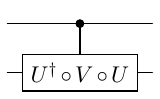}} \right)
        &\text{(by $\approx_E$)}
        \\
        &=
        C \left( \pi_E\left( \scalebox{0.65}{\circimg{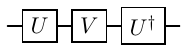}} \right) \right)
        &\text{(by \cref{Thm:HQCCtrl})}
        \\
        &=
        C\left( \pi_E\left( U^\dagger \right) \circ \pi_E( V ) \circ \pi_E( U ) \right)
        \\
        &=
        C\left( \pi_E( U )^{\overline{\dagger}} \circ \pi_E( V ) \circ \pi_E( U ) \right)
        \\
        &=
        C\left( X^{\overline{\dagger}} \circ Y \circ X \right)
        &\text{(by \cref{Thm:HQCInv})}
    \end{align*}
    Since $X$ and $Y$ were arbitrary, then $C( - )$ is conjugated.
\end{proof}

\section{Proofs for Section~\ref{Sect:Completeness}}
%

This section contains all proofs for~\cref{Sect:Completeness}, except for those which appear in \cref{Sect:Completeness:Full}.

\subsection{Proof of Theorem~\ref{Thm:Completeness1}}
\begin{proof}
    In assumptions (1) through to (3), local properties placed on $\encoder( - )$ and $\decoder( - )$ in terms of generators and relations.
    These local properties lift to global properties as follows.
    \begin{enumerate}
    \item \textbf{Derivation (1)}.
          Assumption (1) states that $\sem{\encoder(x)}_\Gamma = \sem{x}_\Sigma$ for all $x \in \Sigma$.
          This means that $\sem{-}_\Gamma \circ \encoder \circ i = \sem{-}_\Sigma \circ i$.
          Then $\sem{-}_\Gamma \circ \encoder = \sem{-}_\Sigma$ by the universal property of free prop categories.
    \item \textbf{Derivation (2)}.
          Let $( X, Y ) \in Q$.
          By assumption (3), there exists an $X^* \in \decoder( X )$ and $Y^* \in \decoder( Y )$ such that $X^* \approx_E Y^*$.
          Since $X^* \in \decoder( X )$, then $\pi_E( X^* ) = \decoder( X )$.
          Since $Y^* \in \decoder( Y )$, then $\pi_E( Y^* ) = \decoder( Y )$.
          Since $X^* \approx_E Y^*$, then $\pi_E( X^* ) = \pi_E( Y^* )$.
          Then in conclusion, $\decoder( X ) = \pi_E( X^* ) = \pi_E( Y^* ) = \decoder( Y )$.
          Since $( X, Y )$ was an arbitrary relation in $Q$, then there exists a unique controlled prop functor $\decoder_Q: P_C( \Gamma ) / Q \to P( \Sigma ) / E$ such that $\decoder_Q \circ \pi_Q = \decoder$.
    \item \textbf{Derivation (3)}.
          Let $x \in \Sigma$.
          Then by assumption (2), there exists a $f \in \decoder( \encoder( g ) )$ such that $x \approx_E f$.
          Since $f \in \decoder( \encoder( x ) )$, then $\pi_E( f ) = \decoder( \encoder( x ) )$.
          Since $x \approx_E f$, then $\pi_E( x ) = \pi_E( f )$.
          Then $\pi_E( x ) = \pi_E( f ) = \decoder( \encoder( x ) )$.
          Since $x$ was as arbitrary generator, then $\pi_E \circ i = \decoder \circ \encoder \circ i$.
          Then $\pi_E = \decoder \circ \encoder$ by the universal property of prop categories.
    \end{enumerate}
     Next, it will be shown that $( \Sigma, E )$ is complete with respect to $\sem{-}_\Sigma$.
     Let $f, g \in P( \Sigma )$ such that $\sem{f}_\Sigma = \sem{g}_\Sigma$.
     Then $\sem{\encoder(f)}_\Gamma = \sem{f}_\Sigma = \sem{g}_\Sigma = \sem{\encoder(g)}_\Gamma$ by derivation (1).
     Since $( \Gamma, Q )$ is complete with respect to $\sem{-}_\Gamma$, then $\encoder( f ) \approx_Q \encoder( g )$.
     This means that $\pi_Q( \encoder( f ) ) = \pi_Q( \encoder( g ) )$, which implies that $\decoder_Q( \pi_Q( \encoder( f ) ) ) = \decoder_Q( \pi_Q( \encoder( g ) ) )$.
     Then $\decoder( \encoder( f ) ) = \decoder( \encoder( g ) )$ by derivation (2).
     Since $\pi_E = \decoder \circ \encoder$ by derivation (3), then $\pi_E( f ) = \pi_E( g )$.
     Then $f \approx_E g$.
     Since $f$ and $g$ were arbitrary, then $( \Sigma, E )$ is complete with respect to $\sem{-}_\Sigma$.
\end{proof}

\subsection{Proof of Theorem~\ref{Thm:Completeness2}}
\begin{proof}
    First it must be shown that for each $f \in P( \Gamma )$, there exists a $g \in P( \Sigma )$ such that $f \approx_E g$.
    The proof follows by structural induction on $P( \Gamma )$.
    \begin{itemize}
    \item \textbf{Identity}.
          Assume that $f = 1_n$.
          Since $P( \Sigma )$ is a subcategory of $P( \Gamma )$, then $f \in P( \Sigma )$.
          Since $( \approx_E )$ is reflexive, then $f \approx_E f$.
    \item \textbf{Symmetries}.
          Assume that $f = \sigma$.
          Since $P( \Sigma )$ is a subcategory of $P( \Gamma )$, then $f \in P( \Sigma )$.
          Since $( \approx_E )$ is reflexive, then $f \approx_E f$.
    \item \textbf{Generators}.
          Assume that $f = x$ for some $x \in \Gamma$.
          Then by assumption, there exists some $g \in P( \Sigma )$ such that $f \approx_E g$.
    \item \textbf{Sequential Composition}.
          Assume that $f = b \circ a$ for $a: n \to s$ and $b: s \to m$.
          Then by the inductive hypothesis, there exists some $g \in P( \Sigma )( n, s )$ and $h \in P( \Sigma )( s, m )$ such that $a \approx_E g$ and $b \approx_E h$.
          Then $f \approx_E h \circ a \approx_E h \circ g$.
          Then $f \approx_E h \circ g$ by the transitivity of $( \approx_E )$.
          Moreover, $h \circ g \in P( \Sigma )( n, m )$.
    \item \textbf{Parallel Composition}.
          Assume that $f = a \boxtimes b$ for $a: n \to m$ and $b: s \to t$.
          Then by the inductive hypothesis, there exists some $g \in P( \Sigma )( n, m )$ and $h \in P( \Sigma )( s, t )$ such that $a \approx_E g$ and $b \approx_E h$.
          Then $f \approx_E g \otimes b \approx_E g \otimes h$.
          Then $f \approx_E g \otimes h$ by the transitivity of $( \approx_E )$.
          Moreover, $g \otimes h \in P( \Sigma )( n + s, m + t )$.
    \end{itemize}
    Then for all $f \in P( \Gamma )$, there exists a $g \in P( \Sigma )$ such that $f \approx_E g$.
    It remains to be shown that $( \Gamma, E )$ is complete with respect to $\sem{-}$.
    Let $f \in P( \Gamma )( n, m )$ and $g \in P( \Gamma )( n, m )$ with $\sem{f} = \sem{g}$.
    By the first result, there exists some $h \in P( \Sigma )( n, m )$ and $k \in P( \Gamma )( n, m )$ such that $f \approx_E h$ and $g \approx_E k$.
    Since $E$ is sound with respect to $E$, then $\sem{f} =\sem{h}$ and $\sem{g} = \sem{k}$.
    Then $\sem{h} = \sem{f} = \sem{g} = \sem{k}$.
    Since $( \Sigma, E )$ is complete with respect to $\sem{-}$, then $h \approx_E k$.
    Since $( \approx_E )$ is transitive, then $f \approx_E g$.
    Since $f$ and $g$ were arbitrary, then $( \Gamma, E )$ is complete with respect to $\sem{-}$.
\end{proof}

\subsection{Proof of Lemma~\ref{Lem:CRestrict}}
\begin{proof}
    Let $U \in \Mor( \Core / E )$.
    Then there exists some $V \in \Mor( \Core )$ such that $\pi_E( V ) = U$.
    Then there exists some $n \in \mathbb{N}$ such that $V \in \Mor( \mathcal{D}_n )$.
    Then $\scalebox{0.65}{\circimg{tikz-cache/dagger__term1.pdf}} \odot V \in \mathcal{G}_{n+1}$ by definition of the gate set.
    It follows that $C( U ) = C( \pi( V ) ) = \pi( \scalebox{0.65}{\circimg{tikz-cache/dagger__term1.pdf}} \odot V ) \in \Mor( \Core / E )$.
    Since $U$ was arbitrary, then $C( U ) \in \Mor( \Core / E )$ for all $U \in \Mor( \Core / E )$.
    In conclusion, $C: \CatHQC / E \to \CatHQC / E$ restricts to $\Core / E$.
\end{proof}

\subsection{Proof of Lemma~\ref{Lem:CoreDag}}
\begin{proof}
    For each $n \in \mathbb{N}$, let $D_n: \Core[n] \to \Core[n]$ denote the unique prop functor such that $D_n \circ i = \ddagger \circ i$.
    The proof follows by induction on $n$.
    \begin{itemize}
    \item \textbf{Base Case}.
          Let $g \in \Sigma_{\Core}^0$.
          Then by definition, $D_0( g ) = g^\dagger$.
          Since $g$ was arbitrary, then $D_0 \circ i = \dagger \circ i$.
          Then by uniqueness, $D_0( U ) = U^\dagger$ for all $U \in \Mor( \Core[0] )$.
          Then by reflexivity, $D_0( U ) \approx_E U^\dagger$ for all $U \in \Mor( \Core[n] )$.
    \item \textbf{Inductive Hypothesis}.
          For some $n \in \mathbb{N}$, if $U \in \Mor( \Core[n] ) $, then $D_0( U ) \approx_E U^\dagger$.
    \item \textbf{Inductive Step}.
          Assume that for an $n \in \mathbb{N}$, if $U \in \Mor( \Core[n] )$, then $D_n( U ) \approx_E U^\dagger$.
          Let $g \in \Sigma_{\Core}^{n+1}$.
          If $g \in \Sigma_{\Core}^n$ as well, then $D_{n+1}( U ) = D_n( U ) \approx_E g$ by the inductive hypothesis.
          Assume instead that $g \not \in \Sigma_{\Core}^n$.
          Then there exists some $U \in \Mor( \Core[n] )$ such that $g = \scalebox{0.65}{\circimg{tikz-cache/dagger__term1.pdf}} \odot D_n( U )$.
          Then by the inductive hypothesis, $D_n( U ) \approx_E U^\dagger$.
          Then the following derivation holds.
          \begin{align*}
              \scalebox{0.65}{\circimg{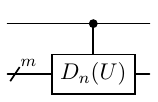}}
              \;\;\;\overunderset{\HQCCtrlCong}{I.H}{\approx}\;\;\;&
              \scalebox{0.65}{\circimg{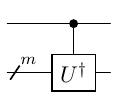}}
              \overset{\text{Thm(\ref{Thm:HQCInv})}}{\approx}
              \scalebox{0.65}{\circimg{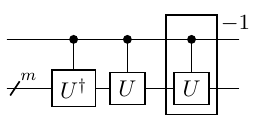}}
              \overunderset{\HqcCtrlComp}{I.H}{\approx}\;\;\;
              \scalebox{0.65}{\circimg{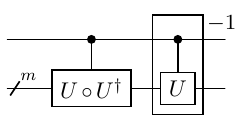}}
              \\
              \overunderset{\HQCCtrlCong}{\text{Thm(\ref{Thm:HQCInv})}}{\approx}&
              \scalebox{0.65}{\circimg{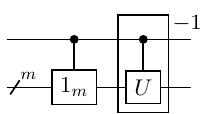}}
              \overset{\HqcCtrlId}{\approx}
              \scalebox{0.65}{\circimg{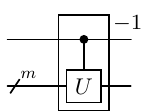}}
          \end{align*}
          In either case, $D_{n+1}( g ) \approx_E g^\dagger$.
          Since $g$ was arbitrary, then $D_{n+1} \circ i = \dagger \circ i$.
          Then by uniqueness, $D_{n+1}( U ) \approx_E U^\dagger$ for all $U \in \Mor( \Core[n] )$.
          Then the inductive step holds.
    \end{itemize}
    Then by the principle of induction, for each $n \in \mathbb{N}$, if $U \in \Mor( \Core[n] )$, then $D_n( U ) \approx_E U^\dagger$.
    Then let $U \in \Mor( \Core )$.
    There there exists some $n \in \mathbb{N}$ such that $U \in \Mor( \Core[n] )$.
    Then $U^\ddagger = D_n( U ) \approx_E U^\dagger$.
    Since $U$ was arbitrary, then $U^\ddagger \approx_E U^\dagger$ for all $U \in \Mor( \Core )$.
\end{proof}

\subsection{Proof of Lemma~\ref{Lem:Encoder}}
\begin{proof}
    The proof follows by induction.
    \begin{itemize}
    \item \textbf{Base Case}.
          Let $g \in \Sigma_{\Core}^0$. There are six cases to consider.
          \begin{enumerate}
              \item Let $g = \scalebox{0.65}{\circimg{tikz-cache/encoder__eqn3_lhs.pdf}}$.
                    Then
                    $
                        \sem{\encoder(g)}_C
                        =
                        \sem*{ \scalebox{0.65}{\circimg{tikz-cache/encoder__eqn3_rhs.pdf}} }_C
                        =
                        e^{i\alpha\pi}
                        =
                        \exp( \alpha \Log( -1 ) )
                        =
                        \sem{g}_H
                    $.
              \item Let $g = \scalebox{0.65}{\circimg{tikz-cache/encoder__eqn6_lhs.pdf}}$.
                    Then
                    $
                        \sem{\encoder(g)}_C
                        =
                        \sem*{ \scalebox{0.65}{\circimg{tikz-cache/encoder__eqn6_lhs.pdf}} }_C
                        =
                        \tfrac{1}{\sqrt{2}}
                        \begin{bsmallmatrix}
                            1 & 1 \\
                            1 & -1
                        \end{bsmallmatrix}
                        =
                        \sem*{g}_H
                    $.
              \item Let $g = \scalebox{0.65}{\circimg{tikz-cache/encoder__eqn1_lhs.pdf}}$.
                    Recall that $\Log( Z ) = i\pi \ket{1}\bra{1}$.
                    Then the following equation holds.
                    \begin{equation*}
                        \sem{\encoder(g)}_C
                        =
                        \sem*{ \scalebox{0.65}{\circimg{tikz-cache/encoder__eqn1_rhs.pdf}} }_C
                        =
                        \ket{0} \bra{0} \otimes \mathbb{I}_1 + \ket{1} \bra{1} \otimes e^{i\alpha\pi}
                        =
                        \exp( i\alpha\pi \ket{1} \bra{1} )
                        =
                        \sem{g}_H
                    \end{equation*}
              \item Let $g = \scalebox{0.65}{\circimg{tikz-cache/encoder__eqn4_lhs.pdf}}$.
                    Recall that $HZH = X$.
                    Then the following equation holds
                    \begin{align*}
                        \sem{\encoder(g)}_C
                        =
                        \sem*{ \scalebox{0.65}{\circimg{tikz-cache/encoder__eqn4_rhs.pdf}} }_C
                        &=
                        H
                        \exp( \alpha \Log( Z ) )
                        H
                        &
                        ( \text{by case (3)} )
                        \\
                        &=
                        \exp( \alpha H\Log( Z )H )
                        &
                        ( \text{by~\cref{Eqn:UnitaryConj}} )
                        \\
                        &=
                        \exp( \alpha \Log( HZH ) )
                        &
                        ( \text{by~\cref{Eqn:UnitaryConj}} )
                        \\
                        &=
                        \exp( \alpha \Log( X ) )
                        =
                        \sem{g}_H
              \end{align*}
              \item Let $g = \scalebox{0.65}{\circimg{tikz-cache/encoder__eqn2_lhs.pdf}}$.
                    Recall that $XZX = -Z$.
                    Then the following equation holds.
                    \begin{align*}
                        \sem*{\encoder(g)}_C
                        &=
                        \sem*{ \scalebox{0.65}{\circimg{tikz-cache/encoder__eqn2_rhs.pdf}} }_C
                        \\
                        &=
                        \exp( \Log(X) ) \exp( \alpha \Log(Z) ) \exp( \Log(X) )
                        &
                        ( \text{by cases (3) and (4)} )
                        \\
                        &=
                        X \exp( \alpha \Log(Z) ) X
                        \\
                        &=
                        \exp( \alpha X \Log(Z) X )
                        &
                        ( \text{by~\cref{Eqn:UnitaryConj}} )
                        \\
                        &=
                        \exp( \alpha \Log( XZX ) )
                        &
                        ( \text{by~\cref{Eqn:UnitaryConj}} )
                        \\
                        &=
                        \exp( \alpha \Log( -Z ) )
                        =
                        \sem{g}_H
                    \end{align*}
              \item Let $g = \scalebox{0.65}{\circimg{tikz-cache/encoder__eqn5_lhs.pdf}}$.
                    Recall that $XZX = -Z$.
                    Then the following equation holds.
                    \begin{align*}
                        \sem*{\encoder(g)}_C
                        &=
                        \sem*{ \scalebox{0.65}{\circimg{tikz-cache/encoder__eqn5_rhs.pdf}} }_C
                        \\
                        &=
                        \exp( \Log(Z) ) \exp( \alpha \Log(X) ) \exp( \Log(Z) )
                        &
                        ( \text{by cases (3) and (4)} )
                        \\
                        &=
                        Z \exp( \alpha \Log(X) ) Z
                        \\
                        &=
                        \exp( \alpha Z \Log(X) Z )
                        &
                        ( \text{by~\cref{Eqn:UnitaryConj}} )
                        \\
                        &=
                        \exp( \alpha \Log( ZXZ ) )
                        &
                        ( \text{by~\cref{Eqn:UnitaryConj}} )
                        \\
                        &=
                        \exp( \alpha \Log( -X ) )
                        =
                        \sem{g}_H
                    \end{align*}
          \end{enumerate}
          Since $g \in \Sigma_{\Core}^0$ was arbitrary, then $\sem{\encoder(g)}_C = \sem{g}_H$ for all $g \in \Sigma_{\Core}^0$.
    \item \textbf{Inductive Hypothesis}.
          For some $n \in \mathbb{N}$, if $g \in \Sigma_{\Core}^n$, then $\sem{\encoder(g)}_C = \sem{g}_H$.
    \item \textbf{Inductive Step}.
          Assume that the inductive hypothesis holds for some $n \in \mathbb{N}$ and let $g \in \Sigma_{\Core}^{n+1}$.
          If $g \in \Sigma_{\Core}^n$ as well, then $\sem{\encoder(g)}_C = \sem{g}_H$ by the inductive hypothesis.
          Assume instead that $g \in \Sigma_{\Core}^{n+1} \setminus \Sigma_{\Core}^n$.
          Then there exists some $U \in \Mor( \Core[n] )$ such that $g = \scalebox{0.65}{\circimg{tikz-cache/dagger__term1.pdf}} \odot U$.
          Then by the inductive hypothesis, since $\sem{\encoder(-)}_C$ and $\sem{-}_H$ agree on all generators, then $\sem{\encoder(U)}_C = \sem{U}_H$.
          Then by \cref{Thm:GenCtrlNF}, the following holds.
          \begin{equation*}
              \sem{g}_H
              =
              Z \odot \sem{U}_H
              =
              Z \odot \sem{\encoder(U)}_C
              =
              \ket{0} \bra{0} \otimes \mathbb{I}_{2^n} + \ket{1} \bra{1} \otimes \sem{\encoder(U)}_C
              =
              \sem{\encoder(g)}_C
          \end{equation*}
          In either case, $\sem{g}_H = \sem{\encoder(g)}_C$.
          Then the inductive step holds.
    \end{itemize}
    Then by the principle of induction, $\sem{\encoder(g)}_H = \sem{g}_C$ for each $g \in \Sigma_{\Core}$.
\end{proof}

\subsection{Proof of Lemma~\ref{Lemma:CNot}}
\label{App:CNOT}
\begin{proof}
    By \cref{Thm:HQCCtrl}, the following theorem holds.
    \begin{align*}
        \decoder\left( \scalebox{0.65}{\circimg{tikz-cache/cnot__class.pdf}} \right)
        &=
        ( 1 \boxtimes \pi_E( H ) )
        \circ
        C( C( \pi_E( \scalebox{0.65}{\circimg{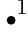}} ) ) )
        \circ
        ( 1 \boxtimes \pi_E( H ) )
        \\
        &=
        \pi_E\left( \scalebox{0.65}{\circimg{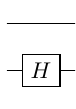}} \right)
        \circ
        C( C( \pi_E( \scalebox{0.65}{\circimg{tikz-cache/cnot__phase.pdf}} ) ) )
        \circ
        \pi_E\left( \scalebox{0.65}{\circimg{tikz-cache/cnot__had.pdf}} \right)
        \\
        &=
        \pi_E\left( \scalebox{0.65}{\circimg{tikz-cache/cnot__had.pdf}} \right)
        \circ
        C\left( \pi_E\left( \scalebox{0.65}{\circimg{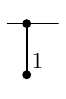}} \right) \right)
        \circ
        \pi_E\left( \scalebox{0.65}{\circimg{tikz-cache/cnot__had.pdf}} \right)
        \\
        &=
        \pi_E\left( \scalebox{0.65}{\circimg{tikz-cache/cnot__had.pdf}} \right)
        \circ
        \pi_E\left( \scalebox{0.65}{\circimg{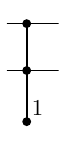}} \right)
        \circ
        \pi_E\left( \scalebox{0.65}{\circimg{tikz-cache/cnot__had.pdf}} \right)
        \\
        &=
        \pi_E\left( \scalebox{0.65}{\circimg{tikz-cache/cnot__rep.pdf}} \right)
    \end{align*}
    In other words, the following equation holds.
    \begin{align*}
        \scalebox{0.65}{\circimg{tikz-cache/cnot__rep.pdf}}
        \in
        \decoder\left( \scalebox{0.65}{\circimg{tikz-cache/cnot__class.pdf}} \right)
    \end{align*}
    Since $H = H^\dagger$, then the following derivation also holds.
    \begin{align*}
        \scalebox{0.65}{\circimg{tikz-cache/cnot__rep.pdf}}
        \overunderset{\HQCCtrlCong}{\HqcFreePhase}{\approx}
        \scalebox{0.65}{\circimg{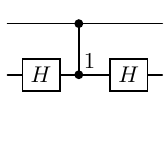}}
        \overunderset{\HQCCtrlCong}{\HqcPowId}{\approx}
        \scalebox{0.65}{\circimg{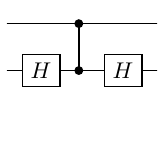}}
        \overset{\HqcCtrlDag}{\approx}
        \scalebox{0.65}{\circimg{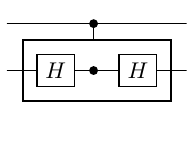}}
        \overunderset{\HQCCtrlCong}{\HqcTargDef}{\approx}
        \scalebox{0.65}{\circimg{tikz-cache/cnot__simp.pdf}}
    \end{align*}
    This completes the proof.
\end{proof}

\subsection{Proof of Lemma~\ref{Lem:Decoder}}
\begin{proof}
    Let $( X, Y ) \in  Q$.
    There are 9 cases to consider.
    \begin{enumerate}
    \item Assume that $( X, Y )$ is rule Q(1).
          Clearly $\scalebox{0.65}{\circimg{tikz-cache/hqc__eqn7_lhs.pdf}} \in \decoder( X )$ and $0 \in \decoder( Y )$.
          It then follows by $\HqcPOrd$ that $\scalebox{0.65}{\circimg{tikz-cache/hqc__eqn7_lhs.pdf}} \approx_E 0$.
    \item Assume that $( X, Y )$ is rule Q(2).
          Clearly $\scalebox{0.65}{\circimg{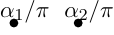}} \in \decoder( X )$ and $\scalebox{0.65}{\circimg{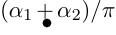}} \in \decoder( Y )$.
          It then follows by $\HqcPowAdd$ that $\scalebox{0.65}{\circimg{tikz-cache/proofs__q2_1.pdf}} \approx_E \scalebox{0.65}{\circimg{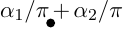}} = \scalebox{0.65}{\circimg{tikz-cache/proofs__q2_3.pdf}}$.
    \item Assume that $( X, Y )$ is rule Q(3).
          Clearly $\sigma \in \decoder( X )$.
          Since $\CatCQC$ is a controlled prop category, then the following equation holds.
          \begin{align*}
              Y
              =
              \scalebox{0.65}{\circimg{tikz-cache/cqc__eqn3_rhs.pdf}}
              &=
              \scalebox{0.65}{\circimg{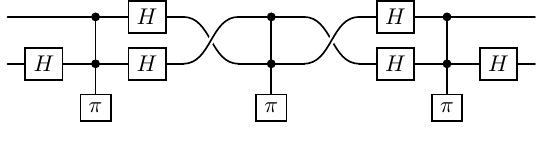}}
              \\
              &=
              \scalebox{0.65}{\circimg{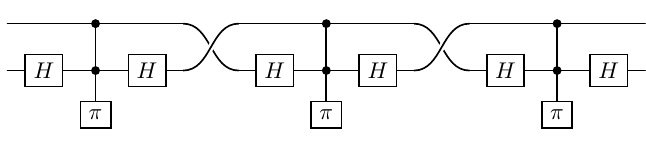}}
          \end{align*}
          Then by \cref{Lemma:CNot}, the following equation also holds.
          \begin{equation*}
              \scalebox{0.65}{\circimg{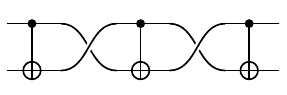}}
              \in
              \decoder( Y )
          \end{equation*}
          Moreover, the following derivation holds.
          \begin{equation*}
              \scalebox{0.65}{\circimg{tikz-cache/proofs__decode_q3_lhs.pdf}}
              \overset{\HqcSwapDef}{\approx}
              \scalebox{0.65}{\circimg{tikz-cache/proofs__decode_q3_mid.pdf}}
              \overset{\HqcCtrlSwap}{\approx}
              \scalebox{0.65}{\circimg{tikz-cache/proofs__decode_q3_rhs.pdf}}
          \end{equation*}
    \item Assume that $( X, Y )$ is rule Q(4).
          Clearly $\scalebox{0.65}{\circimg{tikz-cache/cqc__eqn4_lhs.pdf}} \in \decoder( X )$ and $\scalebox{0.65}{\circimg{tikz-cache/cqc__eqn4_rhs.pdf}} \in \decoder( Y )$.
          Since $H = H^\dagger$, then $\scalebox{0.65}{\circimg{tikz-cache/cqc__eqn4_lhs.pdf}} \approx_E \scalebox{0.65}{\circimg{tikz-cache/cqc__eqn4_rhs.pdf}}$ by \cref{Thm:HQCInv}.
    \item Assume that $( X, Y )$ is rule Q(5).
          Then $\alpha_1 \in \mathbb{R}$, $\alpha_2 \in \mathbb{R}$, and $( \beta_0, \beta_1, \beta_2, \beta_2 ) = \mathsf{Euler}( \alpha_1, \alpha_2 )$.
          Clearly the following equations hold (see~\cref{App:CNOT}).
          \begin{align*}
              \scalebox{0.65}{\circimg{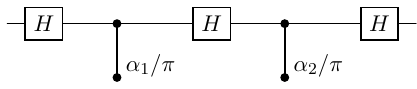}}
              &\in
              \decoder\left( \scalebox{0.65}{\circimg{tikz-cache/cqc__eqn5_lhs.pdf}} \right)
              \\
              \scalebox{0.65}{\circimg{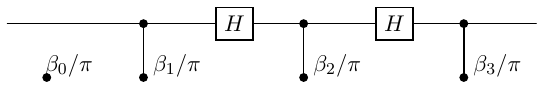}}
              &\in
              \decoder\left( \scalebox{0.65}{\circimg{tikz-cache/cqc__eqn5_rhs.pdf}} \right)
          \end{align*}
          Then the following derivation holds since $H = H^\dagger$.
          \begin{align*}
              \scalebox{0.65}{\circimg{tikz-cache/proofs__euler_1.pdf}}
              \overset{\HqcFreePhase}{\approx}&
              \scalebox{0.65}{\circimg{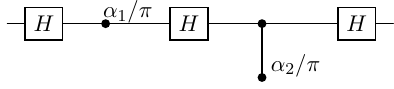}}
              \\
              \overset{\HqcFreePhase}{\approx}&
              \scalebox{0.65}{\circimg{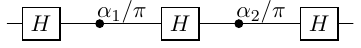}}
              \\
              \overset{\HqcPowDag}{\approx}&
              \scalebox{0.65}{\circimg{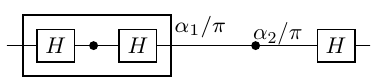}}
              \\
              \overunderset{\HqcPowCong}{\HqcTargMDef}{\approx}&
              \scalebox{0.65}{\circimg{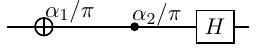}}
              \\
              \overset{\HqcEuler}{\approx}\,&
              \scalebox{0.65}{\circimg{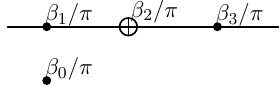}}
              \\
              \overset{\HqcFreePhase}{\approx}&
              \scalebox{0.65}{\circimg{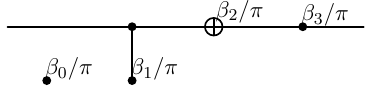}}
              \\
              \overunderset{\HqcPowCong}{\HqcTargMDef}{\approx}&
              \scalebox{0.65}{\circimg{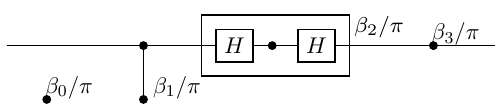}}
              \\
              \overset{\HqcPowDag}{\approx}&
              \scalebox{0.65}{\circimg{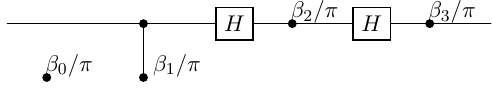}}
              \\
              \overset{\HqcFreePhase}{\approx}&
              \scalebox{0.65}{\circimg{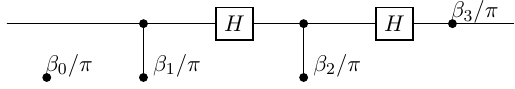}}
              \\
              \overset{\HqcFreePhase}{\approx}&
              \scalebox{0.65}{\circimg{tikz-cache/proofs__euler_11.pdf}}
          \end{align*}
    \item Assume that $( X, Y )$ is rule Q(6).
          As in~\ref{App:CNOT}, the following equations hold where $\beta = \alpha / \pi$.
          \begin{align*}
              \scalebox{0.65}{\circimg{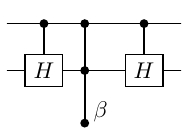}}
              &\in
              \decoder\left( \scalebox{0.65}{\circimg{tikz-cache/cqc__eqn6_lhs.pdf}} \right)
              &
              \scalebox{0.65}{\circimg{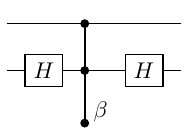}}
              &\in
              \decoder\left( \scalebox{0.65}{\circimg{tikz-cache/cqc__eqn6_rhs.pdf}} \right)
          \end{align*}
          Since $C( - )$ is a conjugated control functor and $( H \boxtimes 0 )^\dagger = H^\dagger \boxtimes 0^\dagger = H \boxtimes 0$, then the following derivation holds.
          \begin{equation*}
              \scalebox{0.65}{\circimg{tikz-cache/proofs__q6_1.pdf}}
              \approx_E
              \scalebox{0.65}{\circimg{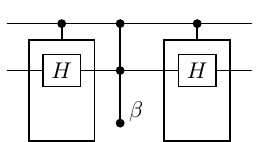}}
              \approx_E
              \scalebox{0.65}{\circimg{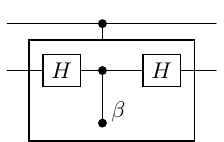}}
              \approx_E
              \scalebox{0.65}{\circimg{tikz-cache/proofs__q6_4.pdf}}
          \end{equation*}
    \item Assume that $( X, Y )$ is rule Q(7).
          As in~\ref{App:CNOT}, the following equations hold where $\beta = \alpha / \pi$.
          \begin{align*}
              \scalebox{0.65}{\circimg{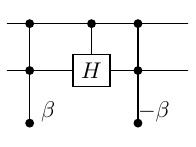}}
              &\in
              \decoder\left( \scalebox{0.65}{\circimg{tikz-cache/cqc__eqn7_lhs.pdf}} \right)
              &
              \scalebox{0.65}{\circimg{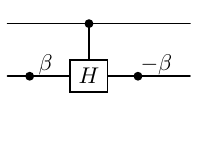}}
              &\in
              \decoder\left( \scalebox{0.65}{\circimg{tikz-cache/cqc__eqn7_rhs.pdf}} \right)
          \end{align*}
          Since $C( - )$ is a conjugated control functor and $( U \exparr \alpha )^\dagger = U \exparr ( -\alpha )$ for each unitary $U$, then the following derivation holds.
          \begin{align*}
              \scalebox{0.65}{\circimg{tikz-cache/proofs__q7_1.pdf}}
              \overunderset{\HQCCtrlCong}{\HqcFreePhase}{\approx}
              \scalebox{0.61}{\circimg{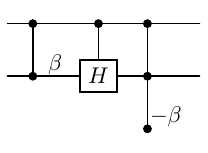}}
              \overunderset{\HQCCtrlCong}{\HqcFreePhase}{\approx}
              \scalebox{0.65}{\circimg{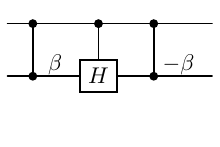}}
              \approx_E
              \scalebox{0.65}{\circimg{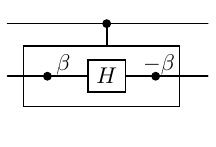}}
              \approx_E
              \scalebox{0.65}{\circimg{tikz-cache/proofs__q7_5.pdf}}
          \end{align*}
    \item Assume that $( X, Y )$ is rule Q(8).
          Clearly the following equations hold (see~\cref{App:CNOT}).
          \begin{align*}
              \scalebox{0.65}{\circimg{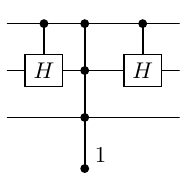}}
              &\in
              \decoder\left( \scalebox{0.65}{\circimg{tikz-cache/cqc__eqn8_lhs.pdf}} \right)
              &
              \scalebox{0.65}{\circimg{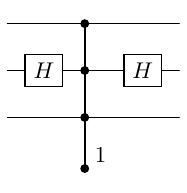}}
              &\in
              \decoder\left( \scalebox{0.65}{\circimg{tikz-cache/cqc__eqn8_rhs.pdf}} \right)
          \end{align*}
          Since $C( - )$ is a conjugated control functor and $( H \boxtimes 1 \boxtimes 0 )^\dagger = H^\dagger \boxtimes 1^\dagger \boxtimes 0^\dagger = H \boxtimes 1 \boxtimes 0$, then the following derivation holds.
          \begin{equation*}
              \scalebox{0.65}{\circimg{tikz-cache/proofs__q8_1.pdf}}
              \approx_E
              \scalebox{0.65}{\circimg{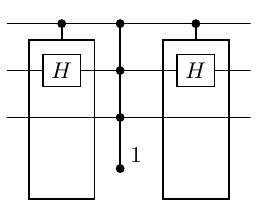}}
              \approx_E
              \scalebox{0.65}{\circimg{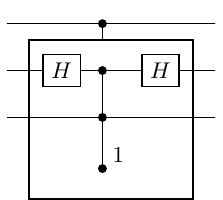}}
              \approx_E
              \scalebox{0.65}{\circimg{tikz-cache/proofs__q8_4.pdf}}
          \end{equation*}
    \item Assume that $( X, Y )$ is rule Q(9).
          Clearly the following equations hold (see~\cref{App:CNOT}).
          \begin{align*}
              \scalebox{0.65}{\circimg{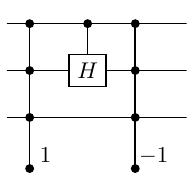}}
              &\in
              \decoder\left( \scalebox{0.65}{\circimg{tikz-cache/cqc__eqn9_lhs.pdf}} \right)
              &
              \scalebox{0.65}{\circimg{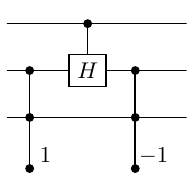}}
              &\in
              \decoder\left( \scalebox{0.65}{\circimg{tikz-cache/cqc__eqn9_rhs.pdf}} \right)
          \end{align*}
          As in the previous proofs, this case will be handled using the property that $C( - )$ is a conjugated control functor.
          First, it must be shown that the third gate on the left-hand side is the adjoint to the first gate on the left-hand side.
          \begin{align*}
              \scalebox{0.65}{\circimg{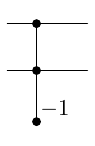}}
              \overunderset{\HQCCtrlCong}{\HqcFreePhase}{\approx}&
              \scalebox{0.65}{\circimg{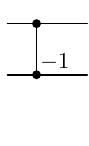}}
              \overset{\text{Thm(\ref{Thm:HQCInv})}}{\approx}
              \scalebox{0.65}{\circimg{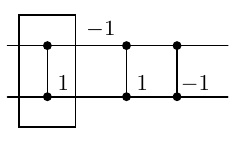}}
              \overset{\HqcCtrlComp}{\approx}
              \scalebox{0.65}{\circimg{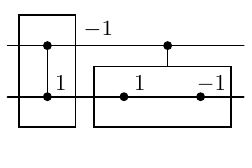}}
              \\
              \overunderset{\HQCCtrlCong}{\HqcPowAdd}{\approx}&
              \scalebox{0.65}{\circimg{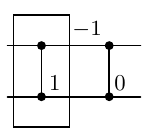}}
              \overunderset{\HQCCtrlCong}{\HqcCtrlId}{\approx}
              \scalebox{0.65}{\circimg{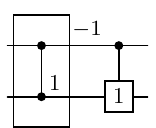}}
              \overset{\HqcCtrlId}{\approx}
              \scalebox{0.65}{\circimg{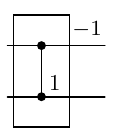}}
          \end{align*}
          Let $U$ denote the final string in this derivation.
          Since $C( - )$ is a conjugated control functor and $C( V )^\dagger = C( V ) \exparr ( -1 )$ for each unitary $V$, then the following derivation holds.
          \begin{align*}
              \scalebox{0.65}{\circimg{tikz-cache/proofs__q9_1.pdf}}
              \overunderset{\HQCCtrlCong}{\HqcFreePhase}{\approx}&
              \scalebox{0.65}{\circimg{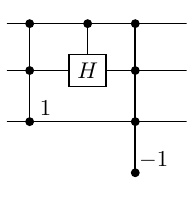}}
              \approx_E
              \scalebox{0.65}{\circimg{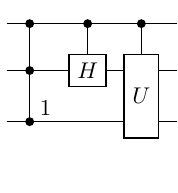}}
              \approx_E
              \scalebox{0.65}{\circimg{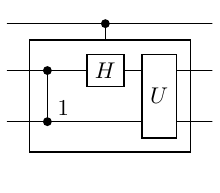}}
              \approx_E
              \scalebox{0.65}{\circimg{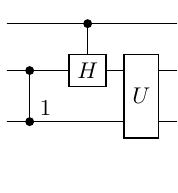}}
              \\
              \approx_E&
              \scalebox{0.65}{\circimg{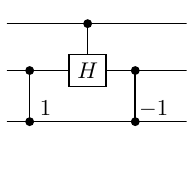}}
              \overunderset{\HQCCtrlCong}{\HqcFreePhase}{\approx}
              \scalebox{0.65}{\circimg{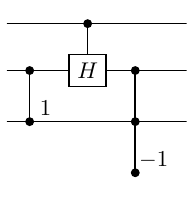}}
              \overunderset{\HQCCtrlCong}{\HqcFreePhase}{\approx}
              \scalebox{0.65}{\circimg{tikz-cache/proofs__q9_8.pdf}}
          \end{align*}
    \end{enumerate}
    In each case, there exists a $U \in \decoder( X )$ and $V \in \decoder( Y )$ such that $U \approx_E V$.
    Since $( X, Y )$ was arbitrary, then this completes the proof.
\end{proof}

\subsection{Proof of Lemma~\ref{Lem:RoundTrip}}
\begin{proof}
    The proof follows by induction.
    \begin{itemize}
    \item \textbf{Base Case}.
          Assume that $g \in \Sigma_{\Core}^0$.
          There are six cases to consider.
          \begin{enumerate}
          \item Assume that $g = \scalebox{0.65}{\circimg{tikz-cache/encoder__eqn6_lhs.pdf}}$.
                Then by definition $\scalebox{0.65}{\circimg{tikz-cache/encoder__eqn6_lhs.pdf}} \in \decoder( \scalebox{0.65}{\circimg{tikz-cache/encoder__eqn6_lhs.pdf}} ) = \decoder( \encoder( \scalebox{0.65}{\circimg{tikz-cache/encoder__eqn6_lhs.pdf}} ) )$.
                Since $( \approx_E )$ is reflexive and $g \in \decoder( \encoder( g ) )$, then there exists some $U \in \decoder( \encoder( g ) )$ such that $g \approx_E U$.
          \item Assume that $g = \scalebox{0.65}{\circimg{tikz-cache/encoder__eqn3_lhs.pdf}}$.
                Then by definition $\scalebox{0.65}{\circimg{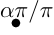}} \in \decoder( \scalebox{0.65}{\circimg{tikz-cache/encoder__eqn3_rhs.pdf}} ) = \decoder( \encoder( \scalebox{0.65}{\circimg{tikz-cache/encoder__eqn3_lhs.pdf}} ) )$.
                Since $\alpha\pi/\pi = \pi$, then $g  \in \decoder( \encoder( g ) )$.
                Since $( \approx_E )$ is also reflexive, then there exists some $U \in \decoder( \encoder( g ) )$ such that $g \approx_E U$.
          \item Assume that $g = \scalebox{0.65}{\circimg{tikz-cache/encoder__eqn1_lhs.pdf}}$.
                Then the following equation holds  (see~\cref{App:CNOT}).
                \begin{align*}
                    \scalebox{0.65}{\circimg{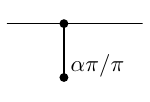}}
                    \in
                    \decoder\left(
                        \scalebox{0.65}{\circimg{tikz-cache/encoder__eqn1_rhs.pdf}}
                    \right)
                    =
                    \decoder\left( \encoder\left(
                        \scalebox{0.65}{\circimg{tikz-cache/encoder__eqn1_lhs.pdf}}
                    \right) \right)
                \end{align*}
                Then the following derivation holds.
                \begin{align*}
                    \scalebox{0.65}{\circimg{tikz-cache/proofs__round_c1_1.pdf}}
                    =
                    \scalebox{0.65}{\circimg{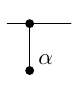}}
                    \overset{\HqcFreePhase}{\approx}
                    \scalebox{0.65}{\circimg{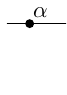}}
                \end{align*}
          \item Assume that $g = \scalebox{0.65}{\circimg{tikz-cache/encoder__eqn4_lhs.pdf}}$.
                It follows from case (3) that following equation holds.
                \begin{align*}
                    \scalebox{0.65}{\circimg{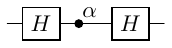}}
                    \in
                    \decoder\left(
                        \scalebox{0.65}{\circimg{tikz-cache/encoder__eqn4_rhs.pdf}}
                    \right)
                    =
                    \decoder\left( \encoder\left(
                        \scalebox{0.65}{\circimg{tikz-cache/encoder__eqn4_lhs.pdf}}
                    \right) \right)
                \end{align*}
                Then the following derivation holds.
                \begin{align*}
                    \scalebox{0.65}{\circimg{tikz-cache/proofs__round_c4_1.pdf}}
                    \overset{\HqcPowDag}{\approx}
                    \scalebox{0.65}{\circimg{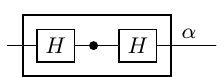}}
                    \overunderset{\HqcPowCong}{\HqcTargDef}{\approx}
                    \scalebox{0.65}{\circimg{tikz-cache/encoder__eqn4_lhs.pdf}}
                \end{align*}
          \item Assume that $g = \scalebox{0.65}{\circimg{tikz-cache/encoder__eqn5_lhs.pdf}}$.
                It follows from cases (3--4) that following equation holds.
                \begin{align*}
                    \scalebox{0.65}{\circimg{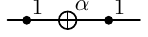}}
                    \in
                    \decoder\left(
                        \scalebox{0.65}{\circimg{tikz-cache/encoder__eqn5_rhs.pdf}}
                    \right)
                    =
                    \decoder\left( \encoder\left(
                        \scalebox{0.65}{\circimg{tikz-cache/encoder__eqn5_lhs.pdf}}
                    \right) \right)
                \end{align*}
                Then the following derivation holds.
                \begin{align*}
                    \scalebox{0.65}{\circimg{tikz-cache/proofs__round_c5_1.pdf}}
                    \overset{\HqcPowId}{\approx}
                    \scalebox{0.65}{\circimg{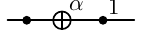}}
                    \overset{\HqcPowId}{\approx}
                    \scalebox{0.65}{\circimg{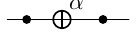}}
                    \overset{\HqcPowDag}{\approx}
                    \scalebox{0.65}{\circimg{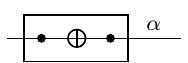}}
                    \overunderset{\HqcPowCong}{\HqcTargMDef}{\approx}
                    \scalebox{0.65}{\circimg{tikz-cache/encoder__eqn5_lhs.pdf}}
                \end{align*}
          \item Assume that $g = \scalebox{0.65}{\circimg{tikz-cache/encoder__eqn2_lhs.pdf}}$.
                It follows from cases (3--4) that following equation holds.
                \begin{align*}
                    \scalebox{0.65}{\circimg{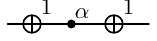}}
                    \in
                    \decoder\left(
                        \scalebox{0.65}{\circimg{tikz-cache/encoder__eqn2_rhs.pdf}}
                    \right)
                    =
                    \decoder\left( \encoder\left(
                        \scalebox{0.65}{\circimg{tikz-cache/encoder__eqn2_lhs.pdf}}
                    \right) \right)
                \end{align*}
                Then the following derivation holds.
                \begin{align*}
                    \scalebox{0.65}{\circimg{tikz-cache/proofs__round_c2_1.pdf}}
                    \overset{\HqcPowId}{\approx}
                    \scalebox{0.65}{\circimg{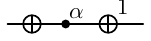}}
                    \overset{\HqcPowId}{\approx}
                    \scalebox{0.65}{\circimg{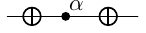}}
                    \overset{\HqcPowDag}{\approx}
                    \scalebox{0.65}{\circimg{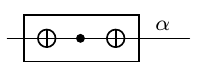}}
                    \overunderset{\HqcPowCong}{\HqcTargMDef}{\approx}
                    \scalebox{0.65}{\circimg{tikz-cache/encoder__eqn2_lhs.pdf}}
                \end{align*}
          \end{enumerate}
          In each case, there exists a $U \in \decoder( \encoder( g ) )$ such that $g \approx_E U$.
    \item \textbf{Inductive Hypothesis}.
          For each $n \in \mathbb{N}$, if $g \in \Sigma_{\Core}^n$, then there exists some circuit $U \in \decoder( \encoder( g ) )$ such that $g \approx_E U$.
    \item \textbf{Inductive Step}.
          Assume that the inductive hypothesis holds for some $n \in \mathbb{N}$ and let $g \in \Sigma_{\Core}^{n+1}$.
          If $g \in \Sigma_{\Core}^n$ as well, then there exists some $U \in \decoder( \encoder( U ) )$ such that $g \approx_E U$ by the inductive hypothesis.
          Assume instead that $g \in \Sigma_{\Core}^{n+1} \setminus \Sigma_{\Core}^n$.
          Then there exists some $V \in \Mor( \Core[n] )$ such that $g = \scalebox{0.65}{\circimg{tikz-cache/dagger__term1.pdf}} \odot V$.
          Then by the definitions of $\encoder( - )$ and $\decoder( - )$, the following equation holds.
          \begin{equation*}
              \decoder( \encoder( g ) )
              =
              \decoder( C( \encoder( V ) ) )
              =
              C( \decoder( \encoder( V ) ) )
          \end{equation*}
          Then by the inductive hypothesis, there exists some $W \in \decoder( \encoder( V ) )$ such that $V \approx_E W$.
          Then the following equations hold  (see~\cref{App:CNOT}).
          \begin{equation*}
              \scalebox{0.65}{\circimg{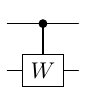}}
              \in
              \decoder\left( \encoder\left(
                \scalebox{0.65}{\circimg{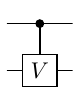}}
              \right) \right)
              \qquad\qquad
              \scalebox{0.65}{\circimg{tikz-cache/proofs__round_ind_1.pdf}}
              \overunderset{\HQCCtrlCong}{I.H.}{\approx}
              \scalebox{0.65}{\circimg{tikz-cache/proofs__round_ind_2.pdf}}
          \end{equation*}
          Then there exists some $U \in \decoder( \encoder( g ) )$ such that $g \approx_E U$
    \end{itemize}
    Then by the principle of inductive, for each $g \in \Sigma_{\Core}$, there exists some $U \in \decoder( \encoder( g ) )$ such that $g \approx_E U$.
\end{proof}

\subsection{Proof of Theorem~\ref{Thm:CoreComplete}}
\begin{proof}
    By \cref{Thm:Soundness}, $( \Sigma_{\Core}, E )$ is sound with respect to $\sem{-}_H$.
    By \cref{Lem:Encoder}, if $g \in \Sigma_{\Core}$, then $\sem{\encoder(g)}_C = \sem{g}_H$.
    By \cref{Lem:Decoder}, if $( X, Y ) \in Q$, then there exists some $X^* \in \decoder( X )$ and $Y^* \in \decoder( Y )$ such that $X^* \approx_E Y^*$.
    By \cref{Lem:RoundTrip}, if $g \in \Sigma_{\Core}$, then there exists some $U \in \decoder( \encoder( g ) )$ such that $g \approx_E U$.
    Then by \cref{Thm:Completeness1}, $( \Sigma_{\Core}, E )$ is complete with respect to the semantic interpretation $\sem{-}_H$.
\end{proof}
\section{Circuit Normalization Routines}
\label{App:Routines}

This section introduces the algorithms used in~\cref{Sect:Completeness}.
The first sub-section establishes the correctness of $\Drop( -, - )$ along with its relevant helper functions.
The second sub-section establishes the correctness of $\Expand( -, - )$ along with its relevant helper functions.
Both cases rely on an oracle function $\Diag( U ) = ( P, \Lambda )$ which returns a diagonalization of $U$.
In particular, if $U \in \Core( n, n )$ and $( P, \Lambda ) = \Diag( U )$, then $P \in \Core( n, n )$ and $\Lambda \in \Core( n, n )$ with $\Lambda$ in diagonal form and $\sem{U}_H = \sem{P^\dagger \circ \Lambda \circ P}_H$.
We note that $\Diag( - )$ need not be computable, and is best thought of as a choice function.
For the interested reader, the underlying choices are elucidated in the proofs of the following theorems.

\begin{remark}
    In this section, we think of each diagonal form $\Lambda$ as a sequence of lambda generators.
    This means that the prefixes of $\Lambda$ are also sequences of lambda generators.
    We write $|\Lambda|$ for the number of lambda generators in $\Lambda$.
    If $\Lambda \in \Core( n, n )$ is the prefix of a diagonal form, then $|\Lambda| \in \{ 0, 1, \ldots, 2^n \}$.
\end{remark}

\begin{proposition}[\cite{BarencoBennett1995,Clement2024Minimal}]
    \label{Prop:Full}
    Let $\mathcal{C}$ be a prop category and $F: \mathcal{C} \to \CatUnitary$ a symmetric monoidal functor such that $\Ob( F )( 1 ) = \mathbb{C}^2$.
    Then $F$ is full if and only if the following conditions hold.
    \begin{enumerate}
    \item For each $\alpha \in (-\pi, \pi]$, there exists a $U \in \mathcal{C}( 0, 0 )$ such that $F( U ) = e^{i\alpha}$.
    \item For each $\alpha \in (-\pi, \pi]$, there exists a $U \in \mathcal{C}( 1, 1 )$ such that $F( U ) = \exp( i \ket{1}\bra{1} \alpha )$.
    \item There exists a $U \in \mathcal{C}( 1, 1 )$ such that $F( U ) = H$.
    \item There exists a $U \in \mathcal{C}( 2, 2 )$ such that $F( U ) = \ket{0} \bra{0} \otimes \mathbb{I}_2 + \ket{1} \bra{1} \otimes X$.
    \end{enumerate}
\end{proposition}

\begin{lemma}
    \label{Lem:Universal}
    The functor $\sem{-}_H$ is full when restricted to $\Core$.
\end{lemma}

\begin{proof}
    By construction, $\sem{-}_H$ is a prop functor such that $\Ob( \sem{-}_H )( 1 ) = \mathbb{C}^2$.
    Then to show that $\sem{-}_H$ is full, it suffices to show that the conditions of \cref{Prop:Full} hold.
    \begin{enumerate}
    \item Let $\alpha \in ( -\pi, \pi ]$.
          Since $\Log( \sem{\scalebox{0.65}{\circimg{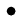}}}_H ) = \Log( e^{i\pi} ) = i\pi$, then $\sem{\scalebox{0.65}{\circimg{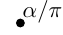}}}_H = e^{i(\alpha/\pi)\pi} = e^{i\alpha}$.
          Since $\alpha$ was arbitrary, then for each $\alpha \in ( -\pi, \pi ]$, there exists $U \in \Core( 0, 0 )$ such that $\sem{U}_H = e^{i\alpha}$.
    \item Let $\alpha \in ( -\pi, \pi ]$.
          Since $\Log( \sem{\scalebox{0.65}{\circimg{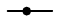}}}_H ) = i \ket{1} \bra{1} \pi$, then the following equation holds.
          \begin{equation*}
              \sem*{\scalebox{0.65}{\circimg{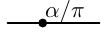}}}_H
              =
              \exp( i ( \alpha / \pi ) \ket{1} \bra{1} \pi )
              =
              \exp( i \ket{1} \bra{1} \alpha )
          \end{equation*}
          Since $\alpha$ was arbitrary, then for each $\alpha \in ( -\pi, \pi ]$, there exists $U \in \Core( 1, 1 )$ such that $\sem{U}_H = \exp( i \ket{1} \bra{1} \alpha )$.
    \item Since $\sem{\scalebox{0.65}{\circimg{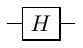}}}_H = H$, then there exists a $U \in \Core( 1, 1 )$ such that $\sem{U}_H = H$.
    \item Clearly $\sem{\scalebox{0.65}{\circimg{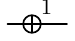}}}_H = X^1 = X$.
          Then by \cref{Thm:GenCtrlNF}, the following equation holds.
          \begin{equation*}
              \sem*{\scalebox{0.65}{\circimg{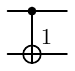}}}_H
              =
              Z \odot X
              =
              \ket{0} \bra{0} \otimes \mathbb{I}_2
              +
              \ket{1} \bra{1} \otimes X
          \end{equation*}
          Then there exists a $U \in \Core( 2, 2 )$ such that $\sem{U} = I \otimes X$.
    \end{enumerate}
    Then $\sem{-}_H$ is full by \cref{Prop:Full}.
\end{proof}

\begin{theorem}
    \label{Thm:Diag}
    There exists a function $\Diag: \Mor( \CatHQC ) \to \Mor( \Core ) \times \Mor( \Core )$ such that for each $U \in \Mor( \CatHQC )$, if $( P, \Lambda ) = \Diag( U )$, then $\sem{U}_H = \sem{P^\dagger \circ \Lambda \circ P}_H$ with $\Lambda$ in diagonal form.
    If $U \in \Mor( \Core )$, then $U \approx_E P^\dagger \circ \Lambda \circ P$.
\end{theorem}

\begin{proof}
    Let $U \in \CatHQC( n, n )$.
    Since $\sem{U}_H$ is unitary (and therefore normal), then by the spectral decomposition theorem there exists a unitary matrix $M \in \CatUnitary( 2^n, 2^n )$ and a diagonal matrix $D \in \CatUnitary( 2^n, 2^n )$ such that $\sem{U}_H = M^\dagger D M$.
    Since $\sem{-}_H$ is full by \cref{Lem:Universal}, there exists a choice of circuit $P \in \Core( n, n )$ such that $\sem{P}_H = M$.
    Since $D$ is diagonal, then for each choice of bijection $f: \{ 1, 2, \ldots, 2^n \} \to \{ 0, 1 \}^n$, there exists a unique diagonal form $\Lambda \in \Core( n, n )$ such that $\sem{\Lambda}_H = D$.
    Then there exists some pair of circuits $( P, \Lambda ) \in \Mor( \Core ) \times \Mor( \Core )$ such that $\sem{U}_H = \sem{P^\dagger \circ \Lambda \circ P}_H$.
    Since $U$ was arbitrary, then for each $U \in \Mor( \CatHQC )$, there exists some $( P, \Lambda ) \in \Mor( \Core ) \times \Mor( \Core)$ such that $\sem{U}_H = \sem{P^\dagger \circ \Lambda \circ P}_H$ with $\Lambda$ in diagonal form.
    Then $\Diag: \Mor( \CatHQC ) \to \Mor( \Core ) \times \Mor( \Core )$ exists.
    Next, assume that $U \in \Core( n, n )$ and let $( P, \Lambda ) = \Diag( U )$.
    Then by assumption $\sem{U}_H = \sem{P^\dagger \circ \Lambda \circ P}_H$.
    Since $P \in \Core( n, n )$ and $\Lambda \in \Core( n, n )$, then $P^\dagger \circ \Lambda \circ P \in \Core( n, n )$.
    Since $( \Sigma_{\Core}, E )$ is complete with respect to the semantic interpretation $\sem{-}_H$ by \cref{Thm:CoreComplete}, then $U \approx_E P^\dagger \circ \Lambda \circ P$.
    Since $U$ was arbitrary, then this completes the proof.
\end{proof}

\subsection{The Drop Routine}

\begin{algorithm}[t]
\caption{$\DiagDrop(\Lambda, \alpha)$}\label{Alg:DiagDrop}
\begin{algorithmic}
\Require $\Lambda \in \CatHQC( k, k )$ in diagonal form and $\alpha \in \mathbb{R}$.
\Ensure A circuit $U \in \Core( k, k )$ such that $U \approx_E \Lambda \exparr \alpha$.
    \If{$|\Lambda| = 0$}
        \Return $1_k$
    \EndIf
    \State $\textbf{let} \; \Lambda' \circ \lambda( x, \beta ) \gets \Lambda$
    \State $\textbf{let} \; U \gets \DiagDrop( \Lambda', \alpha )$
    \State \Return $U \circ \lambda( x, \alpha\beta )$
\end{algorithmic}
\end{algorithm}

\begin{algorithm}[t]
\caption{$\Drop_{n+1}(U, \alpha)$}\label{Alg:Drop}
\begin{algorithmic}
\Require $U \in \CoreExt[n]( k, k )$, and $\alpha \in \mathbb{R}$.
\Ensure A circuit $U \in \Core( k, k )$ such that $U \approx_E U \exparr \alpha$.
    \State $\textbf{let} \; V \gets \Reduce_n( U )$
    \State $\textbf{let} \; ( P, \Lambda ) \gets \Diag( V )$
    \State $\textbf{let} \; W \gets \DiagDrop( \Lambda, \alpha )$
    \State \Return $P^\ddagger \circ W \circ P$
\end{algorithmic}
\end{algorithm}

The special case of diagonal circuits is handled first by~\cref{Alg:DiagDrop}.
The algorithm iterates over each lambda generator in the diagonal form $\Lambda$, and then performs the necessary rewrites to reduce the circuit $( \Lambda \exparr \alpha )$ to $\Core$.
This is then used by~\cref{Alg:Drop} to handle all powers in $\CoreExt[n]$, through subcircuit diagonalization.
Of course, the correctness of $\Drop_{n+1}( -, - )$ relies on the correctness of $\Reduce_n( - )$.

\begin{lemma}
    \label{Lem:DiagDrop}
    If $\Lambda \in \Core( k, k )$ in diagonal form and $\alpha \in \mathbb{R}$, then $\DiagDrop( -, - )$ terminates when applied to $( \Lambda, \alpha )$ with $\DiagDrop( \Lambda, \alpha ) \in \Core( k, k )$ and $( \Lambda \exparr \alpha ) \approx_E \DiagDrop( \Lambda, \alpha )$.
\end{lemma}

\begin{proof}
    The proof follows by induction on $|\Lambda|$ where $\Lambda$ is a prefix of a diagonal form.
    \begin{itemize}
    \item \textbf{Base Case}.
          If $|\Lambda| = 0$, then the first line is reached and $1_k$ is returned.
          In other words, $\DiagDrop( -, - )$ terminates when applied to $( \Lambda, \alpha )$ and $\DiagDrop( \Lambda, \alpha ) = 1_k \in \Core( k, k )$.
          Since $|\Lambda| = 0$, then $\Lambda = 1_k$.
          Then $( \Lambda \exparr \alpha ) = ( 1_k \exparr \alpha ) \approx_E 1_k = \DiagDrop( \Lambda, \alpha )$ by $\HqcPowIdem$.
    \item \textbf{Inductive Hypothesis}.
          For some $\ell \in \mathbb{N}$, if $\Lambda$ is a diagonal form prefix with $|\Lambda| = \ell$, then $\DiagDrop( -, - )$ terminates when applied to $( \Lambda, \alpha )$ with $\DiagDrop( \Lambda, \alpha ) \in \Core( k, k )$ and $( \Lambda \exparr \alpha ) \approx_E \DiagDrop( \Lambda, \alpha )$.
    \item \textbf{Inductive Step}.
          Assume that for some $\ell \in \mathbb{N}$, if $\Lambda$ is a diagonal form prefix with $|\Lambda| = \ell$, then $\DiagDrop( -, - )$ terminates when applied to $( \Lambda, \alpha )$ with $\DiagDrop( \Lambda, \alpha ) \in \Core( k, k )$ and $( \Lambda \exparr \alpha ) \approx_E \DiagDrop( \Lambda, \alpha )$.
          Let $\Lambda$ be a diagonal form prefix with $|\Lambda| = \ell + 1$.
          Since $|\Lambda| = \ell + 1 > 0$, then the if-condition on the first line fails, and the program continues to line two.
          Then after executing line two, $\lambda( x, \beta )$ is assigned to the last symbol in $\Lambda$ and $\Lambda'$ is assigned to the rest of $\Lambda$.
          This means that $\Lambda'$ is also a prefix of a diagonal form with $|\Lambda'| = \ell$.
          Then by the the inductive hypothesis, the third line terminates with $U \gets \DiagDrop( \Lambda', \alpha )$, $U \in \Core( k, k )$, and $(\text{A1}): ( \Lambda' \exparr \alpha ) \approx_E U$.
          Then the fourth line is reached, at which point $U \circ \lambda( x, \alpha\beta )$ is returned.
          In other words, $\Diag( -, - )$ terminates when applied to $ ( \Lambda, \alpha )$ and $\DiagDrop( \Lambda, \alpha ) = U \circ \lambda( x, \alpha\beta ) \in \Core( k, k )$.
          Moreover, the following derivation holds.
          \begin{equation*}
            \scalebox{0.65}{\circimg{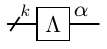}}
            \overset{\HqcPowExpand}{\approx}
            \scalebox{0.65}{\circimg{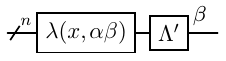}}
            \overunderset{\HqcPowCong}{\text{(A1)}}{\approx}
            \scalebox{0.65}{\circimg{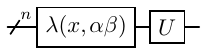}}
          \end{equation*}
          Then $( \Lambda \exparr \alpha ) \approx_E \DiagDrop( \Lambda, \alpha )$ and the inductive step holds.
    \end{itemize}
    It follows by the principle of induction that if $\Lambda \in \Core( k, k )$ is a diagonal form prefix and $\alpha \in \mathbb{R}$, then $\DiagDrop( -, - )$ terminates when applied to $( \Lambda, \alpha )$ with $\DiagDrop( \Lambda, \alpha ) \in \Core( k, k )$ and $( \Lambda \exparr \alpha ) \approx_E \DiagDrop( \Lambda, \alpha )$.
    Since every diagonal form is a prefix of itself, then this completes the proof.
\end{proof}

\begin{theorem}
    \label{Thm:DropDef}
    If $U \in \CoreExt[n]( k, k )$, $\alpha \in \mathbb{R}$, and $\Reduce_n( - )$ terminates when applied to $U$, then $\Drop_{n+1}( -, - )$ terminates when applied to $( U, \alpha )$ with $\Drop_{n+1}( U, \alpha ) \in \Core( k, k )$.
    Moreover, if $U \approx_E \Reduce_n( U )$, then $( U \exparr \alpha ) \approx_E \Drop_{n+1}( U, \alpha )$
\end{theorem}

\begin{proof}
    Let $U \in \CoreExt[n]( k, k )$ and $\alpha \in \mathbb{R}$.
    Assume that $\Reduce_n( - )$ terminates when applied to $U$.
    Then by assumption, the first line terminates and $V \gets \Reduce_n( U )$ with $V \in \Core( k, k )$.
    Then by \cref{Thm:Diag}, the second line terminates and $( P, \Lambda ) \gets \Diag( V )$ with $P \in \Core( k, k )$ and $\Lambda \in \Core( k, k )$ in diagonal form with $V \approx_E P^\dagger \circ \Lambda \circ P$.
    Since $\Lambda$ is in diagonal form, then by \cref{Lem:DiagDrop} the third line also terminates and $W \gets \DiagDrop( \Lambda, \alpha )$ with $W \in \Core( k, k )$ and $(\text{A1}): ( \Lambda \exparr \alpha ) \approx_E W$.
    Then the fourth line is reached, at which point $P^\dagger \circ W \circ P$ is returned. 
    In other words, $\Drop( -, - )$ terminates when applied to $( U, \alpha )$ and $\Drop( U, \alpha ) = P^\ddagger \circ W \circ P$.
    Since $P^\ddagger \in \Core( k, k )$, then $\Drop( U, \alpha ) \in \Core( k, k )$.
    Now assume that $U \approx_E \Reduce_n( U ) = V$.
    Since $V \approx_E P^\dagger \circ \Lambda \circ P$ as well, then $(\text{A2}): U \approx_E P^\dagger \circ \Lambda \circ P$ by the transitivity of $( \approx_E )$.
    Then by \cref{Lem:CoreDag}, $(\text{A3}): P^\ddagger \approx_E P^\dagger$, and the following derivation holds.
    \begin{align*}
        \scalebox{0.65}{\circimg{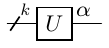}}
        \overunderset{\HqcPowCong}{\text{(A2)}}{\approx}
        \scalebox{0.65}{\circimg{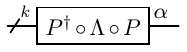}}
        \overset{\HqcPowDag}{\approx}
        \scalebox{0.65}{\circimg{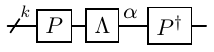}}
        \overunderset{\HqcPowCong}{\text{(A1)}}{\approx}
        \scalebox{0.65}{\circimg{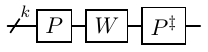}}
        \overunderset{\HqcPowCong}{\text{(A3)}}{\approx}
        \scalebox{0.65}{\circimg{tikz-cache/completeness__drop_5.pdf}}
    \end{align*}
    In conclusion, $( U \exparr \alpha ) \approx_E P^\ddagger \circ W \circ P = \Drop( U, \alpha )$.
\end{proof}

\subsection{The Expand Routine}

\begin{algorithm}[t]
\caption{$\LambdaExpand( x, \alpha, \Gamma )$}\label{Alg:LambdaExpand}
\begin{algorithmic}
\Require $x \in \{ 0, 1 \}^n$, $\alpha \in (-\pi, \pi]$, and $\Gamma \in \Core( m, m )$ the prefix of a diagonal form.
\Ensure A circuit $U \in \Core( n + m, n + m )$ such that $\lambda( x, \alpha ) \odot \Gamma \approx_E U$.
    \If{$|\Gamma| = 0$}
        \Return $1_{n+m}$
    \EndIf
    \State $\Gamma' \circ \lambda( y, \beta ) \gets \Gamma$
    \State $U \gets \LambdaExpand( x, \alpha, \Gamma' )$
    \State \Return $U \circ \lambda( xy, \alpha\beta / \pi )$
\end{algorithmic}
\end{algorithm}

\begin{algorithm}[t]
\caption{$\DiagExpand( \Lambda, \Gamma )$}\label{Alg:DiagExpand}
\begin{algorithmic}
\Require $\Lambda \in \Core( n, n )$ the prefix of a diagonal form and $\Gamma \in \Core( m, m )$ in diagonal form.
\Ensure A circuit $U \in \Core( n + m, n + m )$ such that $\Lambda \odot \Gamma \approx_E U$.
    \If{$|\Lambda| = 0$}
        \Return $1_{n+m}$
    \EndIf
    \State $\Lambda' \circ \lambda( x, \alpha ) \gets \Lambda$
    \State $U \gets \DiagExpand( \Lambda', \Gamma )$
    \State $V \gets \LambdaExpand( x, \alpha, \Gamma )$
    \State \Return $U \circ V$
\end{algorithmic}
\end{algorithm}

\begin{algorithm}[t]
\caption{$\Expand_{n+1}( T )$}\label{Alg:Expand}
\begin{algorithmic}
\Require $T \in \Tower( \CoreExt[n] )$ with $\dom( T ) = k$, $|T| \ge 1$, and $T = T_1 \odot T'$.
\Ensure A circuit $U \in \Core( k, k )$ such that $T \approx_E U$.
    \State $\textbf{let} \; U \gets \Reduce_n( T_1 )$
    \If{$|T| = 1$}
        \Return $U$
    \EndIf
    \State $\textbf{let} \; ( P, \Lambda ) \gets \Diag( U )$
    \State $\textbf{let} \; V \gets \Expand_{n+1}( T' )$
    \State $\textbf{let} \; ( Q, \Gamma ) \gets \Diag( V )$
    \State $\textbf{let} \; W \gets \DiagExpand( \Lambda, \Gamma )$
    \State \Return $( P^\ddagger \boxtimes Q^\ddagger ) \circ W \circ ( P \boxtimes Q )$
\end{algorithmic}
\end{algorithm}

First, ~\cref{Alg:LambdaExpand} is introduced to handle diagonal circuits controlled by lambda generators.
This algorithm iterates over each lambda generator in the diagonal form $\Gamma$, and then performs the necessary rewrites to reduce the circuit $\lambda( x, \alpha ) \odot \Gamma$ to $\Core$.
This is then used by ~\cref{Alg:DiagExpand} to handle the slightly more general case of diagonal circuits controlled by diagonal circuits.
This algorithm iterates over each lambda generator in the diagonal form $\Lambda$, and the performs the necessary rewrites to reduce the circuit $\Lambda \odot \Gamma$ to $\Core$.
This is then used by~\cref{Alg:Expand} to handle all generalized controls in $\CoreExt[n]$, through subcircuit diagonalization.
This algorithm iterates over each subcircuit in the control tower.
Of course, the correctness of $\Expand_{n+1}( - )$ relies on the correctness of $\Reduce_n( - )$.

\begin{lemma}
    \label{Lem:LambdaExpand}
    If $x \in \{ 0, 1 \}^n$, $\alpha \in (-\pi, \pi]$, and $\Gamma$ is in diagonal form, then $\LambdaExpand( -, -, - )$ terminates when applied to the tuple $( x, \alpha, \Gamma )$ with $\LambdaExpand( x, \alpha, \Gamma ) \in \Core( n + m, n + m )$ and $\lambda( x, \alpha ) \odot \Gamma \approx_E \LambdaExpand( x, \alpha, \Gamma )$ where $m = \dom( \Lambda )$.
\end{lemma}

\begin{proof}
    Let $x \in \{ 0, 1 \}^n$ and $\alpha \in (-\pi, \pi]$.
    The proof follows by induction on $|\Gamma|$ where $\Gamma$ is a prefix of a diagonal form.
    \begin{itemize}
    \item \textbf{Base Case}.
          Let $\Gamma \in \Core( m, m )$.
          If $|\Gamma| = 0$, then the first line is reached and $1_{n+m}$ is returned.
          In other words, $\LambdaExpand( -, -, - )$ terminates when applied to $( x, \alpha, \Gamma )$ and $\LambdaExpand( x, \alpha, \Gamma ) = 1_{n+m} \in \Core( n + m, n + m )$.
          Since $|\Gamma| = 0$, then $\Gamma = 1_m$.
          It follows by $\HqcCtrlId$ that $\lambda( x, \alpha ) \odot \Gamma = \lambda( x, \alpha ) \odot 1_m \approx_E 1_{n+m} = \LambdaExpand( x, \alpha, \Gamma )$.
    \item \textbf{Inductive Hypothesis}.
          For some $\ell \in \mathbb{N}$, if $\Gamma \in \Core( m, m )$ is a diagonal form prefix and $|\Gamma| = \ell$, then $\LambdaExpand( -, -, - )$ terminates when applied to $( x, \alpha, \Gamma )$ with $\LambdaExpand( x, \alpha, \Gamma ) \in \Core( n + m, n + m )$ and $\lambda( x, \alpha ) \odot \Gamma \approx_E \LambdaExpand( x, \alpha, \Gamma )$.
    \item \textbf{Inductive Step}.
          Assume that for some $\ell \in \mathbb{N}$, if $\Gamma \in \Core( m, m )$ is a diagonal form prefix and $|\Gamma| = \ell$, then $\LambdaExpand( -, -, - )$ terminates when applied to $( x, \alpha, \Gamma )$ with $\LambdaExpand( x, \alpha, \Gamma ) \in \Core( n + m, n + m )$ and $\lambda( x, \alpha ) \odot \Gamma \approx_E \LambdaExpand( x, \alpha, \Gamma )$.
          Let $\Gamma \in \Core( m, m )$ such that $\Gamma$ is the prefix of a diagonal form with $|\Gamma| = \ell + 1$.
          Since $|\Gamma| = \ell + 1 > 0$, then the if-condition on the first line fails, and the program continues to line two.
          Then after executing line two, $\lambda( y, \beta )$ is assigned to the last symbol in $\Gamma$ and $\Gamma'$ is assigned to the rest of $\Gamma$.
          This means that $\Gamma'$ is a prefix of a diagonal form with $|\Gamma'| = \ell$ and $|y| = m$.
          Then by the inductive hypothesis, the third line terminates and $U \gets \LambdaExpand( x, \alpha, \Gamma' )$ with $U \in \Core( n + m, n + m )$ and $(\text{A1}): \lambda( x, \alpha ) \odot \Gamma' \approx_E U$.
          Then the fourth line is reached, at which point $U \circ \lambda( xy, \alpha\beta / \pi )$ is returned.
          Since $|x| = n$ and $|y| = m$, then $|xy| = n + m$, and therefore $\lambda( xy, \alpha\beta / \pi ) \in \Core( n + m, n + m )$.
          Since $U \circ \lambda( xy, \alpha\beta / \pi )$ was returned, then $\LambdaExpand( -, -, - )$ terminates when applied to $( x, \alpha, \Gamma )$ with $\LambdaExpand( x, \alpha, \Gamma ) \in \Core( n + m, n + m )$.
          Moreover, the following derivation holds.
          \begin{align*}
              \scalebox{0.65}{\circimg{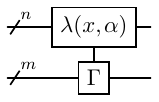}}
              \overset{\HqcCtrlSwap}{\approx}&
              \scalebox{0.65}{\circimg{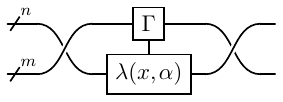}}
              \overset{\HqcCtrlExpand}{\approx}
              \scalebox{0.65}{\circimg{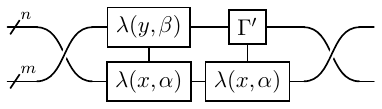}}
              \\
              =\;\;\;&
              \scalebox{0.65}{\circimg{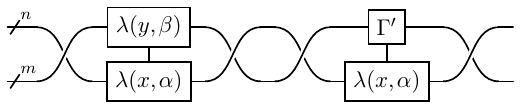}}
              \overset{\HqcCtrlSwap}{\approx}
              \scalebox{0.65}{\circimg{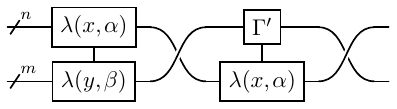}}
              \\
              \overset{\HqcCtrlMerge}{\approx}&
              \scalebox{0.65}{\circimg{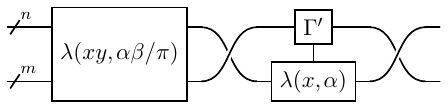}}
              \overset{\HqcCtrlSwap}{\approx}
              \scalebox{0.65}{\circimg{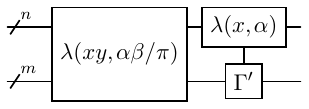}}
              \\
              &\overunderset{\HQCCtrlCong}{(\text{A1})}{\approx}
              \scalebox{0.65}{\circimg{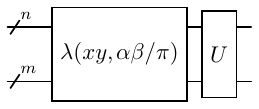}}
          \end{align*}
          Then $\lambda( x, \alpha ) \odot \Gamma \approx_E U \odot \lambda( xy, \alpha\beta / \pi ) = \LambdaExpand( x, \alpha, \Gamma )$ and the inductive step holds.
    \end{itemize}
    It follows by the principle of induction that if $\Gamma \in \Core( m, m )$ is a diagonal form prefix, then $\LambdaExpand( -, -, - )$ terminates when applied to $( x, \alpha, \Gamma )$ with $\LambdaExpand( x, \alpha, \Gamma ) \in \Core( n + m, n + m )$ and $\lambda( x, \alpha ) \odot \Gamma \approx_E \LambdaExpand( x, \alpha, \Gamma )$.
    Since every diagonal form is a prefix of itself, then this establishes that the proof holds for a fixed $x$ and $\alpha$.
    Since $x$ and $\alpha$ were arbitrary, then this completes the proof.
\end{proof}

\begin{lemma}
    \label{Lem:DiagExpand}
    If $\Lambda$ and $\Gamma$ are both in diagonal form, then $\DiagExpand( -, - )$ terminates when applied to $( \Lambda, \Gamma )$ with $\DiagExpand( \Lambda, \Gamma ) \in \Core( k, k )$ and $\Lambda \odot \Gamma \approx_E \DiagExpand( \Lambda, \Gamma )$ where $k = \dom( \Lambda ) + \dom( \Gamma )$.
\end{lemma}

\begin{proof}
    Let $\Gamma \in \Core( m, m )$ in diagonal form and for a given $n \in \mathbb{N}$ write $k = n + m$.
    The proof follows by induction on $|\Lambda|$ where $\Lambda$ is a prefix of a diagonal form.
    \begin{itemize}
    \item \textbf{Base Case}.
          Let $\Lambda \in \Core( n, n )$.
          If $|\Lambda| = 0$, then the first line is reached and $1_{n+m}$ is returned.
          In other words, $\DiagExpand( -, - )$ terminates when applied to $( \Lambda, \Gamma )$ and $\DiagExpand( \Lambda, \Gamma ) = 1_{n+m} \in \Core( k, k )$.
          Since $|\Lambda| = 0$, then $\Lambda = 1_n$ and the following derivation holds.
          \begin{equation*}
              \scalebox{0.65}{\circimg{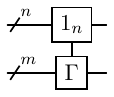}}
              \overset{\HqcCtrlSwap}{\approx}
              \scalebox{0.65}{\circimg{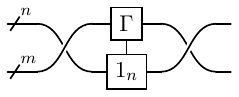}}
              \overset{\HqcCtrlId}{\approx}
              \scalebox{0.65}{\circimg{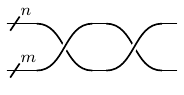}}
              =
              \scalebox{0.65}{\circimg{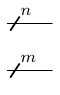}}
          \end{equation*}
          Then $\Lambda \odot \Gamma = 1_n \odot \Gamma \approx_E 1_{n+m} = \DiagExpand( \Lambda, \Gamma )$.
    \item \textbf{Inductive Hypothesis}.
          For some $\ell \in \mathbb{N}$, if $\Lambda \in \Core( n, n )$ is a diagonal form prefix and $|\Lambda| = \ell$, then $\DiagExpand( -, - )$ terminates when applied to $( \Lambda, \Gamma )$ with $\DiagExpand( \Lambda, \Gamma ) \in \Core( k, k )$ and $\Lambda \odot \Gamma \approx_E \DiagExpand( \Lambda, \Gamma )$.
    \item \textbf{Inductive Step}.
          Assume that for some $\ell \in \mathbb{N}$, if $\Lambda \in \Core( n, n )$ is a diagonal form prefix and $|\Lambda| = \ell$, then $\DiagExpand( -, - )$ terminates when applied to $( \Lambda, \Gamma )$ with $\Lambda \odot \Gamma \approx_E \DiagExpand( \Lambda, \Gamma )$ and $\DiagExpand( \Lambda, \Gamma ) \in \Core( k, k )$.
          Let $\Lambda \in \Core( n, n )$ such that $\Lambda$ is the prefix of a diagonal form with $|\Lambda| = \ell + 1$.
          Since $|\Lambda| = \ell + 1 > 0$, then the if-condition on the first line fails, and the program continues to line two.
          Then after executing line two, $\lambda( x, \alpha )$ is assigned to the last symbol in $\Lambda$ and $\Lambda'$ is assigned to the rest of $\Lambda$.
          This means that $\Lambda'$ is a prefix of a diagonal form with $|\Lambda'| = \ell$.
          Then by the inductive hypothesis, the third line terminates and $U \gets \DiagExpand( \Lambda', \Gamma )$ with $(\text{A1}): \Lambda' \odot \Gamma \approx_E U$ and $U \in \Core( k, k )$.
          Then by \cref{Lem:LambdaExpand}, the fourth line terminates and $V \gets \LambdaExpand( x, \alpha, \Gamma )$ with $V \in \Core( k, k )$ and $(\text{A2}): \lambda( x, \alpha ) \odot \Gamma \approx_E V$.
          Then the fourth line is reached, at which point $U \circ V$ is returned.
          That is, $\DiagExpand( -, - )$ terminates when applied to $( \Lambda, \Gamma )$ with $\DiagExpand( \Lambda, \Gamma ) = U \circ V \in \Core( k, k )$.
          Moreover, the following derivation holds.
          \begin{equation*}
              \scalebox{0.65}{\circimg{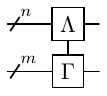}}
              \overset{\HqcCtrlExpand}{\approx}
              \scalebox{0.65}{\circimg{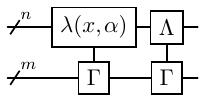}}
              \overunderset{\HQCCtrlCong}{(\text{A1})}{\approx}
              \scalebox{0.65}{\circimg{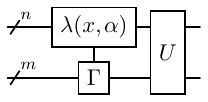}}
              \overunderset{\HQCCtrlCong}{(\text{A2})}{\approx}
              \scalebox{0.65}{\circimg{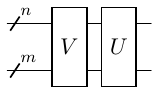}}
          \end{equation*}
          Then $\Lambda \odot \Gamma \approx_E U \circ V \approx_E \DiagExpand( \Lambda, \Gamma )$ and the inductive step holds.
    \end{itemize}
    It follows by the principle of induction that if $\Lambda \in \Core( n, n )$ is a diagonal form prefix, then $\DiagExpand( -, - )$ terminates when applied to $( \Lambda, \Gamma )$ such that $\Lambda \odot \Gamma \approx_E \DiagExpand( \Lambda, \Gamma )$ and $\DiagExpand( \Lambda, \Gamma ) \in \Core( k, k )$.
    Since every diagonal form is a prefix of itself, then this establishes the proof for a fixed $\Gamma$.
    Since $\Gamma$ were arbitrary, then this completes the proof.
\end{proof}

\begin{theorem}
    \label{Thm:ExpandDef}
    Assume that $\Reduce_n( - )$ terminates when applied to each $U \in \Mor( \CoreExt[n] )$.
    If $T \in \Tower( \CoreExt[n] ) \cap \CoreExt[n + 1]( k, k )$, then $\Expand_{n+1}( - )$ terminates when applied to $T$ with $\Expand_{n+1}( U ) \in \Core( k, k )$.
    Moreover, if $U \approx_E \Reduce_n( U )$ for all $U \in \Mor( \CoreExt[n] )$, then $T \approx_E \Expand_{n+1}( T )$.
\end{theorem}

\begin{proof}
    Let $T \in \Tower( \CoreExt[n] )$ with $|T| \ge 1$.
    The proof follows by induction on $|T|$.
    \begin{itemize}
    \item \textbf{Base Case}.
          Let $T \in \Tower( \CoreExt[n] )$ with $k = \dom( T )$ and $|T| = 1$.
          By assumption, the first line terminates and $U \gets \Reduce_n( T_1 )$.
          Since $|T| = 1$, then the if-condition on the second line will succeed and $\Expand_n( - )$ will return $U$.
          In other words, $\Expand_{n+1}( - )$ terminates when applied to $T$ with $\Expand_{n+1}( T ) = U$.
          Moreover, since $|T| = 1$, then $T = T_1$ with $k = \dom( T_1 )$.
          This means that $\Expand_{n+1}( T ) = U \in \Core( k, k )$.
    \item \textbf{Inductive Hypothesis}.
          For some strictly positive integer $\ell$, if $T \in \Tower( \CoreExt[n] )$ with $\dom( T ) = k$ and $|T| = \ell$, then $\Expand_{n+1}( - )$ terminates when applied to $T$ with $\Expand_{n+1}( T ) \in \Core( k, k )$.
    \item \textbf{Inductive Step}.
          Assume that for some strictly positive integer $\ell$, if $T \in \Tower( \CoreExt[n] )$ with $\dom( T ) = k$ and $|T| = \ell$, then $\Expand_{n+1}( - )$ terminates when applied to $T$ with $\Expand_{n+1}( T ) \in \Core( k, k )$.
          Let $T \in \Tower( \CoreExt[n] )$ with $|T| = \ell + 1$ and $k = \dom( T )$.
          Then there exists some $T' \in \Tower( \CoreExt[n] )$ with $|T'| = \ell > 0$ and $T = T_1 \odot T'$.
          Now consider applying $\Expand_{n+1}( - )$ to $T$ and let $k' = \dom( T' )$.
          By assumption, the first line terminates and $U \gets \Reduce_n( T_1 )$ with $U \in \Core( k - k', k - k' )$.
          Since $|T| = \ell + 1 > 1$, then the if-condition on the second line fails, and the program continues to line three. 
          Then by \cref{Thm:Diag}, the third line terminates and $( P, \Lambda ) \gets \Diag( U )$ with $P \in \Core( k - k', k - k' )$ and $\Lambda \in \Core( k - k', k - k' )$ in diagonal form.
          Since $|T'| = \ell > 0$, then by the inductive hypothesis the fourth line terminates and $V \gets \Expand_{n+1}( T' )$ with $V \in \Core( k', k' )$.
          Then by \cref{Thm:Diag}, the fifth line terminates and $( Q, \Gamma ) \gets \Diag( V )$ with $P \in \Core( k', k' )$ and $\Lambda \in \Core( k', k' )$ in diagonal form.
          Then by \cref{Lem:DiagExpand}, the sixth line terminates and $W \gets \DiagExpand( \Lambda, \Gamma )$ with $W \in \Core( k, k )$.
          Then the seventh line is reached, at which point $( P^\ddagger \boxtimes Q^\ddagger ) \circ W \circ ( P \boxtimes Q )$ is returned.
          That is, $\Expand_{n+1}( - )$ terminates when applied to $T$ with $\Expand_{n+1}( T ) = ( P^\ddagger \boxtimes Q^\ddagger ) \circ W \circ ( P \boxtimes Q )$.
          Since $( P^\ddagger \boxtimes Q^\ddagger ) = ( P \boxtimes Q )^\ddagger \in \Core( k, k )$ by definition, then $\Expand_{n+1}( T ) \in \Core( k, k )$.
          Then the inductive step holds.
    \end{itemize}
    Then by the principle of induction, if $T \in \Tower( \CoreExt[n] )$ with $\dom( T ) = k$, then $\Expand_{n+1}( - )$ terminates when applied to $T$ with $\Expand_{n+1}( T ) \in \Core( k, k )$.
    Next, if $U \approx_E \Reduce_n( U )$ for each $U \in \Mor( \CoreExt[n] )$, then it follows by induction on $|T|$ that $T \approx_E \Expand_{n+1}( T )$.
    \begin{itemize}
    \item \textbf{Base Case}.
          Let $T \in \Tower( \CoreExt[n] )$ and $|T| = 1$.
          By the start of this proof, $\Expand_{n+1}( - )$ terminates when applied to $T$.
          Then by the second assumption, $U \gets \Reduce_n( T_1 )$ after the first line with $T_1 \approx_E U$.
          Since $|T| = 1$, then the if-condition on the second line succeeds and $\Expand_n( - )$ is return $U$.
          This means that $T = T_1 \approx_E \Reduce_n( T_1 ) = \Expand_{n+1}( T )$.
    \item \textbf{Inductive Hypothesis}.
          For some strictly positive integer $\ell$, if $T \in \Tower( \CoreExt[n] )$, then $T \approx_E \Expand_{n+1}( T )$.
    \item \textbf{Inductive Step}.
          Assume that for some strictly positive integer $\ell$, if $T \in \Tower( \CoreExt[n] )$ with $|T| = \ell$, then $T \approx_E \Expand_{n+1}( T )$.
          Let $T \in \Tower( \CoreExt[n] )$ with $|T| = \ell + 1$.
          Then there exists some $T' \in \Tower( \CoreExt[n] )$ with $|T| = \ell$ and $T = T_1 \odot T'$.
          Now consider applying $\Expand_{n+1}( - )$ to $T$.
          From the start of this proof, $\Expand_{n+1}( - )$ terminates when applied to $T$.
          By assumption, $U \gets \Reduce_n( T_1 )$ after the first line with $T_1 \approx_E U$.
          Since $|T| = \ell + 1 > 1$, then the if-condition on the second line fails, and the program continues to line three. 
          Then by \cref{Thm:Diag}, $( P, \Lambda ) \gets \Diag( U )$ after the third line with $U \approx_E P^\dagger \circ \Lambda \circ P$.
          Since $T_1 \approx_E U$, then $(\text{A1}): T_1 \approx_E P^\dagger \circ \Lambda \circ P$ by the transitivity of $( \approx_E )$.
          Since $|T'| = \ell$, then by the inductive hypothesis, $V \gets \Expand_{n+1}( T' )$ after the fourth line with $T' \approx_E V$.
          Then by \cref{Thm:Diag}, $( Q, \Gamma ) \gets \Diag( V )$ after the fifth line with $V \approx_E Q^\dagger \circ \Gamma \circ Q$.
          Since $T' \approx_E V$, then $(\text{A2}): T' \approx_E Q^\dagger \circ \Gamma \circ Q$ by the transitivity of $( \approx_E )$.
          Then by \cref{Lem:DiagExpand}, $W \gets \DiagExpand( \Lambda, \Gamma )$ after the sixth line with $(\text{A3}): \Lambda \odot \Gamma \approx_E W$.
          Then the seventh line is reached, at which point $( P^\ddagger \boxtimes Q^\ddagger ) \circ W \circ ( P \boxtimes Q )$ is returned.
          Since $( P^\ddagger \boxtimes Q^\ddagger ) = ( P \boxtimes Q )^\ddagger$ by definition, then by \cref{Lem:CoreDag}, $(\text{A4}): ( P^\dagger \boxtimes Q^\dagger ) = ( P \boxtimes Q )^\dagger \approx_E ( P \boxtimes Q )^\ddagger = ( P^\ddagger \boxtimes Q^\ddagger )$.
          Then the following derivation holds.
          \begin{align*}
              \scalebox{0.65}{\circimg{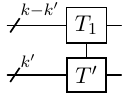}}
              \overunderset{\HQCCtrlCong}{\text{(A2)}}{\approx}&
              \scalebox{0.65}{\circimg{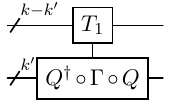}}
              \overset{\HqcCtrlDag}{\approx}
              \scalebox{0.65}{\circimg{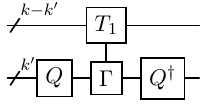}}
              \overset{\HqcCtrlSwap}{\approx}
              \scalebox{0.65}{\circimg{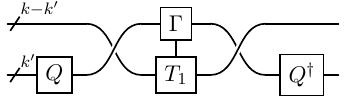}}
              \\
              \overunderset{\HQCCtrlCong}{\text{(A1)}}{\approx}&
              \scalebox{0.65}{\circimg{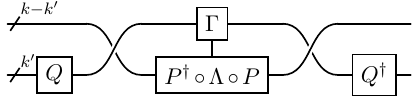}}
              \overset{\HqcCtrlDag}{\approx}
              \scalebox{0.65}{\circimg{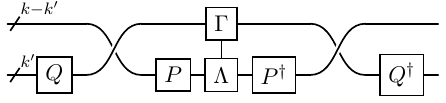}}
              \\
              =\;\;\;&
              \scalebox{0.65}{\circimg{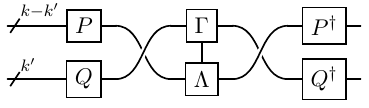}}
              \overset{\HqcCtrlSwap}{\approx}
              \scalebox{0.65}{\circimg{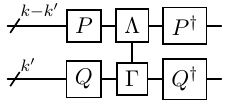}}
              \overunderset{\HQCCtrlCong}{\text{(A3)}}{\approx}
              \scalebox{0.65}{\circimg{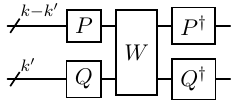}}
              \\
              \overunderset{\HQCCtrlCong}{\text{(A4)}}{\approx}&
              \scalebox{0.65}{\circimg{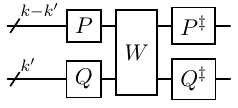}}
          \end{align*}
          Then $T \approx_E ( P^\ddagger \boxtimes Q^\ddagger ) \circ W \circ ( P \boxtimes Q ) = \Expand_{n+1}( T )$ and the inductive step holds.
    \end{itemize}
    Then by the principle of induction, if $T \in \Tower( \CoreExt[n] )$, then $T \approx_E \Expand_{n+1}( T )$.
\end{proof}

\section{Proof of Theorem~\ref{Thm:HQCComplete}}

\begin{proof}
    First, a partitioning will be constructed for $\Sigma_{\CoreExt}^0$.
    Since $\Sigma_{\CoreExt}^0 \subseteq \Sigma_{\CatHQC}$, then the following equation holds by \cref{Lem:FuncDefPart}.
    \begin{equation*}
        \Sigma_{\CoreExt}^0
        =
        \Sigma_{\CoreExt}^0 \cap \Sigma_{\CatHQC}
        =
        \left( \Sigma_{\CoreExt}^0 \cap \Sigma_{\textbf{Prim}} \right)
        \sqcup
        \left( \Sigma_{\CoreExt}^0 \cap \Sigma_{\textbf{Ctrl}} \right)
        \sqcup
        \left( \Sigma_{\CoreExt}^0 \cap \Sigma_{\textbf{Pow}} \right)
    \end{equation*}
    Next, it must be shown that $r_n( - )$ is well-defined for each $n \in \mathbb{N}$.
    This means that $r_n( - )$ assigns each $g \in \Sigma_{\CoreExt}^n$ to a unique word $r_n( g )$ with $r_n( g ) \in \Mor( \Core )$.
    The proof follows by induction on $n$.
    \begin{itemize}
    \item \textbf{Base Case}.
          Let $g \in \Sigma_{\CoreExt}^0$.
          Then there are three cases to consider.
          \begin{enumerate}
          \item Assume that $g \in \Sigma_{\CoreExt}^0 \cap \Sigma_{\textbf{Prim}} = \Sigma_{\textbf{Prim}} \subseteq \Sigma_{\Core}^0$.
                Since $r_0( g ) = g \exparr 1$ by definition, then $r_0( g ) \in \Power( \Sigma_{\Core}^0 ) \subseteq \Sigma_{\Core}^1 \subseteq \Mor( \Core )$.
                Then $r_0( g )$ is well-defined.
          \item Assume that $g \in \Sigma_{\CoreExt}^0 \cap \Sigma_{\textbf{Ctrl}}$ or $g \in \Sigma_{\CoreExt}^0 \cap \Sigma_{\textbf{Pow}}$.
                Then $g \not \in \Sigma_{\CoreExt}^0 \cap \Sigma_{\textbf{Prim}}$.
                Since $\Sigma_{\CoreExt}^0 = \Sigma_{\Core} \cup \Sigma_{\textbf{Prim}}$, then $g \in \Sigma_{\Core} \subseteq \Mor( \Core )$
                Since $r_0( g ) = g$ by definition, then $r_0( g )$ is well-defined.
          \end{enumerate}
          In each case, $r_0( g )$ is well-defined.
          Since $g$ was arbitrary, then $r_0( - )$ is well-defined.
    \item \textbf{Inductive Hypothesis}.
          For some $n \in \mathbb{N}$, $r_n: \Sigma_{\CoreExt}^n \to U( \Core )$ is well-defined.
    \item \textbf{Inductive Step}.
          Assume that for some $n \in \mathbb{N}$, $r_n: \Sigma_{\CoreExt}^n \to U( \Core )$ is well-defined.
          Then there exists a well-defined functor $\Reduce_n = P( r_n )$.
          In particular this means that $\Reduce_n( - )$ terminates when applied to each $U \in \Mor( \CoreExt[n] )$.
          Let $g \in \Sigma_{\CoreExt}^{n+1}$.
          There are three cases to consider.
          \begin{enumerate}
          \item Assume that $g \in \Sigma_{\CoreExt}^n$.
                By assumption, $r_n( - )$ is well-defined, which means that $r( g )$ is unique with $r( g ) \in \Mor( \Core )$.
                Since $r_{n+1}( - )$ assigns $g$ to $r_n( g ) \in \Mor( \Core )$, then $r_{n+1}( g )$ is well-defined.
          \item Assume $g \in \Tower( \CoreExt[n] ) \cap \CoreExt[n+1]( k, k )$ for some $k \in \mathbb{N}$.
                Since $\Reduce_n( - )$ terminates when applied to each $U \in \Mor( \CoreExt[n] )$, then it follows from \cref{Thm:ExpandDef} that $\Expand_{n+1}( - )$ terminates when applied to $g$ with $\Expand_{n+1}( g ) \in \Core( k, k )$.
                Since $r_{n+1}( - )$ assigns $g$ to $\Expand_{n+1}( g )$, then $r_{n+1}( g )$ is well-defined.
          \item Assume that $g \in \Power( \CoreExt[n] ) \cap \CoreExt[n+1]( k, k )$ for some $k \in \mathbb{N}$.
                Then there exists a $U \in \CoreExt[n]( k, k )$ and $\alpha \in \mathbb{R}$ such that $g = U \exparr \alpha$.
                Since $\Reduce_n( - )$ terminates when applied to each $U$, then it follows from \cref{Thm:DropDef} that $\Drop_{n+1}( -, - )$ terminates when applied to $( U, \alpha )$ with $\Drop_{n+1}( U, \alpha ) \in \Core( k, k )$.
                Since $r_{n+1}( - )$ assigns $g$ to $\Drop_{n+1}( U, \alpha )$, then $r_{n+1}( g )$ is well-defined.
          \end{enumerate}
          In each case, $r_{n+1}( g )$ is well-defined.
          Since $g$ was arbitrary, then $r_{n+1}( - )$ is well-defined.
          Then the inductive step holds.
    \end{itemize}
    The by the principle of induction, each $r_n( - )$ is well-defined.
    Since $\bigcup_{n=0}^\infty \Sigma_{\CoreExt}^n = \Sigma_{\CatHQC}$, then this defines a signature morphism $r: \Sigma_{\CatHQC} \to U( \Core )$ such that $r( g ) = r_n ( g )$ for each $n \in \mathbb{N}$ and $g \in \Sigma_{\CoreExt}^n$.
    Define $\Reduce = P( r )$.
    It remains to be shown that if $g \in \Sigma_{\CatHQC}$, then $\Reduce( g ) \approx_E g$.
    The proof follows by induction on $\CoreExt[n] \hookrightarrow \CatHQC$.
    \begin{itemize}
    \item \textbf{Base Case}.
          Let $g \in \Sigma_{\CoreExt}^0$.
          Then there are two cases to consider.
          \begin{enumerate}
          \item Assume that $g \in \Sigma_{\CoreExt}^0 \cap \Sigma_{\textbf{Ctrl}}$.
                Then $r_0( g ) = g \exparr 1$.
                Then $r_0( g ) \approx_E g$ by \HqcPowId.
          \item Assume that $g \in \Sigma_{\CoreExt}^0 \cap \Sigma_{\textbf{Ctrl}}$ or $g \in \Sigma_{\CoreExt}^0 \cap \Sigma_{\textbf{Pow}}$.
                Then $r_0( g ) = g$ by definition.
                Since $( \approx_E )$ is reflexive, then $r_0( g ) \approx_E g$.
          \end{enumerate}
          In each case, $\Reduce( g ) = r_0( g ) \approx_E g$.
          Then $\pi_E( \Reduce( g ) ) = \pi_E( g )$.
          Since $g$ was arbitrary, then $\pi_E \circ \Reduce \circ i = \pi_E \circ i$ where $i: \Sigma_{\CoreExt}^0 \to \CoreExt[0]$.
          Then by the universal property of free prop categories, $\pi_E \circ \Reduce = \pi_E$ when restricted to $\CoreExt[0]$.
          Let $U \in \Mor( \CoreExt[0] )$.
          Then $\pi_E( \Reduce( U ) ) = \pi_E( U )$, which implies that $\Reduce( U ) \approx_E U$.
          Since $U$ was arbitrary, then $\Reduce( U ) \approx_E U$ for all $U \in \Mor( \CoreExt[0] )$.
    \item \textbf{Inductive Hypothesis}.
          For some $n \in \mathbb{N}$, if $U \in \Mor( \CoreExt[n] )$, then $\Reduce( U ) \approx_E U$.
    \item \textbf{Inductive Step}.
          Assume that for some $n \in \mathbb{N}$, if $U \in \Mor( \CoreExt[n] )$, then the derivation $\Reduce( U ) \approx_E U$ holds.
          Let $g \in \Sigma_{\CoreExt}^{n+1}$.
          There are three cases to consider.
          \begin{enumerate}
          \item Assume that $g \in \Sigma_{\CoreExt}^n$.
                Then $\Reduce( g ) \approx_E g$ by the inductive hypothesis.
          \item Assume that $g \in \Tower( \CoreExt[n] )$.
                Then $\Expand( g ) = r_{n+1}( g ) = r( g ) = \Reduce( g )$.
                By the inductive hypothesis, $U \approx_E \Reduce( U ) = \Reduce_n( U )$ for each $U \in \Mor( \CoreExt[n] )$.
                It then follows from \cref{Thm:ExpandDef} that $g \approx_E \Expand( g ) = \Reduce( g )$.
          \item Assume that $g \in \Power( \CoreExt[n] )$.
                Then there exists some $U \in \Mor( \CoreExt[n] )$ and $\alpha \in \mathbb{R}$ such that $g = U \exparr \alpha$.
                Then by definition $\Drop( U, \alpha ) = r_{n+1}( g ) = r( g ) = \Reduce( g )$.
                Then by the inductive hypothesis, $U \approx_E \Reduce( U ) = \Reduce_n( U )$.
                It then follows from \cref{Thm:DropDef} that $g = U \exparr \alpha \approx_E \Drop( U, \alpha ) = \Reduce( g )$.
          \end{enumerate}
          In each case, $\Reduce( g ) \approx_E g$, which implies that $\pi_E( \Reduce_{n+1}( g ) ) = \pi_E( g )$.
          Since $g$ was arbitrary, then $\pi_E \circ \Reduce \circ i = \pi_E \circ i$ for $i: \Sigma_{\CoreExt}^{n+1} \hookrightarrow \CoreExt[n+1]$.
          Then by the universal property of free prop categories, $\pi_E \circ \Reduce = \pi_E$ when restricted to $\CoreExt[n+1]$.
          Let $U \in \CoreExt[n+1]$.
          Then $\pi_E( \Reduce( U ) ) = \pi_E( U )$.
          Then $\Reduce( U ) \approx_E U$.
          Since $U$ was arbitrary, then $\Reduce_{n+1}( U ) \approx_E U$ for all $U \in \CoreExt[n+1]$.
          Then the inductive step holds.
    \end{itemize}
    Then by the principle of induction, $\Reduce( U ) \approx_E U$ for each $n \in \mathbb{N}$ and $U \in \Mor( \CoreExt[n] )$.
    This implies that $\Reduce( g ) \approx_E g$ for each $g \in \Sigma_{\CatHQC}$.
    Let $g \in \Sigma_{\CatHQC}$.
    Since $\Sigma_{\CatHQC} = \bigcup_{n=0}^\infty \Sigma_{\CoreExt}^n$, then there exists an $n \in \mathbb{N}$ such that $g \in \Sigma_{\CoreExt}^n \subseteq \Mor( \CoreExt[n] )$.
    Then $\Reduce( g ) \approx_E g$.
    Since $g$ was arbitrary, then $\Reduce( g ) \approx_E g$ for each $g \in \Sigma_{\CatHQC}$ with $\Reduce: \CatHQC \to \Core$.
    Moreover, $( \Sigma_{\Core}, E )$ is complete with respect to $\sem{-}_H$ by \cref{Thm:CoreComplete}.
    Then $( \Sigma_{\CatHQC}, E )$ is complete with respect to $\sem{-}_H$ by \cref{Thm:Completeness2}.
\end{proof}

\section{Redundant Relations in the HQC Equational Theory}
\label{App:Redundancies}

Two presentations $( \Sigma, E )$ and $( \Gamma, Q )$ are equivalent if $P( \Sigma ) / E \cong P( \Gamma ) / Q$ as prop categories.
It can be shown that $( \Sigma, E )$ and $( \Gamma, Q )$ are equivalent if and only if $( \Gamma, Q )$ can be obtained from $( \Sigma, E )$ via a sequence of Tietze transformations~(see \cite{Clerc2015} for a similar case).
The transformations are as follow.
\begin{itemize}
\item \textbf{Removing a Generator}.
      Let $f \in \Sigma$, $\Gamma = \Sigma \setminus \{ f \}$, and $g \in P( \Gamma )( \dom( f ), \cod( f ) )$.
      If $f \approx_E g$, then $( \Sigma, E )$ is equivalent to $( \Gamma, Q )$ where $Q$ is obtained from $E$ by replacing every instance of $f$ by $g$.
\item \textbf{Removing a Relation}.
      Let $( f, g ) \in E$ and $Q = E \setminus \{ ( f, g ) \}$.
      If $f \approx_{Q} g$, then $( \Sigma, E )$ is equivalent to $( \Sigma, Q )$.
\end{itemize}
These transformations can then be used to prove that specific generators and relations in $( \Sigma_{\CatHQC}, E )$ are redundant.

Using these transformations, this section derives several minimal generating sets for $\CatHQC / E$.
It is well-known that all single-qubit circuits can be generated (up-to global phase) using any two of the three generators $\{ X( \theta ), Z( \theta ), H \}$.
This corresponds to single-qubit hierarchical circuits generated from
$\{
        \scalebox{0.65}{\circimg{tikz-cache/dagger__term1.pdf}},
        \scalebox{0.65}{\circimg{tikz-cache/dagger__term3.pdf}},
        \scalebox{0.65}{\circimg{tikz-cache/dagger__term5.pdf}}
\}$.
Regardless of the choice of generators, it is then possible to find circuits $U$ and $V$ such that $\sem{U}_H = Z$ and $\sem{V}_H = X$.
The circuit $U \odot V$ can then act as a controlled-not gate, and all multi-qubit circuits can be generated (up-to global phase).
Finally, the global phase gate can be used to generate all multi-qubit circuits exactly, as described in~\cref{Prop:Full}.
These minimal gate sets can be obtained via Tietze transformations.

First, it will be shown that $g_1 = \scalebox{0.65}{\circimg{tikz-cache/dagger__term4.pdf}}$ can be eliminated from $( \Sigma_{\CatHQC}, E )$.
First, note that $g_1$ appears as the left-hand side of \HqcTargMDef, and does not appear on the right-hand side of \HqcTargMDef.
This means that $g_1$ and \HqcTargMDef can be removed from $( \Sigma_{\CatHQC}, E )$ via a Tietze transformation.
Moreover, since $g_1$ does not appear in any other relations, then the relations in $E$ need not be modified.

Next, it will be shown that $g_2 = \scalebox{0.65}{\circimg{tikz-cache/dagger__term2.pdf}}$ can be eliminated from $( \Sigma_{\CatHQC}, E )$.
As in the case of $g_1$, there exists a relation \HqcOPhaseDef which can be used to eliminate $g_2$ via a Tietze transformation.
However, $g_2$ also appears in \HqcCtrlPair, and must therefore be modified.
To simplify the process, we will manually replace \HqcCtrlPair before removing $g_2$.
Observe that the following derivation holds for each $U \in \CatHQC( n, n )$, since $H$ is self-adjoint.
\begin{equation*}
    \scalebox{0.65}{\circimg{tikz-cache/hqc__eqn15_lhs.pdf}}
    \overunderset{\HQCCtrlCong}{\HqcOPhaseDef}{=}
    \scalebox{0.65}{\circimg{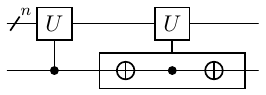}}
    \overset{\HqcCtrlDag}{=}
    \scalebox{0.65}{\circimg{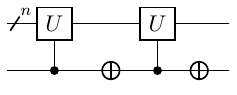}}
\end{equation*}
Using the derivation, it is then possible to introduce a new relation \text{E'(13)}, which can then be used to eliminate \HqcCtrlPair.
\begin{equation*}
    \scalebox{0.65}{\circimg{tikz-cache/hqc__eqn15_lhs_alt.pdf}}
    \overset{\text{E'(13)}}{=}
    \scalebox{0.65}{\circimg{tikz-cache/hqc__eqn15_rhs.pdf}}
\end{equation*}
Since $g_2$ does not appear in any instance of \text{E'(13)}, then $g_2$ can be removed using \HqcOPhaseDef without any further modifications to the relation set.
This yields a new presentation, $( \Gamma, Q )$ where $\Gamma$ and $Q$ are defined as follows.
\begin{align*}
    \Gamma &= \{
        \scalebox{0.65}{\circimg{tikz-cache/dagger__term1.pdf}},
        \scalebox{0.65}{\circimg{tikz-cache/dagger__term3.pdf}},
        \scalebox{0.65}{\circimg{tikz-cache/dagger__term5.pdf}},
        \;\scalebox{0.65}{\circimg{tikz-cache/dagger__term8.pdf}}
    \}
    &
    Q &= E \setminus \{
        \HqcTargMDef, \HqcOPhaseDef, \HqcCtrlPair
    \} \cup \{
        \text{E'(13)}
    \}
\end{align*}

It remains to be shown that $\Gamma$ can be reduced to the global phase gate with only two of the three single-qubit gates.
Before proceeding, it will be shown that \HqcHDef is redundant, and can be removed from the presentation.
In principle, \HqcHDef can be derived via several applications of \HqcEuler, though this process is tedious.
Instead, it should be noted that in the proof of completeness, \HqcHDef is used precisely once, to prove that the Hadamard gate is self-inverse.
Then, it suffices to prove that the Hadamard gate is self-inverse without the use of \HqcHDef.
It then follows from completeness, together with the soundness of $E$, that \HqcHDef is derivable from $Q \setminus \{ \HqcHDef \}$.
As a first step, \HqcEuler is used to derive an alternative definition for the Hadamard gate.
\begin{equation}
    \scalebox{0.65}{\circimg{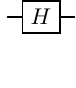}}
    \overset{\HqcPowZero}{=}
    \scalebox{0.65}{\circimg{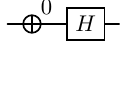}}
    \overset{\HqcPowZero}{=}
    \scalebox{0.65}{\circimg{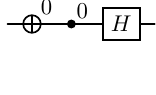}}
    \overset{\HqcEuler}{=}
    \scalebox{0.65}{\circimg{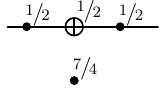}}
    \label{Eqn:HadDef}
\end{equation}
This alternative definition can then be used to prove that the Hadamard gate is self-inverse.
\begin{align*}
    \scalebox{0.65}{\circimg{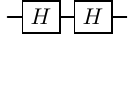}}
    \overset{\text{\cref{Eqn:HadDef}}}{=}&
    \scalebox{0.65}{\circimg{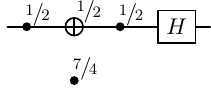}}
    \overset{\HqcEuler}{=}
    \scalebox{0.65}{\circimg{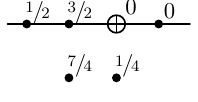}}
    \overset{\HqcPowZero}{=}
    \scalebox{0.65}{\circimg{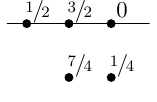}}
    \overset{\HqcPowZero}{=}
    \scalebox{0.65}{\circimg{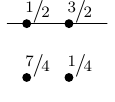}}
    \\
    \overset{\HqcPowAdd}{=}&
    \scalebox{0.65}{\circimg{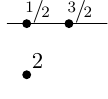}}
    \overset{\HqcPowAdd}{=}
    \scalebox{0.65}{\circimg{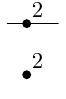}}
    \overset{\HqcPOrd}{=}
    \scalebox{0.65}{\circimg{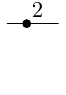}}
    \overset{\HqcFreePhase}{=}
    \scalebox{0.65}{\circimg{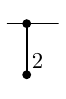}}
    \overunderset{\HQCCtrlCong}{\HqcPOrd}{=}
    \scalebox{0.65}{\circimg{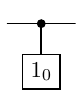}}
    \overset{\HqcCtrlId}{=}
    \scalebox{0.65}{\circimg{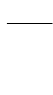}}
\end{align*}
In conclusion, $( \Gamma, Q )$ is equivalent to $( \Gamma, R )$ where $R = Q \setminus \{ \HqcHDef \}$.
It remains to be shown that $\Gamma$ can be further reduced.

\begin{figure}[t]
    \input{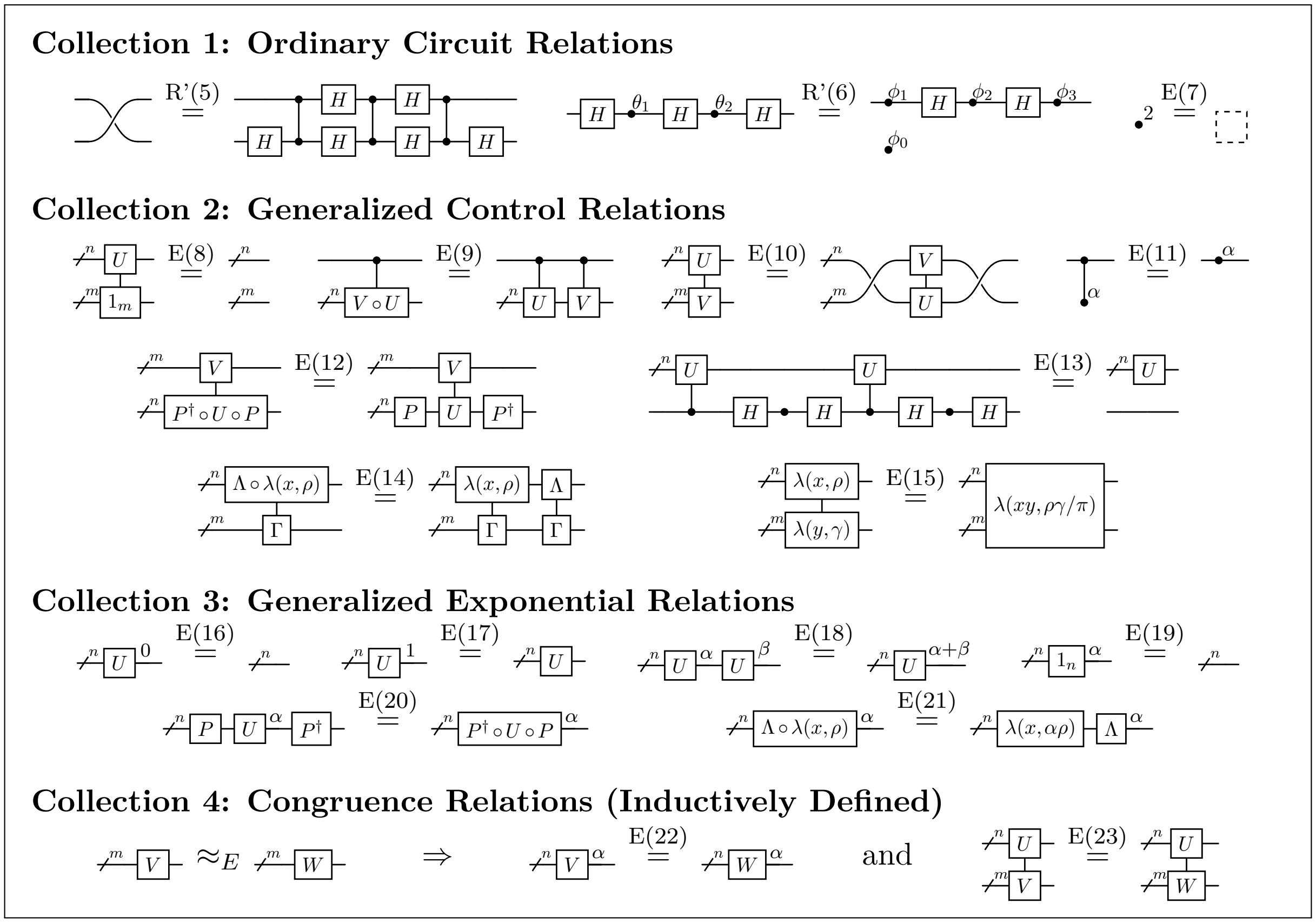}
    \caption{An alternative equational theory for $\CatHQC$, using only the generators in $\Gamma_X$.}
    \label{Fig:RXRelations}
\end{figure}

\vspace{1em}
\noindent
\textbf{Eliminating the Pauli-X Gate}.
Let $g_X = \scalebox{0.65}{\circimg{tikz-cache/dagger__term3.pdf}}$.
Since $g_X$ appears as the left-hand side of \HqcTargDef, and does not appear on the right-hand side of \HqcTargDef, then $g_X$ and \HqcTargDef can be removed via a Tietze transformation. 
Unfortunately, $g_X$ still appears in relations $\{ \HqcSwapDef, \HqcEuler, \text{E'(13)} \}$, meaning that some care must be taken before $g_X$ can be fully removed from $( \Gamma, R )$.
To simplify the process, we will manually replace $\{ \HqcSwapDef, \HqcEuler, \text{E'(13)} \}$ before removing $g_X$.
This is possible since $H$ is self-adjoint.
To replace \HqcSwapDef, the following derivation suffices.
\begin{align*}
    \scalebox{0.65}{\circimg{tikz-cache/proofs__decode_q3_mid.pdf}}
    \overset{\HqcCtrlSwap}{=}&
    \scalebox{0.65}{\circimg{tikz-cache/proofs__decode_q3_rhs.pdf}}
    \overunderset{\HQCCtrlCong}{\HqcTargDef}{=}
    \scalebox{0.65}{\circimg{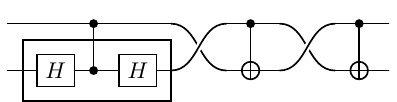}}
    \\
    \overset{\HqcCtrlDag}{=}&
    \scalebox{0.65}{\circimg{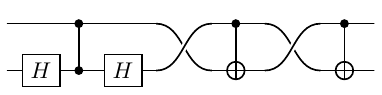}}
    \overunderset{\HQCCtrlCong}{\HqcTargDef}{=}
    \scalebox{0.65}{\circimg{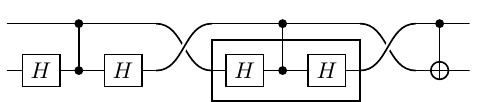}}
    \\
    \overset{\HqcCtrlDag}{=}&
    \scalebox{0.65}{\circimg{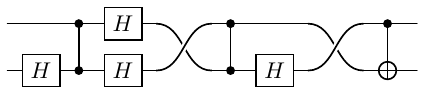}}
    \overunderset{\HQCCtrlCong}{\HqcTargDef}{=}
    \scalebox{0.65}{\circimg{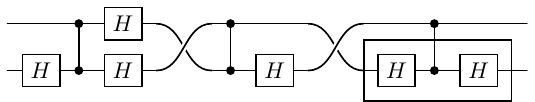}}
    \\
    \overset{\HqcCtrlDag}{=}&
    \scalebox{0.65}{\circimg{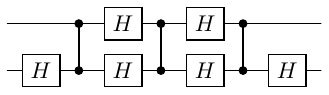}}
\end{align*}
To replace \HqcEuler, the following derivations suffice.
\begin{align*}
    \scalebox{0.65}{\circimg{tikz-cache/hqc__euler_lhs.pdf}}
    &\overunderset{\HqcPowCong}{\HqcTargDef}{=}
    \scalebox{0.65}{\circimg{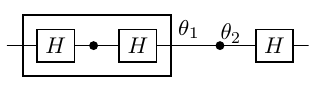}}
    \overset{\HqcPowDag}{=}
    \scalebox{0.65}{\circimg{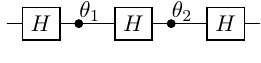}}
    \\
    \scalebox{0.65}{\circimg{tikz-cache/hqc__euler_rhs.pdf}}
    &\overunderset{\HqcPowCong}{\HqcTargDef}{=}
    \scalebox{0.65}{\circimg{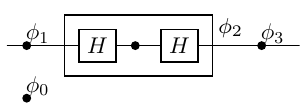}}
    \overset{\HqcPowDag}{=}
    \scalebox{0.65}{\circimg{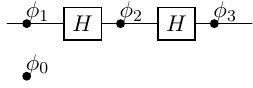}}
\end{align*}
To replace \text{E'(13)}, the following derivation suffices.
\begin{align*}
    \scalebox{0.65}{\circimg{tikz-cache/hqc__eqn15_lhs_alt.pdf}}
    \overset{\HqcTargDef}{=}
    \scalebox{0.65}{\circimg{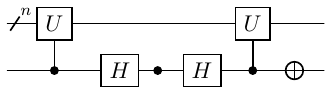}}
    \overset{\HqcTargDef}{=}
    \scalebox{0.65}{\circimg{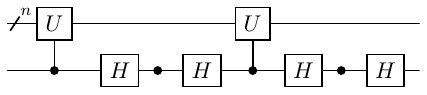}}
\end{align*}
Using these derivations, relations $\{ \text{R'(5)}, \text{R'(6)}, \text{R'(13)} \}$ can be introduced (see~\cref{Fig:RXRelations}), which can ten be used to eliminate $\{ \HqcSwapDef, \HqcEuler, \text{E'(13)} \}$.
Since $g_X$ does not appear in any instance of \text{E'(5)} or \text{E'(6)}, then $g_X$ can be removed using \HqcTargDef without any further modifications to the relation set.
This yields a new presentation, $( \Gamma_X, R_X )$ where $\Gamma_X$ and $R_X$ are defined as follows.
\begin{align*}
    \Gamma &= \{
        \scalebox{0.65}{\circimg{tikz-cache/dagger__term1.pdf}},
        \scalebox{0.65}{\circimg{tikz-cache/dagger__term5.pdf}},
        \;\scalebox{0.65}{\circimg{tikz-cache/dagger__term8.pdf}}
    \}
    &
    R_X &= R \setminus \{
        \HqcTargDef,
        \HqcSwapDef,
        \HqcEuler
    \} \cup \{
        \text{E'(5)},
        \text{E'(6)}
    \}
\end{align*}
The relations in $Q_X$ are depicted in \cref{Fig:RXRelations}.

\vspace{1em}
\noindent
\textbf{Eliminating the Pauli-Z Gate}.
Let $g_Z = \scalebox{0.65}{\circimg{tikz-cache/dagger__term1.pdf}}$.
A-priori, there does not exist a relation to eliminate $g_Z$.
However, the following derivation can be used to introduce such a relation, via a Tietze transformation.
\begin{align*}
    \scalebox{0.65}{\circimg{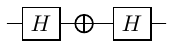}}
    \overset{\HqcTargDef}{=}
    \scalebox{0.65}{\circimg{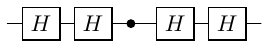}}
    \approx_E
    \scalebox{0.65}{\circimg{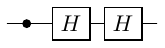}}
    \approx_E
    \scalebox{0.65}{\circimg{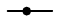}}
\end{align*}
Since the generalized control relations in $( \Sigma_{\CatHQC}, E )$ are written in terms of the $X$-basis, then obtaining concise relations for $\Gamma \setminus \{ g_Z \}$ is somewhat tedious.
For this reason, the equational theory has been omitted.
Intuitively, this theory corresponds to writing the generalized control relations with respect to the $Z$-basis, as opposed to the $X$-basis.

\begin{figure}[t]
    \input{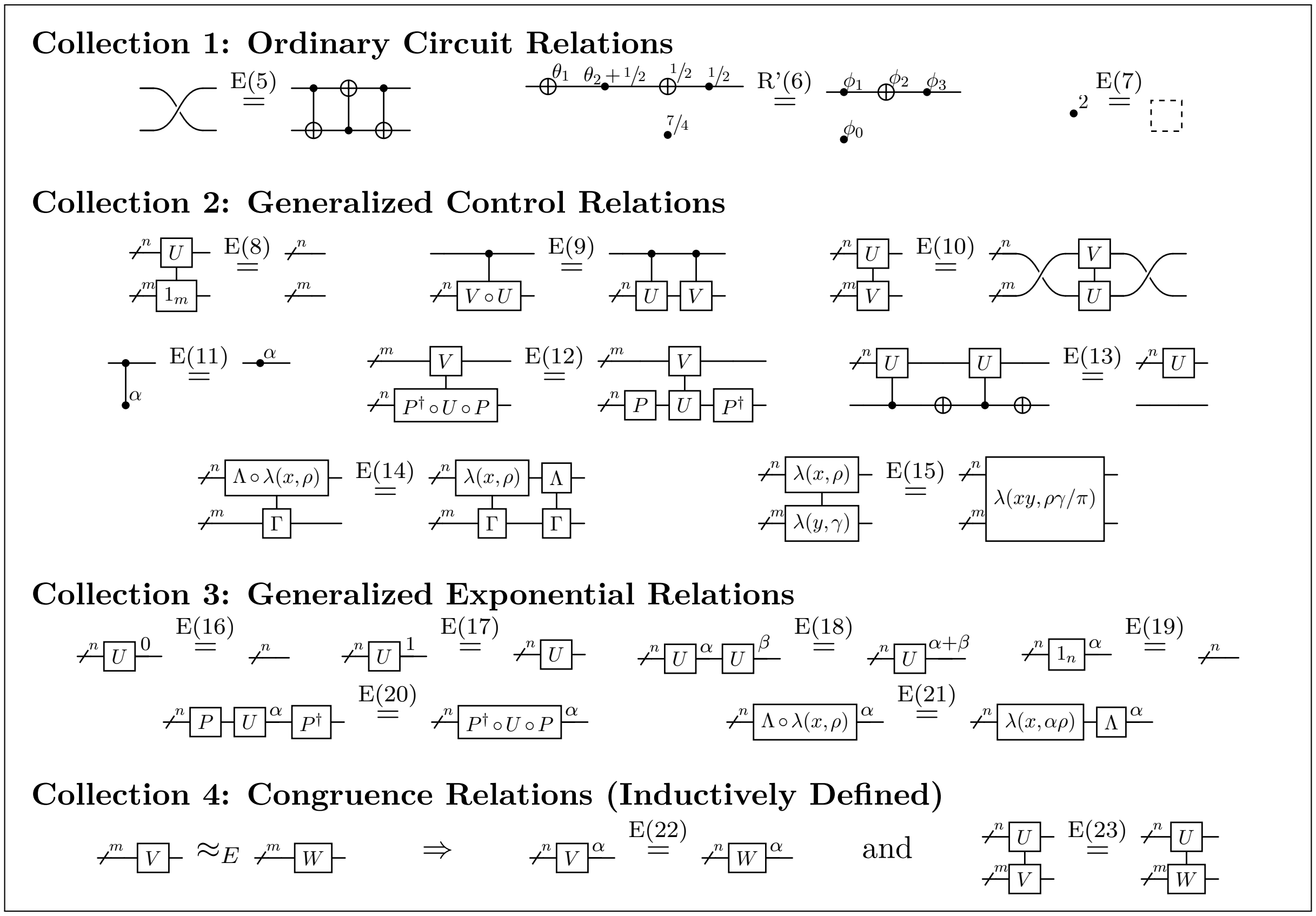}
    \caption{An alternative equational theory for $\CatHQC$, using only the generators in $\Gamma_H$.}
    \label{Fig:RHRelations}
\end{figure}

\vspace{1em}
\noindent
\textbf{Eliminating the Hadamard Gate}.
Let $g_H = \scalebox{0.65}{\circimg{tikz-cache/dagger__term5.pdf}}$.
To eliminate $g_H$, it is necessary to first introduce \cref{Eqn:HadDef} as a relation \text{R'(1)} via a Tietze transformation.
Since $g_H$ appears as the left-hand side of \text{R'(1)}, and does not appear on the right-hand side of \text{R'(1)}, then $g_H$ and \text{R'(1)} can be eliminated via a Tietze transformation. 
Unfortunately, $g_H$ still appears in relations $\{ \HqcTargDef, \HqcEuler \}$, meaning that some care must be taken before $g_H$ can be fully removed from $( \Gamma, R )$.
In turns out that the relation \HqcTargDef is redundant, as illustrated by the following derivation.
\begin{align*}
    \scalebox{0.65}{\circimg{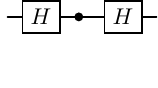}}
    \overset{\text{R'(1)}}{=}&
    \scalebox{0.65}{\circimg{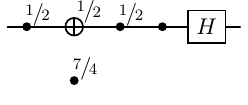}}
    \overset{\HqcPowZero}{=}
    \scalebox{0.65}{\circimg{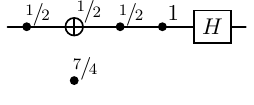}}
    \overset{\HqcPowAdd}{=}
    \scalebox{0.65}{\circimg{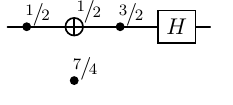}}
    \\
    \overset{\HqcEuler}{=}&
    \scalebox{0.65}{\circimg{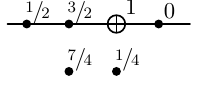}}
    \overset{\HqcPowZero}{=}
    \scalebox{0.65}{\circimg{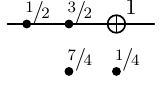}}
    \overset{\HqcPowAdd}{=}
    \scalebox{0.65}{\circimg{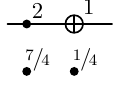}}
    \overset{\HqcPowAdd}{=}
    \scalebox{0.65}{\circimg{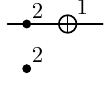}}
    \\
    \overset{\HqcPOrd}{=}&
    \scalebox{0.65}{\circimg{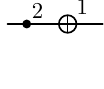}}
    \overset{\HqcPowId}{=}
    \scalebox{0.65}{\circimg{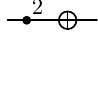}}
    \overset{\HqcFreePhase}{=}
    \scalebox{0.65}{\circimg{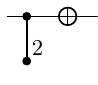}}
    \overunderset{\HQCCtrlCong}{\HqcPOrd}{=}
    \scalebox{0.65}{\circimg{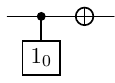}}
    \overset{\HqcCtrlId}{=}
    \scalebox{0.65}{\circimg{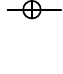}}
\end{align*}
To replace \HqcEuler, the following derivation suffices.
\begin{equation*}
    \scalebox{0.65}{\circimg{tikz-cache/hqc__euler_lhs.pdf}}
    \overset{\text{R'(1)}}{=}
    \scalebox{0.65}{\circimg{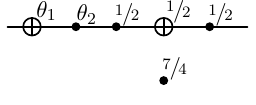}}
    \overset{\HqcPowAdd}{=}
    \scalebox{0.65}{\circimg{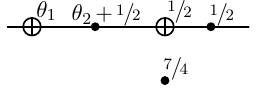}}
\end{equation*}
Using this derivation, it is then possible to introduce a new relation \text{R'(6)} as in \cref{Fig:RHRelations}, which is then used to eliminate \HqcEuler.
Since \HqcTargDef has been eliminated and $g_H$ does not appear in any instance of \text{R'(6)}, then $g_H$ can be removed using \text{R'(1)} without any further modifications to the relation set.
This yields a new presentation, $( \Gamma_H, R_H )$ where $\Gamma_H$ and $R_H$ are defined as follows.
\begin{align*}
    \Gamma &= \{
        \scalebox{0.65}{\circimg{tikz-cache/dagger__term1.pdf}},
        \scalebox{0.65}{\circimg{tikz-cache/dagger__term3.pdf}},
        \;\scalebox{0.65}{\circimg{tikz-cache/dagger__term8.pdf}}
    \}
    &
    R_H &= R \setminus \{
        \HqcTargDef,
        \HqcEuler
    \} \cup \{
        \text{E'(6)}
    \}
\end{align*}
The relations in $Q_H$ are depicted in \cref{Fig:RHRelations}.

\end{document}